\documentclass[aps,prd,reprint,longbibliography,twocolumn,amsmath,amssymb,amsfonts,showpacs,superscriptaddress]{revtex4-1}%

\usepackage{ifpdf}
\usepackage{dcolumn}
\usepackage{enumerate}
\usepackage{enumitem}
\usepackage{bm}
\usepackage{dsfont}
\usepackage{here}
\usepackage{float}
\floatstyle{plaintop}
\restylefloat{table}
\ifpdf

      \usepackage[pdftex]{graphicx}  
      \usepackage[pdftex]{epsfig}

\usepackage[usenames]{color}
\definecolor{navyblue}{rgb}{0.0, 0.0, 0.5}
\definecolor{ferrarired}{rgb}{1.0, 0.11, 0.0}
\definecolor{persianblue}{rgb}{0.11, 0.22, 0.73}

      \usepackage[pdftex]{hyperref}
\hypersetup{
		colorlinks=true,
		citecolor=blue,%
		linkcolor=ferrarired,%
		urlcolor=persianblue,%
        filecolor=blue,
        linktoc=page
	}

\else

      \usepackage[dvips]{graphicx}  
      \usepackage[dvips]{epsfig}

      \usepackage[dvipdfm]{hyperref}
\hypersetup{
		colorlinks=true,
		citecolor=blue,%
		linkcolor=ferrarired,%
		urlcolor=persianblue,%
        filecolor=blue,
        linktoc=page
	}

\fi

\usepackage{xfrac}

\usepackage{amsmath, amsfonts}

\DeclareMathAlphabet{\mathpzc}{OT1}{pzc}{m}{it}

\usepackage{tablefootnote}

\begin{document}

\title{Scattering and conversion of electromagnetic and gravitational waves by Reissner-Nordström black holes: The Regge pole description}

\author{Mohamed \surname{Ould~El~Hadj}}

\email{med.ouldelhadj@gmail.com}

\affiliation{No current affiliation, France}
%
%

\date{\today}

\begin{abstract}

We investigate the problem of scattering and conversion of monochromatic planar gravitational and electromagnetic waves impinging upon a Reissner-Nordström black hole using a Regge pole description, i.e., a complex angular momentum approach. For this purpose, we first compute numerically the Regge pole spectrum for various charge-to-mass ratio configurations. We then derive an asymptotic expressions for the lowest Regge poles, and by considering Bohr-Sommerfeld-type quantization conditions, obtain the spectrum of weakly damped quasinormal frequencies from the Regge trajectories. Next, we construct the scattering and conversion amplitudes as well as the total differential cross sections for different processes using both a complex angular momentum representation and a partial wave expansion method.  Finally, we provide an analytical approximation of the scattering and conversion cross sections of different processes from asymptotic expressions for the lowest Regge poles and the associated residues based on the correspondence Regge poles, ``surface waves'' propagating close to the photon (graviton) sphere. This allows us to extract the physical interpretation encoded in the partial wave expansions in the high-frequency regime (i.e., in the short-wavelength regime), and to describe semiclassically with very good agreement both black hole glory and a large part of the orbiting oscillations, thus unifying these two phenomena from a purely wave point of view.

\end{abstract}

\maketitle

\tableofcontents

\section{Introduction}
\label{sec_1}

The study of perturbations of charged spherically symmetric black holes described by the
Reissner-Nordström metric, an electrovacuum solution of the Einstein-Maxwell equations, has aroused great interest and was widely studied in the $1970$s (see, e.g.,~\cite{Zerilli:1974ai, Moncrief-1974a,Moncrief-1974b,Olson:1974nk,Moncrief-1975, Chitre:1975ew,Chandrasekhar:1979iz} and references therein). It was shown that electromagnetic and gravitational perturbations are nontrivially coupled, suggesting that in a strong electromagnetic field, gravitational waves can be converted into electromagnetic waves of the same frequency and vice versa, involving the existence of the conversion cross section. In other words, an intense electromagnetic field can act as a ``catalyst'' in this process. Many authors have contributed to the study of this conversion process from different aspects (see, \emph{e.g},~\cite{Johnston:1974vf,Gerlach:1974zz,Sibgatullin:1974jq,Matzner:1976kj,Fabbri:1977,DeLogi:1977qe,DeLogi:1977qe,
Breuer:1981kd,Gunter:1980,Castillo:1987,Castillo:1996jm}).

At the origin of this enthusiasm for this purely classical (i.e., nonquantum) phenomenon, is the pioneering work
of Gertsenshte\u{\i}n  in 1961~\cite{gertsenshtein1962wave}, which was later expanded upon by Zel'dovich in a $1973$ paper~\cite{zel1973electromagnetic}. The Gertsenshte\u{\i}n-Zel'dovich (GZ) effect involves the conversion of an electromagnetic wave into a gravitational wave as it passes through a strong magnetic field in a curved space, and vice versa. Recently, the GZ effect has sparked renewed interest. Despite being extremely weak, it could be responsible for generating distortions in the cosmic microwave background (CMB)~\cite{Domcke:2020yzq} and thus have significant consequences for the early universe~\cite{Marklund:1999sp, Dolgov:2012be, Fujita:2020rdx}. Furthermore, it is considered as a potential model for explaining the origin of fast radio bursts (FRBs)~\cite{Kushwaha:2022twx}.

In this article, we extend our previous results in the framework of the ``\emph{Reggeization project}'' of black hole physics by investigating the problem of monochromatic planar gravitational waves and electromagnetic waves impinging upon a Reissner-Nordström black hole using complex angular momentum (CAM) techniques (analytic continuation of partial wave expansions in the CAM plane, effective resummation of $S$-matrix poles (i.e., the so-called Regge poles) and associated residues, the semiclassical interpretation of Regge pole contributions,\ldots). Typically, the scattering of various waves (scalar, electromagnetic, and gravitational) by Reissner-Nordström black holes is studied using partial wave expansions~\cite{Castillo:1996jm,Crispino:2009ki,Crispino:2014eea,Crispino:2015gua} (see also Ref.~\cite{Cotaescu:2016aty,Sporea:2017zxe} for scattering of massless Dirac fermions and references therein). Our previous work~\cite{OuldElHadj:2021fqi} also examined the scattering and conversion differential cross sections of an incident planar wave on a Reissner-Nordström black hole, using both the partial-wave method and a (numerical) geometric-optics approximation. Specifically, we focused on the differential cross section for converting an incoming electromagnetic wave (EW) into an outgoing gravitational wave (GW), which is identical to the cross section for the inverse process.

Although the partial wave expansion approach is a natural method for studying scattering problems involving black
holes (BHs)~\cite{Futterman:1988ni}, it has certain limitations. In particular, it can suffer from a lack of convergence due to the long range nature of the fields propagating on a Reissner-Nordström black hole. Additionally, interpreting physical results described in terms of partial wave expansions can be rather difficult in general. One can circumvent these issues by employing the CAM approach. This has been highlighted in our previous articles dealing with the scattering of scalar, electromagnetic and gravitational waves by the Schwarzschild BH using CAM techniques~\cite{Folacci:2019cmc,Folacci:2019vtt} (see also~\cite{Folacci:2018sef} where the gravitational radiation has
been described by CAM). We have also extended it to the scattering problem by compact objects~\cite{OuldElHadj:2019kji} and dirty BHs~\cite{Torres:2022fyf}.

The CAM approach has already shown its interpretive power in BH physics. In the case of the Schwarzschild BH, this approach has enabled us to obtain analytical approximations for both the glory and orbiting oscillations using asymptotic expansions for the Regge poles and associated residues in the Regge pole sums~\cite{Folacci:2019cmc,Folacci:2019vtt}. These expansions are physically related to the excitation of surface waves and diffraction effects due to the Schwarzschild photon (graviton) sphere in the short-wavelength regime~\cite{Andersson1994bis,Decanini2003,Decanini2010,Dolan:2009,Decanini2010bis}. Traditionally, glory and orbiting
oscillations scattering are considered as two distinct effects and described by different formulas, but we have shown that
it is possible to describe them with a unique analytical formula. Additionally, Decanini~\emph{et
al.}~\cite{Decanini2003,Decanini2010} have demonstrated that the complex frequencies of weakly damped quasinormal modes
(QNMs) are Breit-Wigner-type resonances generated by surface waves near the photon (graviton) sphere, and that the spectrum of complex QNM frequencies has been constructed using Regge trajectories, thus establishing an intuitive interpretation of
Schwarzschild BH QNMs (see Refs.~\cite{Decanini:2009,Dolan:2009,Decanini2010bis,Decanini:2011,Folacci:2021uld} for results concerning other BHs and massive fields). Furthermore, in~\cite{Decanini:2011,Decanini:2011bis,Dolan:2012}, an analytical description in terms of Regge trajectories was given for the absorption cross sections of BHs endowed with a photon (graviton) sphere in the high frequency regime. This has allowed for an explanation of the oscillations of absorption cross sections in terms of the orbital period and Lyapunov exponent of the null unstable geodesics lying on the photon (graviton) sphere.

Our article is organized as follows. In Sec.~\ref{sec_2}, we review the theory of electromagnetic and gravitational
waves on Reissner-Nordström spacetime. Here, we recall the radial wave equations governing the gravitational and electromagnetic perturbation-types (Sec.~\ref{sec_2_1}) and define the $S$-matrix elements of the different processes using the appropriate physical boundary conditions (Sec.~\ref{sec_2_2}). In Sec.~\ref{sec_3}, we focus on the poles of the $S$-matrix elements and boundary conditions. Here, we describe the numerical algorithm  and we present the new numerical results of Regge pole (RP) spectrum for $Q<M$ (Sec.~\ref{sec_3_1}) and for maximally charged  Reissner-Nordström BH case ($Q=M$) (Sec.~\ref{sec_3_2}). By using the Dolan-Ottewill approach~\cite{Dolan:2009,Dolan:2011fh}, we derive an asymptotic expressions for the lowest RPs and by considering Bohr-Sommerfeld-type quantization conditions, we derive the spectrum of weakly damped QNMs frequencies from the Regge trajectories (Sec.~\ref{sec_3_3}). In Sec.~\ref{sec_4}, we construct the CAM representation of the scattering and conversion amplitudes and cross sections. Here, we review the partial wave expansion method for the scattering and conversion amplitudes and cross sections of different processes (\ref{sec_4_1}), we construct the exact CAM representation by using the Sommerfeld-Watson transform~\cite{Newton:1982qc,Watson18,Sommerfeld49} and Cauchy’s residue theorem from their partial wave expansions (\ref{sec_4_2}). We provide a purely analytical approximation of the scattering and conversion cross sections of different processes based on the asymptotic expressions of the RPs and their associated residues. In Sec.~\ref{sec_5}, we outline the numerical computational method (\ref{sec_5_1}), and we display a selection of numerical results of the scattering and conversion cross sections of the different processes (\ref{sec_5_2}). We show the asymptotic results describing with very good agreement both the Reissner-Nordström BH glory and a large part of the orbiting oscillations (\ref{sec_5_3}). We conclude with a discussion in Sec.~\ref{sec_6}.

Throughout this article, we adopt units such that $G=c=4\pi\epsilon_0=1$. We furthermore consider that the exterior of the Reissner-Nordström BH is defined by the line element~\cite{Chandrasekhar:1985kt},
\[ds^2 = -f(r) dt^2 + f(r)^{-1}dr^2 +r^2 \left(d\theta^2+\sin^2\theta d\varphi^2\right),\]
where $f(r) = 1-2M/r+Q^2/r^2 = \left(1-r_+/r\right)\left(1-r_-/r\right)$, with $r_\pm = M \pm \sqrt{M^2-Q^2}$. Here, $r_-$ and $r_+$  represent respectively  the inner (Cauchy) and outer (event) horizons of the black hole, while $M$ and $Q$ denote its mass and charge. We finally assume a time dependence $\text{exp}(-i\omega t)$ for the plane monochromatic waves considered.

\section{Waves on a Reissner-Nordström spacetime} \label{SecII}

\label{sec_2}

\subsection{The radial equations}
\label{sec_2_1}

In the Reissner-Nordström (RN) spacetime, gravitational and electromagnetic-perturbation types are governed by partial modes $\phi_{s\omega\ell}^{\text{(e/o)}}$, which are solutions of Moncrief's odd- and even-parity equations \cite{Moncrief:1974gw,Matzner:1976kj}
\begin{equation}\label{Z-M_R-W_homogene_RN}
  \left[\frac{d^2}{dr_{*}^2}+\omega^2- V_{s\ell}^{(e/o)}(r)\right] \phi_{s\omega\ell}^{(e/o)}(r) = 0,
\end{equation}
where the symbols $(e)$ and $(o)$ are respectively associated with even (polar) and odd (axial) objects, according to whether they are even or odd parity in the antipodal transformation on the $2$-sphere unit $S^2$ which respectively satisfy Eq.~\eqref{Z-M_R-W_homogene_RN}. Here, $s \in {1,2}$, and it is worth noting that $s=1$ is associated with the purely electromagnetic field, while $s=2$ is associated with the purely gravitational field in the Schwarzschild BH limit (i.e., when $Q \to 0$). We also introduced $r_*$, the so-called Regge-Wheeler coordinate or \textit{the tortoise coordinate}, which is defined by
\begin{equation}\label{Coord_RW_RN}
  f(r)\frac{d}{dr} = \frac{d}{dr_*},
\end{equation}
therefore given by
\begin{equation}
\label{Coord_RW_RN_bis}
 r_*(r)=r + \frac{r_{+}^2}{r_{+} - r_{-}} \ln \bigg{|}\frac{r - r_{+}}{2M}\bigg{|}
       -\frac{r_{-}^2}{r_{+} - r_{-}} \ln \bigg{|}\frac{r - r_{-}}{2M}\bigg{|} + \mathrm{C},
\end{equation}
where $ \mathrm{C}$ is an integration constant. We are furthermore working outside the RN BH, i.e., we consider $r > r_{+}$, and we have $r_*(r)$ which is a bijection of $r \in ]r_{+},+\infty[$ into $r_* \in ]-\infty,+\infty[$.

In Eq.~\eqref{Z-M_R-W_homogene_RN}, Moncrief's odd-parity potential is given by
\begin{equation}\label{Potentiel_RW_RN}
V_{s\ell}^{(o)}(r) = f(r)\left(\frac{\Lambda+2}{r^2}- \frac{q_{s}}{r^3}+4 \frac{Q^2}{r^4}\right),
\end{equation}
and Moncrief's even-parity potential can be written according to the odd-parity potential as (see Chandrasekhar Refs.~\cite{Chandrasekhar:1979iz,Chandrasekhar:1985kt})
\begin{equation}\label{Potentiel_ZM_RN}
V_{s\ell}^{(e)}(r) = V_{s\ell}^{(o)}(r) + 2q_{s}\frac{d}{dr_*}\bigg[\frac{f(r)}{r(q_{s}+\Lambda r)} \bigg],
\end{equation}
where $\Lambda =(\ell-1)(\ell+2) =\ell(\ell+2)-2$, and
\begin{eqnarray}\label{fr}
  &&q_{1} = 3M -\sqrt{9M^2+4\Lambda Q^2},\nonumber\\
  &&q_{2} = 3M +\sqrt{9M^2+4\Lambda Q^2}
\end{eqnarray}
Note that we have chosen to define $q_1$ and $q_2$ in the opposite order as in the Refs~\cite{Chandrasekhar:1979iz,Gunter:1980} in order to simplify the subsequent expressions.

\subsection{Scattering and boundary conditions}
\label{sec_2_2}

Due to the behavior of Moncrief's odd-parity and even-parity potentials near the horizon $r_+$  and at spatial infinity, we therefore consider the functions $\phi_{s\omega\ell}^{\text{in (e/o)}}(r)$ which are defined by their purely ingoing behavior at $r \rightarrow r_+$ (i.e., for $r_* \rightarrow -\infty$), while at spatial infinity $r \rightarrow +\infty$ (i.e., for $r_* \rightarrow +\infty$), they have an asymptotic behavior. We then have
\begin{equation}\label{boundary_conditions}
\phi_{s\omega\ell}^{\text{in (e/o)}}(r) \sim \!\! \left\{
    \begin{array}{ll}
     \!\! e^{-i\omega r_\ast}, & \!\! r_\ast \to -\infty,\\
     \!\! A^{(-,\text{e/o})}_{s\omega\ell} e^{-i\omega r_\ast} + A^{(+,\text{e/o})}_{s\omega\ell} e^{+i\omega r_\ast},& \!\! r_\ast \to +\infty,
    \end{array}
\right.
\end{equation}
where the coefficients $ A^{(\pm,\text{e/o})}_{s\omega\ell}$ are complex amplitudes such that
$| A^{(-,\text{e/o})}_{s\omega\ell}|^2 - | A^{(+,\text{e/o})}_{s\omega\ell}|^2 = 1 $. It is also important to recall that the solutions of the homogeneous Moncrief odd- and even-parity equations \eqref{Z-M_R-W_homogene_RN} are
related by the Chandrasekhar-Detweiler transformation~\cite{Chandrasekhar:1979iz,Chandrasekhar:1985kt}. Indeed, the even partial modes $\phi_{s\omega\ell}^{\text{(e)}}$ can be constructed from the odd partial modes $\phi_{s\omega\ell}^{\text{(o)}}$
\begin{equation}\label{Relation_Chandra-Detweiler}
\begin{aligned}
  \left[\vphantom{\frac{d}{dr}}\right.&\left.\Lambda(\Lambda+2) \mp 2 i \omega q_{s}\vphantom{\frac{d}{dr}}\right]\phi_{s\omega\ell}^{\text{(e/o)}} = \\ &\left[\Lambda(\Lambda+2)+\frac{2q_{s}^2}{r\left(\Lambda r+q_{s}\right)}f(r)\pm 2\,q_{s}f(r)\frac{d}{dr}\right]\phi_{s\omega\ell}^{\text{(o/e)}},
\end{aligned}
\end{equation}
where the upper and lower signs are respectively associated with the first and second choice of parity in the superscript.
According to Eq.~\eqref{Relation_Chandra-Detweiler}, the coefficients $A^{(-,\text{e/o})}_{s\omega\ell}$ does not depend on parity
\begin{equation}\label{Rel_1}
A^{(-,\text{e})}_{s\omega\ell} = A^{(-,\text{o})}_{s\omega\ell} \equiv A^{(-)}_{s\omega\ell},
\end{equation}
while the coefficients $A^{(+,\text{e/o})}_{s\omega\ell}$ are parity dependent
\begin{equation}\label{Rel_2}
\begin{aligned}
\left[\Lambda(\Lambda+2)-2i\omega \right.& \left.q_{s}\right] A^{(+,\text{e})}_{s\omega\ell} = \\
              & \left[\Lambda(\Lambda+2)+2i\omega\right. \left.q_{s}\right] A^{(+,\text{o})}_{s\omega\ell}.
\end{aligned}
\end{equation}
From the coefficients  $A^{(\pm,\text{e/o})}_{s\omega\ell}$, we define the reflection coefficients $\mathcal{R}_s^{(\text{e/o})}$
\begin{equation}\label{reflec_coeffs}
 \mathcal{R}_s^{(\text{e/o})} \equiv \frac{ A^{(+,\text{e/o})}_{s\omega\ell}}{ A^{(-)}_{s\omega\ell}}.
\end{equation}

The electromagnetic $(H)$ and the gravitational $(Q)$ perturbations are derived from the even- and odd-parity radial functions  ~\cite{Moncrief-1974a,Moncrief-1974b,Moncrief-1975} (see also \cite{Olson:1974nk,Matzner:1976kj})
\begin{subequations}\label{EW_GW_Radiations}
\begin{align}
H^{(e/o)}&\equiv \phi_{1\ell\omega}^{(e/o)} \cos\psi - \mathcal{P}\, \phi_{2\ell\omega}^{(e/o)}\sin\psi, \\
Q^{(e/o)}&\equiv \mathcal{P} \, \phi_{1\ell\omega}^{(e/o)} \sin\psi + \phi_{2\ell\omega}^{(e/o)}\cos\psi,
\end{align}
\end{subequations}
where $ \mathcal{P} = +1$ and $-1$ associated with even and odd parity respectively.

In expressions~\eqref{EW_GW_Radiations}, we have
\begin{equation}\label{psi_def}
\begin{aligned}
   &\cos^2\psi =\frac{q_2}{q_2-q_1}, \qquad \sin^2\psi=\frac{-q_1}{q_2-q_1},\\
  &\sin(2\psi) = \frac{-2\sqrt{-q_1q_2}}{q_2-q_1} = -2\, Q \frac{\Lambda^{1/2}}{\sqrt{9M^2+4\Lambda Q^2}}.
\end{aligned}
\end{equation}

The conversion coefficient $\mathcal{C}^{(e/o)}$, which determines the fraction of the incident wave converted from
electromagnetic to gravitational and vice versa, is defined from the reflection coefficients $\mathcal{R}_s^{(\text{e/o})}$ (i.e., from the coefficients $A^{(\pm,\text{e/o})}_{s\omega\ell}$) \cite{Chandrasekhar:1985kt,Crispino:2009zza}
\begin{equation}\label{Conv_Coeff}
  \mathcal{C}^{(e/o)} = \bigg{|}\frac{1}{2} \sin(2\psi)(\mathcal{R}_1^{(\text{e/o})} - \mathcal{R}_2^{(\text{e/o})}) \bigg{|}^2.
\end{equation}

To conclude this section, we define the $S$-matrix elements $S_{\ell}^{(e/o,s_i s_f)}$ from the reflection coefficients $\mathcal{R}_s^{(\text{e/o})}$. These $S$-matrix elements play a key role in the scattering processes, and will be discussed later. Here, the superscripts $s_i$ and $s_f$ refer to the spins of the initial and final fields, with $s=1$ for an EW and $s=2$ for a GW. Thus, we have
\begin{subequations}\label{Matrice_S_si_sf}
\begin{equation}\label{Matrice_S_11}
  S_{\ell}^{(e/o,11)}  =
    e^{i(\ell+1)\pi} \left[\mathcal{R}_1^{(\text{e/o})}\cos^2\psi + \mathcal{R}_2^{(\text{e/o})}\sin^2\psi\right],
\end{equation}
\begin{equation}\label{Matrice_S_22}
  S_{\ell}^{(e/o,22)}  =
    e^{i(\ell+1)\pi} \left[ \mathcal{R}_1^{(\text{e/o})}\sin^2\psi + \mathcal{R}_2^{(\text{e/o})}\cos^2\psi\right],
\end{equation}
and
\begin{equation}\label{Matrice_S_12}
  S_{\ell}^{(e/o,12)}  =S_{\ell}^{(e/o,21)}  =
    e^{i(\ell+1)\pi} \frac{\sin(2\psi)}{2}\left[ \mathcal{R}_1^{(\text{e/o})} - \mathcal{R}_2^{(\text{e/o})}\right].
\end{equation}
\end{subequations}

Before proceeding to the construction  of the scattering and conversion cross sections, we will first focus on the poles of the $S$-matrix in the complex plane, which are an important ingredient, that will allow us to construct the CAM representations.

\section{The poles of $S$-matrix elements}
\label{sec_3}

The solutions of the problem~\eqref{Z-M_R-W_homogene_RN}--\eqref{Potentiel_ZM_RN} that govern its resonant modes, the so-called \emph{quasinormal modes} and \emph{Regge modes}, are defined by the purely ingoing and outgoing boundary conditions at the horizon and at spatial infinity, respectively. A spectrum of resonant modes is defined by the boundary conditions \eqref{boundary_conditions} with 
\begin{equation}\label{Spetrum_condition}
  A^{(-,\text{e/o})}_{s\omega\ell} = 0.
\end{equation}

The \emph{quasinormal frequency spectrum} $\omega_{\ell n}^{(s)}$, which characterizes the QNMs, is a set of frequencies in the complex-$\omega$ plane at which the $S$-matrix elements $S_{\ell}^{(e/o,s_i s_f)}$ have simple poles for $\ell \in \mathds{N}$, i.e., simple zeros of \eqref{Spetrum_condition}. 

The \textit{Regge pole spectrum} $\lambda_n^{(s)}(\omega) \equiv \ell_n^{(s)}(\omega) + 1/2$ in the complex-$\lambda$ plane is the set of angular momenta at which the $S$-matrix elements $S_{\ell}^{(e/o,s_i s_f)}$ have simple poles for $\omega \in \mathds{R}$, i.e., simple zeros of the coefficients $A^{(-,\text{e/o})}_{s\omega\ell}$ that depict Regge modes. These spectra are defined with respect to each $s$. According to relation~\eqref{Rel_1}, the zeros of the coefficients  $A^{(-,\text{e})}_{s\omega\ell}$ and $A^{(-,\text{o})}_{s\omega\ell}$ are identical. Here $n = 1,2,3,\ldots$ is a number of overtones that enumerates the discrete spectrum of poles.

It is worth noting that the QNMs of the RN BH and their associated quasinormal frequency spectrum have been extensively studied in the literature (see, for example, the nonexhaustive list of Refs.~\cite{Gunter1980,Kokkotas1988,Leaver:1990zz,Andersson1993,Andersson1994,Kokkotas:1999bd,Berti:2003zu} and references therein). We will therefore focus on its Regge pole spectrum which will be investigated here for the first time.

\subsection{The Regge poles : $Q<M$}
\label{sec_3_1}

As mentioned earlier, the RPs of the matrices $S^{(e,11)}_{\lambda -1/2}$ and $S^{(o,11)}_{\lambda -1/2}$ in the complex $\lambda$ plane are identical, due to the relation~\eqref{Rel_1}. These poles are defined with respect to each spin $s=1$ and $2$ and are located in the first and third quadrants, symmetrically distributed with respect to the origin $O$.

The spectrum associated with electromagnetic-type perturbations ($s=1$) corresponds to the zeros of the coefficients $A^{(-)}_{1\omega,\lambda-1/2}$, i.e., the values $\lambda_n^{(1)}(\omega)$ with $n=1,2,3,\ldots$ such that 
\begin{equation}\label{PR_EW}
A^{(-)}_{1\omega,\lambda_n^{(1)}(\omega)-1/2}=0,
\end{equation}
while the spectrum associated with gravitational-type perturbations ($s=2$) corresponds to the zeros of the coefficients $A^{(-)}_{2\omega,\lambda-1/2}$, i.e., the values $\lambda_n^{(2)}(\omega)$ with $n=1,2,3,\ldots$  such that \begin{equation}\label{PR_Grav}
A^{(-)}_{2\omega,\lambda_n^{(2)}(\omega)-1/2}=0.
\end{equation}

\subsubsection{Numerical method}
\label{sec_3_1_1}

There are several numerical methods to calculate the quasinormal frequency spectrum of BHs, and each method has its own advantages and disadvantages (see Ref.~\cite{Berti:2004md} for a summary of the different methods as well as
Ref.~\cite{Kokkotas:1999bd}).

This being said, the \textit{continued fraction method} introduced by Leaver~\cite{Leaver:1985ax,leaver1986solutions,Leaver:1990zz} remains the most popular and widely used due to its robustness and accuracy. It is also adaptable to the computation of the RP spectrum. However, we will use a slightly different method introduced by Majumdar and Panchapakesan~\cite{mp}, which involves finding the zeros of a given determinant (known as the \emph{Hill determinant}). This method has already been employed to compute the RPs for Schwarzschild and Kerr BHs (see, e.g., Refs.~\cite{Folacci:2018sef,Folacci:2019cmc,Folacci:2019vtt,Folacci:2021uld}), as well as for other cases, such as compact objects~\cite{OuldElHadj:2019kji} and dirty BHs~\cite{Torres:2022fyf}. We will now outline the key steps involved in applying this method to the RN BH case.

Following Leaver~\cite{Leaver:1990zz}, we will look for the resonant mode solutions which satisfy the boundary conditions discussed before by a series expansion of the form
\begin{eqnarray}\label{QNM_sol}
  \phi_{s\omega\ell}^{(o)}(r)& = & e^{-i2\omega r_{+}} \left(\frac{r_{+}}{r}\right) \left(\frac{r_{+}-r_{-}}{2M}\right)^{-i2(2M\omega)-1} \nonumber \\
 & & \times \left(\frac{r-r_{-}}{2M}\right)^{i(2M\omega)+1}
   \left(\frac{r-r_{+}}{r-r_{-}}\right)^{-i\omega \left(\frac{r_{+}^2}{r_{+} - r_{-}}\right)} \nonumber\\
 & &\times e^{i\omega r} \sum_{n=0}^{+\infty}a_n  \left(\frac{r-r_{+}}{r-r_{-}}\right)^n,
\end{eqnarray}
where the coefficients $a_n$ satisfy a four-term recurrence relation:
\begin{equation}\label{Recurrence_4_terms}
\alpha_n a_{n+1} + \beta_n a_{n} +\gamma_{n} a_{n-1} +\delta_{n} a_{n-2}  = 0,
\quad \forall n\geq 2,
\end{equation}
with
\begin{subequations}
\begin{eqnarray}
 && \alpha_n = (n+1)\Big[(n+1)(r_{+}-r_{-})-2 i\omega\, r_{+}^2\Big] r_{+}, \\
 && \beta_n = - (r_{+} - r_{-}) (2 r_{+} + r_{-}) n^2 - \Big[(r_{+} - r_{-})  \nonumber\\
 &&\quad\quad (2 r_{+} - 4 r_{-}) - 2 i \omega (+ 4 r_{+} - r_{-}) r_{+}^2 \Big] n \nonumber\\
 &&\quad\quad - (r_{+} - r_{-}) \Big[(\Lambda + 2) r_{+} - q_s + 3 r_{-}\Big]  \nonumber\\
 &&\quad\quad+ 2 i \omega (2 r_{+} - 3 r_{-}) rp^2  + 8 \omega^2 r_{+}^4,          \\
 && \gamma_n = (r_{+} - r_{-}) (r_{+} + 2 r_{-}) n^2 - \Big[10 r_{-} (r_{+} - r_{-})\nonumber\\
 &&\quad\quad + 2 i \omega r_{+} (2 r_{+}^2 + 4 r_{-} r_{+} - 3 r_{-}^2)\Big] n \nonumber\\
 &&\quad\quad+ (r_{+} - r_{-}) \Big[(\Lambda + 2) r_{-} - q_s + 12 r_{-} - r_{+}\Big] \nonumber\\
 &&\quad\quad + 10 i \omega r_{+} r_{-} (2 r_{+} - r_{-}) - 4 \omega^2 r_{+}^3 (r_{+} + 3 r_{-}),\\
 && \delta_n = - (r_{+} - r_{-}) r_{-} n^2 + \Big[6 (r_{+} - r_{-}) r_{-} \nonumber \\
 &&\quad\quad + 2 i \omega (2 r_{+}^2 - r_{-}^2) r_{-} \Big] n \nonumber \\
 &&\quad\quad - 9 (r_{+} - r_{-}) r_{-} - 6 i \omega (2 r_{+}^2 - r_{-}^2) r_{-} \nonumber\\
 &&\quad\quad + 4 \omega^2 r_{+}^2 (r_{-} + r_{+}) r_{-}.
\end{eqnarray}
\end{subequations}

We then find solutions to the recurrence formula \eqref{Recurrence_4_terms} using the Hill determinant approach. Nontrivial solutions to Eq. \eqref{Recurrence_4_terms} exist when the Hill determinant vanishes:
\begin{equation}
\label{Determinant_Hill_4_termes}
\small
D   =  \begin{vmatrix}
\beta_0 &  \alpha_0 &  0 & 0 & 0 &  \ldots  & \ldots  &  \ldots \\
\gamma_1 & \beta_1 & \alpha_1 & 0  &  0  &  \ldots  &  \ldots  &  \ldots  \\
\delta_2 &  \gamma_2 &  \beta_2  &  \alpha_2 &  0  &  \ldots  &  \ldots  &  \ldots  \\
\vdots & \ddots & \ddots  &  \ddots  & \ddots  &  \ddots  &  \ldots  &  \ldots \\
\vdots & \vdots & \delta_{n-1} & \gamma_{n-1} & \beta_{n-1} & \alpha_{n-1} & \ddots & \ldots \\
\vdots & \vdots & \vdots & \delta_n & \gamma_{n} & \beta_{n} & \alpha_{n} & \ddots   \\
\vdots & \vdots & \vdots & \vdots & \ddots & \ddots & \ddots & \ddots
\end{vmatrix} = 0.
\end{equation}
Considering $D_n$ as the determinant of the $n \times n$ submatrix of $D$
\begin{equation}
\label{derecurrence_4_termes}
D_n=\beta_n D_{n-1} - \gamma_{n}\alpha_{n-1}D_{n-2}
+ \delta_n \alpha_{n-1} \alpha_{n-2} D_{n-3},
\end{equation}
with the initial conditions
\begin{equation}
\label{Determinant_initial_conds}
\begin{split}
D_0 &=\beta_0, \\
D_1 &=\beta_1\beta_0-\gamma_1\alpha_0 , \\
D_2 &=\beta_0(\beta_1 \beta_2 - \alpha_1 \gamma_2)
- \alpha_0(\alpha_1 \delta_2 -\gamma_1 \beta_2),
\end{split}
\end{equation}
or, equivalently
\begin{align}
\label{recurrence_Hill_RN_4_termes}
D_n    & =\left(\prod_{k=1}^{n+1} k^2\right) P_{n+1} \nonumber\\
       &  = 1\times 2^2 \ldots (n-2)^2(n-1)^2n^2(n+1)^2 P_{n+1},
\end{align}
where
\begin{eqnarray}
\label{recurrence_Hill_RN_4_termes_bis}
P_n=
&&\left(\frac{\beta_{n-1}}{n^2}\right)P_{n-1}-\left(\frac{\gamma_{n-1}}{n^2}\right)\left(\frac{\alpha_{n-2}}{(n-1)^2}\right)P_{n-2} \nonumber\\
& &+ \left(\frac{\delta_{n-1}}{n^2}\right) \left(\frac{\alpha_{n-2}}{(n-1)^2}\right)
\left(\frac{\alpha_{n-3}}{(n-2)^2}\right) P_{n-3},
\end{eqnarray}
with the following initial conditions
 \begin{equation}
 \begin{split}
    P_0&= 1,\\
    P_1&=\beta_0, \\
    P_2&=\frac{\beta_0\beta_1-\gamma_1\alpha_0}{4}.
 \end{split}
 \end{equation}
The RPs (QNM frequencies) are found by setting $\omega \in \mathds{R}$ ($\lambda \equiv \ell+1/2$ with $\ell \in \mathds{N}$) and numerically finding the roots $\lambda_n^{(s)}(\omega)$ ($\omega_{\ell n}^{(s)}$) of $D_n(\lambda_n,\omega)= 0$ ($D_n(\lambda,\omega_n)= 0$), i.e., of $P_n(\lambda_n,\omega)= 0$ ($P_n(\lambda,\omega_n)= 0$).

\subsubsection{Numerical results and comments}
\label{sec_3_1_2}

\begin{figure*}[htbp]
 \includegraphics[scale=0.60]{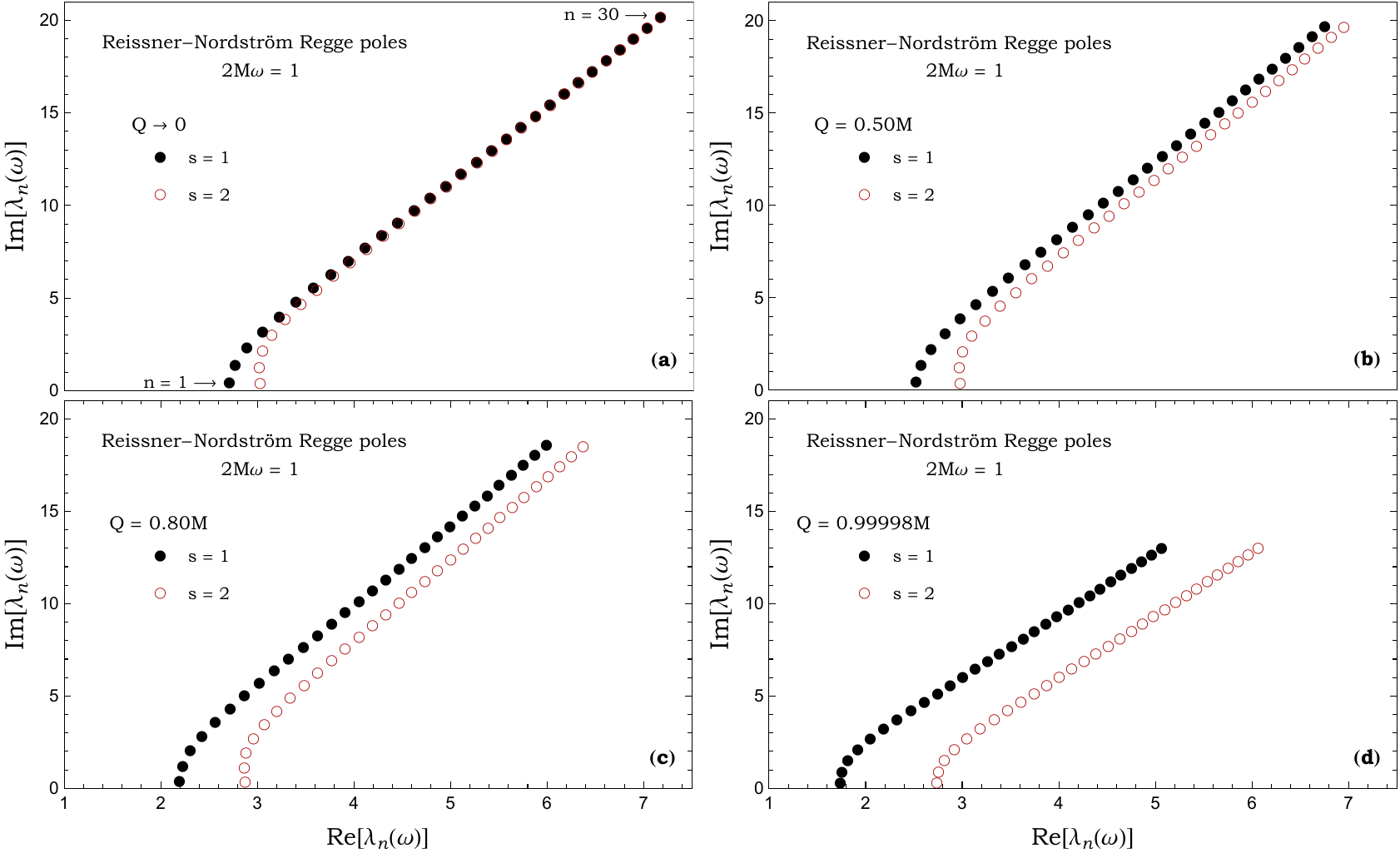}
\caption{\label{PR_2Mw_1_Diff_Q} The Reissner-Nordström Regge poles for the electromagnetic-type perturbations $\lambda_n^{(1)}$ (filled markers) and for the gravitational-type perturbations $\lambda_n^{(2)}$ (unfilled markers). The results are obtained for charge-to-mass ratios $Q/M =0, 0.50, 0.80$ and $0.99998$ at $2M\omega = 1$. Here, we took $n=1,2,3,4,\ldots,30$.}
\end{figure*}

\begin{figure*}[htbp]
 \includegraphics[scale=0.60]{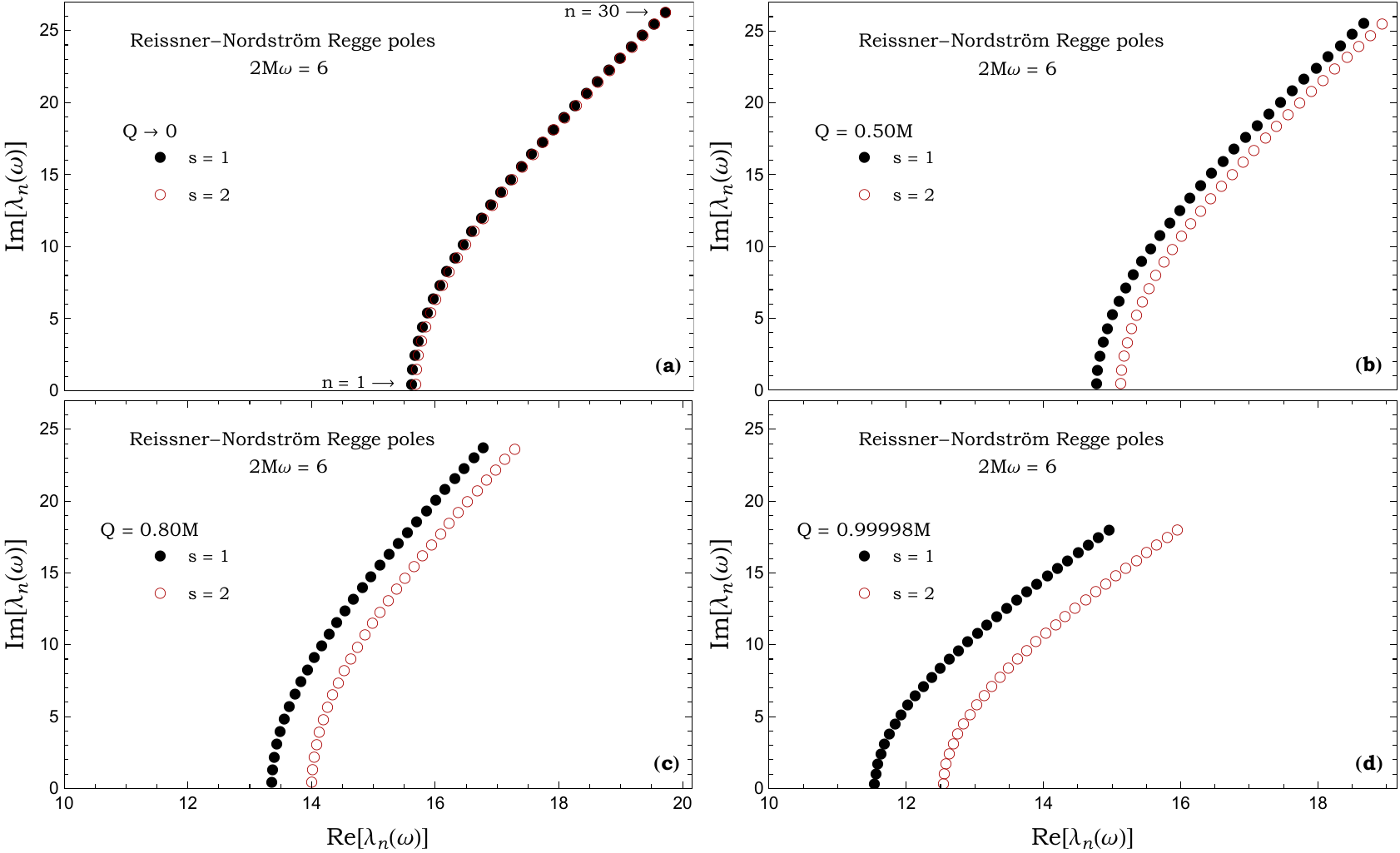}
\caption{\label{PR_2Mw_6_Diff_Q}  The Reissner-Nordström Regge poles for the electromagnetic-type perturbations $\lambda_n^{(1)}$ (filled markers) and for the gravitational-type perturbations $\lambda_n^{(2)}$ (unfilled markers). The results are obtained for charge-to-mass ratios $Q/M =0, 0.50, 0.80$ and $0.99998$ at $2M\omega = 6$. Here, we took $n=1,2,3,4,\ldots,30$. }
\end{figure*}

In Figs.~\ref{PR_2Mw_1_Diff_Q}~and~\ref{PR_2Mw_6_Diff_Q}, we present numerical results of the RP spectra for electromagnetic-type perturbations $[\lambda_n^{(1)}]$ ($s=1$) and gravitational-type perturbations $[\lambda_n^{(2)}]$ ($s=2$) for different configurations of the RN BH.

Figure~\ref{PR_2Mw_1_Diff_Q} shows the first $30$ RPs of both branches $(\lambda_n^{(1)})$ and $(\lambda_n^{(2)})$ at reduced frequency $2M\omega=1$ for charge-to-mass ratios $Q/M=0, 0.50, 0.80$, and $0.99998$. Note that the splitting of the two branches occurs as the charge-to-mass ratio $Q/M$ increases, and the difference between these two branches tends to unity as $Q \to M$ and becomes exactly equal to unity in the extremal charged RN BH case (cf. Sec.~\ref{sec_3_2}).

Figure~\ref{PR_2Mw_6_Diff_Q} shows the first $30$ Regge poles of both branches $(\lambda_n^{(1)})$ and $(\lambda_n^{(2)})$ for the ``high''-frequency case at $2M\omega=6$ for charge-to-mass $Q/M=0, 0.50, 0.80$, and $0.99998$. We observe that the structure of the two branches remains the same.

Table~\ref{tab:table1} presents the lowest Regge poles for the RN BH at reduced frequencies $2M\omega=1$ and $2M\omega=6$ for charge-to-mass ratios $Q/M=0, 0.50, 0.80$, and $0.99998$.

\begingroup
\squeezetable
\begin{table}[htbp]
\caption{\label{tab:table1} Lowest Regge poles $\lambda_{n}^{(1)}(\omega)$ correspond to electromagnetic-type perturbations ($s = 1$), and $\lambda_{n}^{(2)}(\omega)$ correspond to gravitational-type perturbations ($s = 2$). We assume $2M=1$.}
\smallskip
\centering
\begin{ruledtabular}
\begin{tabular}{ccccc}
 $Q$ &$2M\omega$  &  $n$ & $\lambda_n^{(1)}(\omega)$ & $\lambda_n^{(2)}(\omega)$
 \\ \hline
$\mathbf{0}$ & $1$  & $1$  &$\phantom{1}2.705358 + 0.475454 i      $  & $\phantom{1}3.031094 + 0.411206 i   $   \\[-1ex]
             &      & $2$  &$\phantom{1}2.769262 + 1.419171 i      $  & $\phantom{1}3.018269 + 1.260871 i   $   \\[-1ex]
             &      & $3$  &$\phantom{1}2.892291 + 2.330092 i      $  & $\phantom{1}3.050451 + 2.151728 i   $   \\[-1ex]
             &      & $4$  &$\phantom{1}3.049988 + 3.194709 i      $  & $\phantom{1}3.149417 + 3.032799 i   $   \\[-1ex]
             &      & $5$  &$\phantom{1}3.222920 + 4.017260 i      $  & $\phantom{1}3.289281 + 3.877087 i   $   \\[-1ex]
             &      & $6$  &$\phantom{1}3.401342 + 4.805736 i      $  & $\phantom{1}3.448514 + 4.684410 i   $   \\[+1ex]

             & $6$  & $1$  &$15.607716 + 0.499228 i                $  & $15.671479 + 0.496524 i             $   \\[-1ex]
             &      & $2$  &$15.625210 + 1.496753 i                $  & $15.688070 + 1.488756 i             $   \\[-1ex]
             &      & $3$  &$15.659860 + 2.491532 i                $  & $15.720985 + 2.478570 i             $   \\[-1ex]
             &      & $4$  &$15.711016 + 3.481878 i                $  & $15.769709 + 3.464449 i             $   \\[-1ex]
             &      & $5$  &$15.777770 + 4.466296 i                $  & $15.833503 + 4.445015 i             $   \\[-1ex]
             &      & $6$  &$15.859025 + 5.443533 i                $  & $15.911449 + 5.419065 i             $   \\[+1.5ex]
\hline
$\mathbf{0.50M}  $ & $1$& $1$ &$\phantom{1}2.523634 + 0.453499 i   $  & $\phantom{1}2.985198 + 0.401517 i   $   \\[-1ex]
                 &      & $2$ &$\phantom{1}2.576162 + 1.357438 i   $  & $\phantom{1}2.973672 + 1.229947 i   $   \\[-1ex]
                 &      & $3$ &$\phantom{1}2.682486 + 2.237176 i   $  & $\phantom{1}3.006106 + 2.098412 i   $   \\[-1ex]
                 &      & $4$ &$\phantom{1}2.823439 + 3.077972 i   $  & $\phantom{1}3.104273 + 2.956203 i   $   \\[-1ex]
                 &      & $5$ &$\phantom{1}2.981750 + 3.881305 i   $  & $\phantom{1}3.241485 + 3.776757 i   $   \\[-1ex]
                 &      & $6$ &$\phantom{1}3.147877 + 4.652962 i   $  & $\phantom{1}3.396215 + 4.560892 i   $   \\[+1ex]

                 &  $6$ & $1$ &$14.775517 + 0.483943 i             $  & $15.134550 + 0.479904 i             $   \\[-1ex]
                 &      & $2$ &$14.792059 + 1.450965 i             $  & $15.150065 + 1.438960 i             $   \\[-1ex]
                 &      & $3$ &$14.824835 + 2.415434 i             $  & $15.180854 + 2.395787 i             $   \\[-1ex]
                 &      & $4$ &$14.873253 + 3.375773 i             $  & $15.226446 + 3.348985 i             $   \\[-1ex]
                 &      & $5$ &$14.936484 + 4.330580 i             $  & $15.286167 + 4.297277 i             $   \\[-1ex]
                 &      & $6$ &$15.013516 + 5.278667 i             $  & $15.359178 + 5.239547 i             $   \\[+1.5ex]
\hline
$\mathbf{0.80M} $ & $1$ & $1$ &$\phantom{1}2.200260 + 0.411372 i   $  & $\phantom{1}2.878223 + 0.370251 i   $  \\[-1ex]
                 &      & $2$ &$\phantom{1}2.232046 + 1.239437 i   $  & $\phantom{1}2.867525 + 1.129486 i   $   \\[-1ex]
                 &      & $3$ &$\phantom{1}2.311737 + 2.061194 i   $  & $\phantom{1}2.889105 + 1.924910 i   $   \\[-1ex]
                 &      & $4$ &$\phantom{1}2.430953 + 2.853595 i   $  & $\phantom{1}2.966166 + 2.716062 i   $   \\[-1ex]
                 &      & $5$ &$\phantom{1}2.571183 + 3.611330 i   $  & $\phantom{1}3.078581 + 3.478546 i   $   \\[-1ex]
                 &      & $6$ &$\phantom{1}2.7206111 + 4.3387083 i $  & $\phantom{1}3.208461 + 4.211212 i   $   \\[+1.5ex]

                 &  $6$ & $1$ &$13.365832 + 0.446058 i             $  & $14.011119 + 0.440124 i             $   \\[-1ex]
                 &      & $2$ &$13.379999 + 1.337458 i             $  & $14.023988 + 1.319780 i             $   \\[-1ex]
                 &      & $3$ &$13.408085 + 2.226743 i             $  & $14.049543 + 2.197674 i             $   \\[-1ex]
                 &      & $4$ &$13.449616 + 3.112599 i             $  & $14.087424 + 3.072699 i             $   \\[-1ex]
                 &      & $5$ &$13.503920 + 3.993846 i             $  & $14.137118 + 3.943834 i             $   \\[-1ex]
                 &      & $6$ &$13.570178 + 4.869475 i             $  & $14.197983 + 4.810182 i             $   \\[0ex]
\hline
$\mathbf{0.99998M}$& $1$& $1$ &$\phantom{1}1.737232 + 0.304722 i   $  & $\phantom{1}2.737171 + 0.304701 i  $  \\[-1ex]
                 &      & $2$ &$\phantom{1}1.759495 + 0.916949 i   $  & $\phantom{1}2.759438 + 0.916884 i  $   \\[-1ex]
                 &      & $3$ &$\phantom{1}1.820494 + 1.535097 i   $  & $\phantom{1}2.820443 + 1.534989 i   $   \\[-1ex]
                 &      & $4$ &$\phantom{1}1.925439 + 2.131438 i   $  & $\phantom{1}2.925398 + 2.131285 i   $   \\[-1ex]
                 &      & $5$ &$\phantom{1}2.053919 + 2.693727 i   $  & $\phantom{1}3.053892 + 2.693529 i   $   \\[-1ex]
                 &      & $6$ &$\phantom{1}2.191582 + 3.225730 i   $  & $\phantom{1}3.191571 + 3.225486 i  $   \\[+1.5ex]

                 &  $6$ & $1$ &$11.545607 + 0.351631 i             $  & $12.545547 + 0.351627 i              $   \\[-1ex]
                 &      & $2$ &$11.560499 + 1.054091 i             $  & $12.560439 + 1.054081 i             $   \\[-1ex]
                 &      & $3$ &$11.589995 + 1.754186 i             $  & $12.589935 + 1.754169 i              $   \\[-1ex]
                 &      & $4$ &$11.633546 + 2.450454 i             $  & $12.633487 + 2.450429 i              $   \\[-1ex]
                 &      & $5$ &$11.690378 + 3.141593 i             $  & $12.690319 + 3.141561 i             $   \\[-1ex]
                 &      & $6$ &$11.759546 + 3.826502 i             $  & $12.759488 + 3.826462 i              $   \\[0ex]

\end{tabular}
\end{ruledtabular}
\end{table}
\endgroup

\subsection{The Regge poles of extremal charged RN: $Q=M$}
\label{sec_3_2}

\subsubsection{Numerical algorithm}
\label{sec_3_2_1}

To calculate the RP spectrum in the extremal RN BH case (i.e., for $Q = M$), we will use the method proposed by Onozawa \emph{et al}.~\cite{Onozawa:1995vu}. In fact, Leaver's method, or more precisely, the slightly modified version presented above that we used to calculate the RP spectrum in the case $Q < M$, does not work in the extremal case. Indeed, the associated wave equations for $Q=M$ have an irregular singular point at the horizon and at infinity, which means that the series expansion of the solution~\eqref{QNM_sol} is not valid. This makes the use of the continued fraction (or Hill's determinant method) impossible. To overcome this, Onozawa \emph{et al}. showed that we can still use the continued fraction method if we choose the right regular point around which to expand the solution. To achieve this, we will review some of the essential steps described in~\cite{Onozawa:1995vu}.

We recall that the \textit{tortoise coordinate} $r_*$ for the extremal charged RN BH case ($Q = M$) becomes
\begin{equation}\label{Coord_RW_RN_Extrem}
  r_*(r) = r + 2M \ln\left(\frac{r}{M} - 1\right)-\frac{M^2}{r-M},
\end{equation}
and the wave equation~\eqref{Z-M_R-W_homogene_RN} has an \textit{essential singularity} at $r=M$ and $r \to \infty$. Following Onozawa \emph{et al.}, we expand the solution around the ordinary point $r = 2M$. Thus, we look for resonant mode solutions that satisfy the boundary conditions on both sides simultaneously, using a series expansion of the form
\begin{eqnarray}\label{QNM_sol_Extrem}
  \phi_{s\omega\ell}^{(o)}(r)& = & e^{-i2 M \omega} \left(\frac{r}{M}\right)^{2i(2M\omega)}\left(\frac{r}{M}-1\right)^{-i2M\omega} \nonumber \\
 & & \times e^{i\omega\left(\frac{M^2}{r-M}\right)} e^{i\omega r} \sum_{n=0}^{+\infty}a_n \left(1-\frac{2M}{r}\right)^n,
\end{eqnarray}
where the coefficients $a_n$ verify a five-term recurrence relation:
\begin{subequations}
\begin{align}
&\alpha_1 a_2 +\beta_1 a_1 + \gamma_1 a_0 = 0, \\
&\alpha_2 a_3 +\beta_2 a_2 + \gamma_2 a_1 + \delta_2 a_0 = 0, \\
&\alpha_n a_{n+1} + \beta_n a_{n} +\gamma_n a_{n-1} +\delta_n a_{n-2} + \epsilon_n a_{n-3}  = 0, \forall n\geq 3,\nonumber\\
\end{align}
\end{subequations}
with
\begin{subequations}
\begin{align}
  \alpha_n &= n(n+1), \\
  \beta_n  &= 0, \\
  \gamma_n &= 16(2M\omega)^2 +6i(2M\omega)(2n-1)-2n(n-1) \nonumber\\
                & +2\frac{q_s}{M}-4\Lambda-12, \\
  \delta_n &= 2\left(4-\frac{q_s}{M}\right), \\
  \epsilon_n &= -4(2M\omega)^2 -2i(2M\omega)(2n-3)+n(n-3)-4. \nonumber\\
\end{align}
\end{subequations}

The trick used by Onozawa \emph{et al.} was to separate the five-term recurrence relation into two distinct even and odd five-term recurrence relations, $\sum c_n = a_{2n}$ and $\sum b_n = a_{2n+1}$ respectively, which satisfy the convergence conditions of the even and odd coefficients. Thereafter, the two even and odd five-term recurrence relations are reduced to two three-term recurrence relations using two successive iterations of Gaussian elimination steps. Thus, the method of continued fractions can be applied to determine the RP spectrum, which are the minimal solution of both even and odd three-term recurrence relations simultaneously (see Ref.~\cite{Onozawa:1995vu} for more details).

\subsubsection{Numerical results}
\label{sec_3_2_2}

\begin{figure}[htbp]
 \includegraphics[scale=0.60]{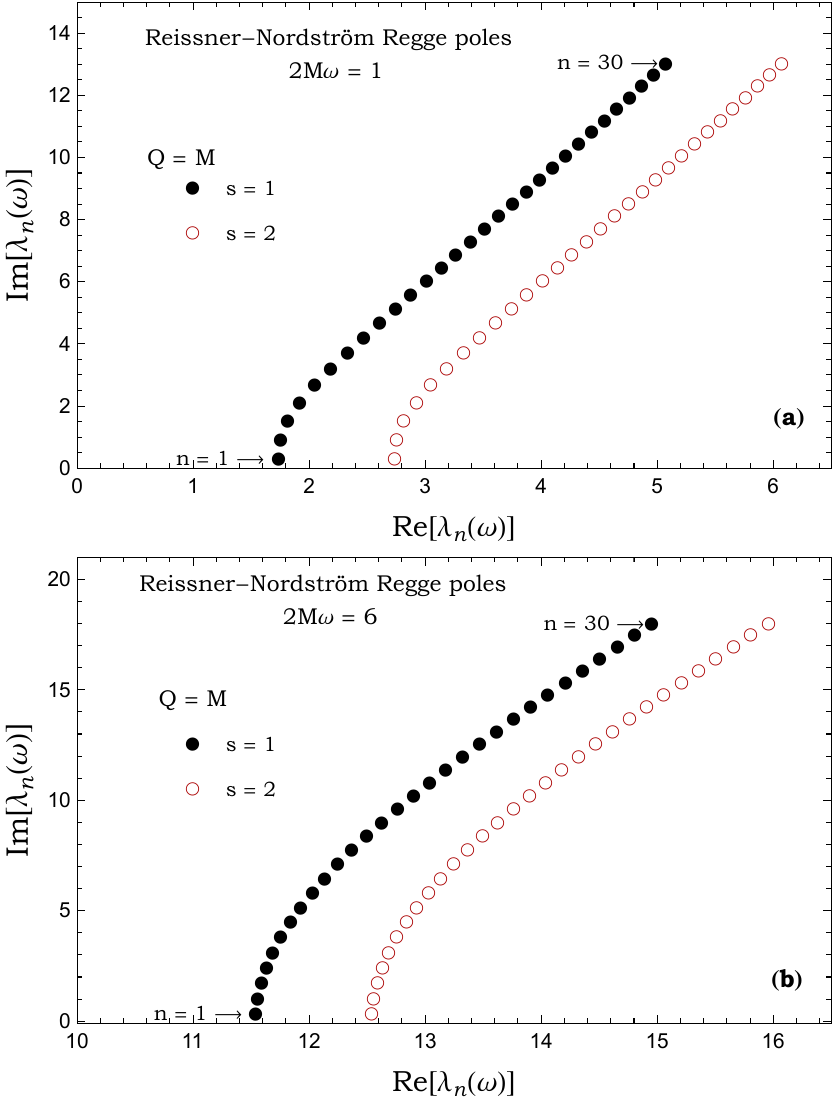}
\caption{\label{PR_2Mw_1_and_6_Q_equal_M}  The Reissner-Nordström Regge poles for the electromagnetic-type perturbations $\lambda_n^{(1)}$ (filled markers) and for the gravitational-type perturbations $\lambda_n^{(2)}$ (unfilled markers). The results are obtained for charge-to-mass $Q/M = 1$ at a) $2M\omega = 1$ and b) $2M\omega =  6$. The results are in agreement with Eq.~\eqref{Relation_PRs}. Here, we took $n=1,2,3,4,\ldots,30$.}
\end{figure}

In Fig.~\ref{PR_2Mw_1_and_6_Q_equal_M}, we display  the first $30$ RPs of the two branches $\lambda_n^{(1)}$ and $\lambda_n^{(2)}$ for extremal charged RN BH case at reduced frequencies $2M\omega = 1$ and $6$. The lowest RPs for this configuration are given in Table~\ref{tab:table_2}. We observe that the spectrum for $Q<M$ smoothly approaches the spectrum for $Q=M$, and no evidence of new modes. Interestingly, even though the continued fraction method should not be effective, the RPs obtained by the Leaver's algorithm for near extremal charged RN BH (i.e., for charge-to-mass ratio $Q/M=0.99998$) are remarkably similar to those obtained by the Onozawa \emph{et al.}'s method for extremal charged RN BH case (i.e., for charge-to-mass ratio $Q/M=1$),  with a difference of less than $0.1\%$ for the real and imaginary parts.

\begingroup
\squeezetable
\begin{table}[htbp]
\caption{\label{tab:table_2} Lowest Regge poles $\lambda_{n}^{(1)}(\omega)$ for electromagnetic-type perturbations ($s = 1$) and $\lambda_{n}^{(2)}(\omega)$ for gravitational-type perturbations ($s = 2$) for the extremely charged RN BH. We assume $2M=1$.}
\smallskip
\centering
\begin{ruledtabular}
\begin{tabular}{ccccc}
 $Q$ &$2M\omega$  &  $n$ & $\lambda_n^{(1)}(\omega)$ & $\lambda_n^{(2)}(\omega)$
 \\ \hline
$\mathbf{M}$     & $1$  & $1$ &$\phantom{1}1.737154 + 0.304689 i   $  & $\phantom{1}2.737154 + 0.304689 i  $  \\[-1ex]
                 &      & $2$ &$\phantom{1}1.759438 + 0.916848 i   $  & $\phantom{1}2.759438 + 0.916848 i  $   \\[-1ex]
                 &      & $3$ &$\phantom{1}1.820478 + 1.534929 i   $  & $\phantom{1}2.820478 + 1.534929 i  $   \\[-1ex]
                 &      & $4$ &$\phantom{1}1.925487 + 2.131205 i   $  & $\phantom{1}2.925487 + 2.131205 i  $   \\[-1ex]
                 &      & $5$ &$\phantom{1}2.054052 + 2.693430 i   $  & $\phantom{1}3.054052 + 2.693430 i  $   \\[-1ex]
                 &      & $6$ &$\phantom{1}2.191821 + 3.225372 i   $  & $\phantom{1}3.191821 + 3.225372 i  $   \\[+1.5ex]

                 &  $6$ & $1$ &$11.545336 + 0.351607 i             $  & $12.545336 + 0.351607 i             $   \\[-1ex]
                 &      & $2$ &$11.560231 + 1.054022 i             $  & $12.560231 + 1.054022 i             $   \\[-1ex]
                 &      & $3$ &$11.589734 + 1.754071 i             $  & $12.589734 + 1.754071 i             $   \\[-1ex]
                 &      & $4$ &$11.633296 + 2.450292 i             $  & $12.633296 + 2.450293 i             $   \\[-1ex]
                 &      & $5$ &$11.690142 + 3.141386 i             $  & $12.690142 + 3.141386 i             $   \\[-1ex]
                 &      & $6$ &$11.759327 + 3.826248 i             $  & $12.759327 + 3.826248 i             $   \\[0ex]

\end{tabular}
\end{ruledtabular}
\end{table}
\endgroup
It is worth noting that as the parameter $Q$ increases, the numerical results indicate that the RPs $\lambda_n^{(2)}$ approach $\lambda_n^{(1)}-1$ and coincide in the extremal charged case. In other words, the RPs $\lambda_n^{(1)}$ associated with electromagnetic-type perturbations ($s=1$) can be obtained from those associated with gravitational-type perturbations ($s=2$), and vice versa. Consequently, we can express this relationship as follows
\begin{equation}\label{Relation_PRs}
  \lambda_n^{(2)} - \lambda_n^{(1)}  = 1.
\end{equation}

In fact, Onozawa \emph{et al.}~\cite{Onozawa:1995vu} observed this property in the case of QNM frequencies. They found that numerically, QNM frequencies associated with gravitational-type perturbations with a multipolar index $\ell$ coincide with those associated with electromagnetic-type perturbations with a multipolar index of $(\ell-1)$ in the extreme limit. It was subsequently observed that this phenomenon is linked to the fact that the extremal charged RN BH preserves super-symmetry, and therefore responds similarly to fields of different spin~\cite{Onozawa:1996ba}. However, this effect was not observed in the case of higher-dimensional charged BHs~\cite{Konoplya:2003dd}.

\subsection{The semiclassical approach of Regge poles}
\label{sec_3_3}

\begingroup
\squeezetable
\begin{table*}[t]
\caption{\label{tab:table_3} The lowest Regge poles $\lambda_{n}^{(1)}(\omega)$ for electromagnetic-type perturbations ($s = 1$) and $\lambda_{n}^{(2)}(\omega)$ for gravitational-type perturbations ($s = 2$) \textit{versus} semiclassical results given by~\eqref{RP_s} for the charge-to-mass ratios (Q/M=0.50) and~\eqref{PR_s_extrem} for extremal charged case (Q/M=1) and the relative error. We assume $2M=1$.}
\smallskip
\centering
\begin{ruledtabular}
\begin{tabular}{ccccccc}
     &             &      &  \text{(exact)}           &  \text{(semiclassical)}  &  \text{(exact)}           &
\text{(semiclassical)}
\\
 $Q$ & $2M\omega$  &  $n$ & $\lambda_n^{(1)}(\omega)$ & $\lambda_n^{(1)}(\omega)$ & $\lambda_n^{(2)}(\omega)$ &
$\lambda_n^{(2)}(\omega)$
 \\
\hline
$\mathbf{0.50M}$ &  $6$ & $1$ &$14.775517 + 0.483943 i$  & $14.778154 + 0.485909 i$  & $15.134550 + 0.47990406 i$  &  $15.132402 + 0.482210 i$\\[-1ex]
    &             &      &                                &  ($0.0178\%, 0.4061\%$)   &                             &
($2.36 \times 10^{-5}\%, 0.0071\%$)\\[+0.25ex]

                 &      & $2$ &$14.792059 + 1.450965 i$  & $14.794958 + 1.457726 i$  & $15.150065 + 1.4389603 i$&
$15.149207 + 1.446631 i$  \\[-1ex]
     &            &      &                                & ($0.0196\%, 0.4659\%$)    &                          &
($0.0002\%, 0.01581\%$) \\[+0.25ex]

                 &      & $3$ &$14.824835 + 2.415433 i$  & $14.828568 + 2.429543 i$  & $15.180854 + 2.3957873 i$  &
$15.182817 + 2.411051 i$  \\[-1ex]
      &           &      &                                & ($0.0251\%, 0.5841\%$)    &                           &
($0.0012\%, 0.0352\%$)\\ [+0.25ex]

                 &      & $4$ &$14.873253 + 3.375772 i$  & $14.878983 + 3.401360 i$  & $15.226446 + 3.3489852 i$  &
$15.233231 + 3.375472 i$ \\[-1ex]
       &          &      &                                & ($0.0385\%, 0.7580\%$)    &                           &
($0.0041\%, 0.0694\%$) \\ [+0.25ex]

                 &      & $5$ &$14.936484 + 4.330579 i$  & $14.946202 + 4.373177 i$  & $15.286167 + 4.2972773 i$  &
$15.300451 + 4.339892 i$ \\[-1ex]
        &         &      &                                & ($0.0650\%, 0.9836\%$)    &                            &
($0.0112\%, 0.1242\%$) \\  [+0.25ex]

\hline
$\mathbf{M}$      &  $6$ & $1$ &$11.545336 + 0.35160757 i$  & $11.545333 + 0.351582 i$  & $12.545336 + 0.35160757 i$  &  $12.545333 + 0.3515825 i$\\[-1ex]
         &        &      &                                &  ($2.57 \times 10^{-5}\%, 0.0071\%$)   &                             &
($0.01419\%, 0.4806\%$)\\[+0.25ex]

                 &      & $2$ &$11.560231 + 1.0540222 i$  & $11.560200 + 1.053855 i$  & $12.560231 + 1.0540222 i$&
$12.560200 + 1.0538555 i$  \\[-1ex]
         &        &      &                                & ($0.0002\%, 0.01581\%$)    &                          &
($0.005664\%, 0.5331\%$) \\[+0.25ex]

                 &      & $3$ &$11.589734 + 1.7540710 i$  & $11.589581 + 1.7534528 i$  & $12.589734 + 1.7540710 i$  &
$12.589581 + 1.7534528 i$  \\[-1ex]
          &       &      &                                & ($0.0013\%, 0.0352\%$)    &                           &
($0.01293\%, 0.6371\%$)\\[+0.25ex]

                 &      & $4$ &$11.633296 + 2.4502928 i$  & $11.632772 + 2.4485903 i$  & $12.633296 + 2.4502928 i$  &
$12.632772 + 2.4485903 i$ \\[-1ex]
           &      &      &                                & ($0.0045\%, 0.0694\%$)    &                           &
($0.04456\%, 0.7909\%$) \\[+0.25ex]

                 &      & $5$ &$11.690142 + 3.1413859 i$  & $11.688716 + 3.1374842 i$  & $12.690142 + 3.1413859 i$  &
$12.688716 + 3.1374842 i$ \\[-1ex]
            &     &      &                                & ($0.0122\%, 0.1242\%$)    &                            &
($0.09344\%, 0.9917\%$) \\[+0.25ex]

\end{tabular}
\end{ruledtabular}
\end{table*}
\endgroup

It is possible to derive an analytical approximations for the lowest RPs for the RN BH by extending the approach developed by Dolan and Ottewill~\cite{Dolan:2009}, which enabled them to derive analytical expressions for the lowest QNM frequencies of spherically symmetric spacetimes and the RPs of the Schwarzschild BH. It is noteworthy that the key idea of their approach is based on an ansatz for the resonant modes (i.e., the QNMs or the Regge modes), which relates high-$\ell$ or high-$\omega$ resonant modes to null geodesics that start at infinity and end in perpetual orbit on the photon (graviton) sphere. Accordingly, we obtained expressions for the lowest Regge poles
\begin{widetext}
\begin{eqnarray}\label{RP_s}
   \lambda_n^{(s)}(\omega)&=&\, b_c \,\omega
   +\left(\frac{i(2n-1)\alpha}{\sqrt{2}\sqrt{\alpha(3+\alpha)}}\mp\frac{\sqrt{3-\alpha}}{\sqrt{2}\sqrt{3+\alpha}}\right)
   + \frac{(1+\alpha)^{3/2}}{32\sqrt{2}\alpha^2(3+\alpha)^{5/2}}\, \left(\pm i 12(2 n - 1)(\alpha - 1) \sqrt{\alpha(3 - \alpha)}\right.\nonumber \\
  &+&\left. 6 n (n - 1)(4\alpha^2-7\alpha+5)+ 84\alpha^2-25\alpha+11 \vphantom{ \frac{(1+\alpha)^{3/2}}{32\sqrt{2}\alpha^2(3+\alpha)^{5/2}}}\right)\frac{1}{(M\omega)}
  + \underset{\omega \to \infty}{\mathcal{O}}\left(\frac{1}{\left(2M\omega\right)^2}\right),
\end{eqnarray}
\end{widetext}
where
\begin{equation}\label{bc}
  b_c= \frac{r_c}{\sqrt{f(r_c)}},
\end{equation}
is the critical impact parameter for the ray that asymptotes to the photon (graviton) orbit at radius $r_c$ which is given by
\begin{equation}\label{rc}
  r_c =  \frac{1}{2}M \left(3+\alpha\right),
\end{equation}
and
\begin{equation}\label{alpha}
  \alpha  = \sqrt{9-8\left(\frac{Q}{M}\right)^2}.
\end{equation}

It should be emphasized that the approximation~\eqref{RP_s} remains valid even at the critical charge-to-mass ratio $Q/M = 1$, when the two horizons coalesce and disappear. The approximation only breaks down when the unstable circular orbit disappears, which occurs at $Q/M = 3/(2\sqrt{2})$. This holds true for the formula derived in Ref.~\cite{Dolan:2009}, which provides an analytical expression for the QNM frequencies of the RN BH.

For the extremal charged RN BH case, $Q/M = 1$, we obtain
\begin{widetext}
\begin{eqnarray}\label{PR_s_extrem}
   \lambda_n^{(s)}(\omega) &=&   b_c \,\omega +\left(\frac{i(2n-1)}{2\sqrt{2}}\mp \frac{1}{2}\right) +\frac{6n(n-1)+35}{256}\frac{1}{(M\omega)} \nonumber \\
   &-&\frac{i(2n-1)\left[31n(n-1)+411\right]}{16384\sqrt{2}}\frac{1}{\left(M\omega\right)^2}
   -\frac{3 n (n - 1) \left[277 n (n - 1) + 6603\right] + 13585}{2097152}\frac{1}{\left(M\omega\right)^3} \nonumber \\
   &+& \underset{\omega \to \infty}{\mathcal{O}}\left(\frac{1}{(2M\omega)^4}\right).
\end{eqnarray}
\end{widetext}
Now, by using approximation~\eqref{PR_s_extrem}, while considering the difference between $\lambda_n^{(2)}$ and $\lambda_n^{(1)}$, we asymptotically obtain relation~\eqref{Relation_PRs}.

Finally, it is noteworthy that the results~\eqref{RP_s}~and~\eqref{PR_s_extrem} allow us to associate the lowest RPs with surface waves propagating close to the photon (graviton) sphere~\cite{Decanini:2002ha,Andersson1994bis,Folacci:2021uld}. These waves are dispersive and damped. Furthermore, this association establishes a partial link between the RP spectrum and the QNM frequency spectrum. Using the Regge trajectories, we can semiclassically construct the complex frequencies
\begin{equation}\label{QNMS_expres}
 \omega_{\ell n} ^{(s)}= {\omega_{\ell n}^{(\Re)}}^{(s)} - i\,  {\omega_{\ell n}^{(\Im)}}^{(s)},
\end{equation}
of the weakly damped QNMs by considering Bohr-Sommerfeld-type quantization conditions. Here, ${\omega_{\ell n}^{(\Re)}}^{(s)} > 0$ and ${\omega_{\ell n}^{(\Im)}}^{(s)} > 0$ as well as ${\omega_{\ell n}^{(\Re)}}^{(s)} \gg {\omega_{\ell n}^{(\Im)}}^{(s)}$.

The first of these semiclassical formulas provides the location of the excitation frequencies ${\omega_{\ell n}^{(\Re)}}^{(s)}$ of the resonances generated by $n$th RP:
\begin{equation}\label{Re_w}
\text{Re}\left[\lambda_n^{(s)}\left({\omega_{\ell n}^{(\Re)}}^{(s)}\right)\right] = \ell +\frac{1}{2}, \quad\quad \ell \in \mathbb{N}.
\end{equation}
The widths of these resonances can be obtained via a second semiclassical formula
\begin{equation}\label{Im_w}
  {\omega_{\ell n}^{(\Im)}}^{(s)}= \frac{\text{Im}\left[\lambda_n^{(s)}(\omega)\right]\left(\frac{d}{d\omega} \text{Re}\left[\lambda_n^{(s)}(\omega)\right]\right)}{\left(\frac{d}{d\omega}\text{Re}\left[\lambda_n^{(s)}(\omega)\right]\right)^2 +\left(\frac{d}{d\omega}\text{Im}\left[\lambda_n^{(s)}(\omega)\right]\right)^2} 
\end{equation}
at $\omega={\omega_{\ell n}^{(\Re)}}^{(s)}$. In the frequency range where the condition $\big{|}d/d\omega \text{Re} \lambda_n^{(s)} (\omega)\big{|} \gg \big{|}d/d\omega \text{Im} \lambda_n^{(s)} (\omega)\big{|}$ is fulfilled, this relation is reduced to
\begin{equation}\label{Im_w_bis}
 {\omega_{\ell n}^{(\Im)}}^{(s)} = \frac{\text{Im}\left[\lambda_n^{(s)}(\omega)\right]}{\frac{d}{d\omega} \text{Re}\left[\lambda_n^{(s)}(\omega)\right]}\Biggr{|}_{\omega={\omega_{\ell n}^{(\Re)}}^{(s)}}.
\end{equation}

Using Eqs.~\eqref{RP_s}, \eqref{Re_w}, and \eqref{Im_w_bis}, we can derive the asymptotic behavior of QNM frequencies for a RN BH ($Q<M$). We then obtain
\begin{align}\label{QNM_s}
   &\omega_{\ell n}^{(s)} =\, \frac{\ell+1/2}{b_c}\nonumber\\
   &-\left(\frac{i(2n-1)\sqrt{\alpha(1 +\alpha)}}{M (\alpha+3)^2}\mp \frac{\sqrt{(3-\alpha)(1+\alpha)}}{M (\alpha+3)^2}\right)\nonumber \\
   &- \frac{(1+\alpha)^{3/2}}{32 M \sqrt{2}\alpha^2(3+\alpha)^{5/2}}\, \left(\pm i 12(2 n - 1)(\alpha - 1) \sqrt{\alpha(3 - \alpha)}\right.\nonumber \\
  &+\left. 6 n (n - 1)(4\alpha^2-7\alpha+5)+ 84\alpha^2-25\alpha+11 \vphantom{ \frac{(1+\alpha)^{3/2}}{32\sqrt{2}\alpha^2(3+\alpha)^{5/2}}}\right)\frac{1}{\ell} \nonumber\\
  &+ \underset{\ell \to \infty}{\mathcal{O}}\left(\frac{1}{\ell^2}\right)
\end{align}
and for extremal charged RN BH case ($Q=M$), we have
\begin{align}\label{QNMs_extrem}
 \quad\quad\omega_{\ell n}^{(s)} &=\, \frac{\ell+1/2}{b_c}- \frac{1}{b_c}\left(\frac{i(2n-1)}{2\sqrt{2}}\mp \frac{1}{2}\right) \nonumber\\
  &-\frac{6n(n-1)+35}{256 M}\frac{1}{\ell} + \underset{\ell \to \infty}{\mathcal{O}}\left(\frac{1}{\ell^2}\right)
\end{align}
with $\ell \in \mathbb{N}$ and $n = 1, 2, 3,\ldots$.

In expressions~\eqref{RP_s},~\eqref{PR_s_extrem},~\eqref{QNM_s}, and~\eqref{QNMs_extrem}, the upper symbol of $\mp$ ($\pm$) is associated with electromagnetic-type perturbations ($s=1$), and the lower one is associated with gravitational-type perturbations ($s=2$). It should be noted that by setting $n=1$ in \eqref{QNM_s}~and~\eqref{QNMs_extrem}, i.e., the fundamental QNM frequency, the results presented by Dolan and Ottewill in Ref.~\cite{Dolan:2009} are obtained.

In Table~\ref{tab:table_3}, we compare the lowest Regge poles $\lambda_{n}^{(1)}(\omega)$ for electromagnetic-type perturbations ($s=1$) and $\lambda_{n}^{(2)}(\omega)$ for gravitational-type perturbations ($s=2$) obtained through numerical methods with the semiclassical approximations outlined in Eqs.~\eqref{RP_s} and \eqref{PR_s_extrem} for $2M\omega=6$ and charge-to-mass ratios $Q/M=0.50$ and $1$. The data suggests that the leading-order terms of semiclassical approximations for the ``high''-frequency regime are in agreement with the numerical results, accurately capturing the fundamental characteristics of the RPs.

\section{Scattering and conversion cross sections and their CAM representations}
\label{sec_4}

\subsection{The partial wave expansion}
\label{sec_4_1}

The cross section of different processes are given by (see Refs.~\cite{Matzner:1976kj,Futterman:1988ni} for details)
\begin{equation}\label{Diff_Cross_Sec_si_sf}
\frac{d\sigma}{d\Omega}^{(s_i \to s_f)} =\big{|}\mathfrak{f}^{(+,s_i s_f)}(\theta)\big{|}^2+\big{|}\mathfrak{f}^{(-,s_i s_f)}(\theta)\big{|}^2,
\end{equation}
with
\begin{equation}\label{fpm_si_sf}
  \mathfrak{f}^{(\pm,s_i s_f)}(\theta) =\mathfrak{D}^{(\pm, s_i s_f)}_\theta \, \tilde{\mathfrak{f}}^{(\pm,s_i s_f)}(\theta),
\end{equation}
where $\mathfrak{f}^{(+,s_i s_f)}(\theta)$ and $\mathfrak{f}^{(-,s_i s_f)}(\theta)$ are respectively the helicity-preserving and helicity-reversing scattering amplitudes, while $\mathfrak{D}^{(\pm, s_i s_f)}_\theta$ are the differential operators associated with each process.

\subsubsection{Scattering amplitude of $EW \to EW$}
\label{sec_4_1_1}

The scattering amplitudes for incoming EW ($s_i = 1$) to an outgoing EW ($s_f = 1$) are given by~\eqref{fpm_si_sf}
\begin{equation}\label{fpm_1_1}
  \mathfrak{f}^{(\pm,11)}(\theta) =\mathfrak{D}^{(\pm,11)}_\theta \, \tilde{\mathfrak{f}}^{(\pm,11)}(\theta),
\end{equation}
with 
\begin{eqnarray}\label{fpm_tild_EW_to_EW}
  \tilde{\mathfrak{f}}^{(\pm,1 1)}(\theta) &=&\frac{1}{2i\omega} \sum_{\ell = 1}^{+\infty}\,  \frac{2\ell+1}{\ell(\ell+1)} \nonumber\\
    & &\times \left[\frac{1}{2} \left(S_\ell^{(e,11)}\pm S_\ell^{(o,11)}\right) -\frac{1\pm 1}{2}\right] \nonumber\\
    & &\times P_\ell(\cos\theta),
\end{eqnarray}
and the associated differential operator is given by
  \begin{equation}\label{Opera_Diff_EW}
  \mathfrak{D}^{(\pm,1 1)}_\theta =-\left(  \frac{d^2}{d\theta^2}\pm  \frac{1}{\sin\theta} \frac{d}{d\theta}\right),
\end{equation}
while the $S$-matrix elements $S_{\ell}^{(e/o,11)}(\omega)$ appearing in Eq.~\eqref{fpm_tild_EW_to_EW} are given by~\eqref{Matrice_S_11}.

\subsubsection{Scattering amplitude of  $GW \to GW$}
\label{sec_4_1_2}

The scattering amplitudes for incoming GW ($s_i = 2$) to an outgoing GW ($s_f = 2$) are given by~\eqref{fpm_si_sf}
\begin{equation}\label{fpm_2_2}
  \mathfrak{f}^{(\pm,22)}(\theta) =\mathfrak{D}^{(\pm,22)}_\theta \, \tilde{\mathfrak{f}}^{(\pm,22)}(\theta),
\end{equation}
with 
\begin{eqnarray}\label{fpm_tild_GW_to_GW}
  \tilde{\mathfrak{f}}^{(\pm,2 2)}(\theta) &=& \frac{1}{2i\omega} \sum_{\ell = 2}^{\infty}\, \frac{(2\ell+1)}{(\ell-1)\ell(\ell+1)(\ell+2)} \nonumber \\
 & &\times \left[\frac{1}{2}\left(S_{\ell}^{(e,22)} \pm S_{\ell}^{(o,22)}\right)-\left(\frac{1\pm 1}{2}\right)\right] \nonumber \\
 & &\times P_{\ell}(\cos\theta),
\end{eqnarray}
and the differential operator $\mathfrak{D}^{(\pm, 2 2)}_x$ which act on the partial wave series \eqref{fpm_tild_GW_to_GW} is given by
\begin{equation}\label{Opera_Diff_Grav}
\mathfrak{D}^{(\pm, 2 2)}_x = (1\pm x)^2 \frac{d}{dx} \left\{(1\mp x)\frac{d^2}{dx^2}\left[(1\mp x)\frac{d}{dx}\right]\right\},
\end{equation}
where the variable $x$ is linked to the scattering angle $\theta$ by $x=\cos\theta$. In Eq.~\eqref{fpm_tild_GW_to_GW}, the $S$-matrix elements $S_{\ell}^{(e/o,22)}$ are given by~\eqref{Matrice_S_22}.

\subsubsection{Conversion amplitude of  $EW \to GW$}
\label{sec_4_1_3}

The conversion scattering amplitudes for incoming EW ($s_i = 1$) to an outgoing GW ($s_f = 2$) are given by~\eqref{fpm_si_sf}
\begin{equation}\label{fpm_1_2}
  \mathfrak{f}^{(\pm,12)}(\theta) =\mathfrak{D}^{(\pm,12)}_\theta \, \tilde{\mathfrak{f}}^{(\pm,22)}(\theta),
\end{equation}
with
\begin{eqnarray}\label{fpm_tild_EW_to_GW}
  \tilde{\mathfrak{f}}^{(\pm,1  2)}(\theta) &=& -\frac{1}{2i\omega} \sum_{\ell = 2}^{+\infty}\,  \frac{2\ell+1}{\ell(\ell+1) \sqrt{(\ell+2)(\ell-1)}} \nonumber\\
   & &\times \left[\frac{1}{2} \left(S_\ell^{(e,1 2)}\pm S_\ell^{(o,1 2)}\right)\right] \nonumber\\
   & &\times  P_\ell(\cos\theta),
\end{eqnarray}
and the associated differential operator $\mathfrak{D}^{(\pm, 1 2)}_\theta$  is given by
\begin{equation}\label{Opera_Diff_EW_to_GW}
   \mathfrak{D}^{(\pm, 1 2)}_\theta =\pm \frac{1\pm \cos\theta}{1\mp \cos\theta} \frac{d}{d\theta} \left\{\frac{(\cos\theta \mp 1)^2}{\sin\theta}\frac{d}{d\theta}\left[\frac{1}{\sin\theta} \frac{d}{d\theta}\right]\right\},
\end{equation}
while the $S$-matrix elements  $S_{\ell}^{(e/o,1 2)}$ appearing in Eq.~\eqref{fpm_tild_EW_to_GW} are given by~\eqref{Matrice_S_12}.

It should be noted that the $EW \to GW$ and $GW\to EW$ conversion cross section are equal, i.e., 
${d\sigma/d\Omega}^{(1 \to 2)} = {d\sigma/d\Omega}^{(2 \to 1)}$.

It is worth noting that the notations we used for the scattering amplitudes~\eqref{fpm_tild_EW_to_EW}~and~\eqref{fpm_tild_GW_to_GW}, as well as the conversion amplitudes~\eqref{fpm_tild_EW_to_GW}, differ slightly from those presented in our previous article~\cite{OuldElHadj:2021fqi} and generally from those found in the literature. Specifically, we employed the Legendre polynomials $P_\ell(\cos\theta)$ \cite{AS65}, rather than the spin-weighted spherical harmonics ${}_sY_\ell^m(\theta)$. This notation is more tractable and allows us to extract the exact CAM representations from their partial wave expansions.

\subsection{CAM representations}
\label{sec_4_2}

\subsubsection{Scattering amplitude of $EW \to EW$}
\label{sec_4_2_1}

\begin{figure}[htp!]
 \includegraphics[scale=0.45]{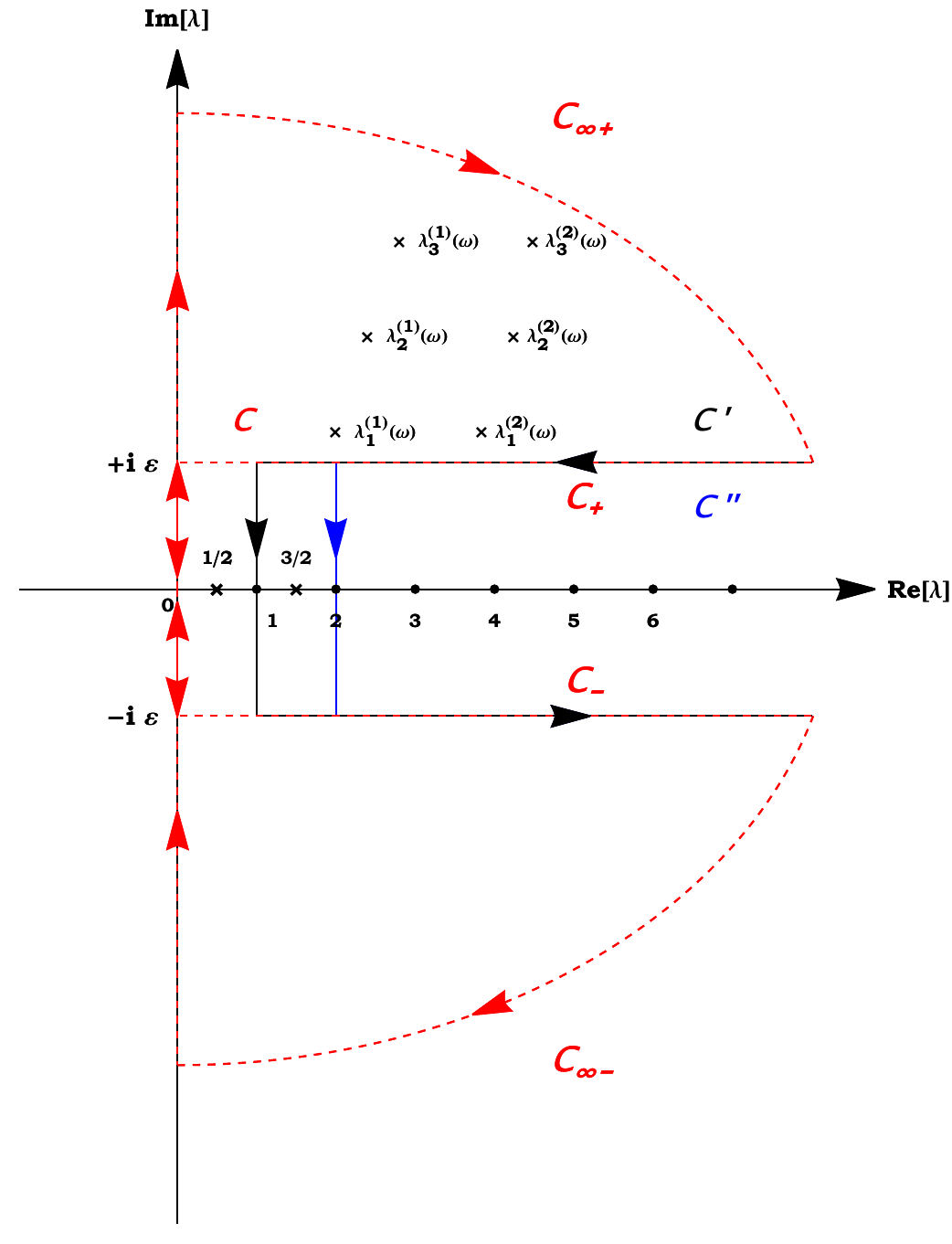}
\caption{\label{Fig:Contour_Shift_s1_s2} Integration contours in the complex $\lambda$ plane  : i) ${\cal C'}$~(black line)   associated with scattering amplitude of electromagnetic field $\eqref{SW_EW_Scattering_amp} $ and ii) ${\cal C''}$~(blue line) the one associated with scattering amplitude of gravitational field $\eqref{SW_Grav_Scattering_amp}$ and the conversion amplitude~\eqref{SW_EW_to_GW_Scattering_amp}. Here, ${\cal C} = \cal{C_+} \cup \cal{C_-}$ defines~\eqref{SW_EW_Scattering_amp_bis_1}~,~\eqref{SW_Grav_Scattering_amp_bis_1} as well as~\eqref{SW_EW_to_GW_Scattering_amp_bis_2} and its deformations allow to collect the contributions of the Regge poles  $\lambda_n^{(1)}$ and  $\lambda_n^{(2)}$.}
\end{figure}

To construct the CAM representation of the scattering amplitudes \eqref{fpm_1_1} and  \eqref{fpm_tild_EW_to_EW}, we use the Sommerfeld-Watson transformation in the form \cite{Watson18,Sommerfeld49,Newton:1982qc}
\begin{equation}\label{Sommerfeld_Watson_1}
\sum_{\ell=1}^{+\infty}(-1)^{\ell}F(\ell) = \frac{i}{2}\int_{C'}d\lambda\frac{F(\lambda-1/2)}{\cos(\pi\lambda)},
\end{equation}
following the steps in Sec.~II of Ref.~\cite{Folacci:2019cmc}, with some subtleties to take into account. Here, $F(\cdot)$ has no singularities on the real $\lambda$ axis, and ${\cal C'}=]+\infty +i\epsilon,1+i\epsilon] \cup [1+i\epsilon,1-i\epsilon] \cup [1-i\epsilon, +\infty -i\epsilon[$ with $\epsilon\to 0_+$ (see Fig.~\ref{Fig:Contour_Shift_s1_s2}).

Now, by replacing the discrete sum over the ordinary angular momentum $\ell$ with a contour integral in the complex $\lambda$ plane (that is, in the complex $\ell$ plane with $\lambda = \ell+1/2$), and noting that $P_\ell(\cos\theta)= (-1)^\ell P_\ell(-\cos\theta)$, we obtain
\begin{eqnarray}\label{SW_EW_Scattering_amp}
\tilde{\mathfrak{f}}^{(\pm,11)}(\theta) &=&  \frac{1}{2 \omega}  \int_{\cal C'} d\lambda \, \frac{\lambda}{(\lambda^2-1/4)\cos (\pi \lambda)} \nonumber \\
&& \times \left[ \frac{1}{2} \left(S^{(e,11)}_{\lambda -1/2} \pm S^{(o,11)}_{\lambda -1/2} \right) - \left( \frac{1 \pm 1}{2} \right) \right].\nonumber \\ && \times \, P_{\lambda -1/2} (-\cos\theta).
\end{eqnarray}

It should be noted that the Legendre function of the first kind, $P_{\lambda -1/2} (-z)$, denotes the analytic extension of the Legendre polynomials $P_{\ell} (z)$ and is defined in terms of hypergeometric functions~\cite{AS65}
\begin{equation}\label{Legendre_Function}
  P_{\lambda -1/2} (z)= F[1/2-\lambda,1/2+\lambda;1;(1-z)/2],
\end{equation}
while $S^{(e,11)}_{\lambda -1/2}$ and $S^{(o,11)}_{\lambda -1/2}$ denote the analytic extensions of the matrices $S^{(e,11)}_{\ell}$ and $S^{(o,11)}_{\ell}$, they can be defined from the complex amplitudes $A_{s\omega,\lambda-1/2}^{(\pm,e/o)}$ [cf. \eqref{Matrice_S_11} and \eqref{reflec_coeffs}]
\begin{eqnarray}\label{Matrice_S_EW_CAM}
  & &S_{\lambda-1/2}^{(e/o,11)} =  e^{i(\lambda+1/2)\pi} \nonumber\\
  & & \quad\quad \times \left[\frac{A_{1\omega,\lambda-1/2}^{(+,\text{e/o})}}{A_{1\omega,\lambda-1/2}^{(-,\text{e/o})}} \cos^2\psi + \frac{A_{2\omega,\lambda-1/2}^{(+,\text{e/o})}}{A_{2\omega,\lambda-1/2}^{(-,\text{e/o})}}\sin^2\psi\right],
\end{eqnarray}
which are the analytical extensions of $A_{s\omega\ell}^{(\pm,e/o)}$ obtained from the analytic extension of the modes $\phi_{s\omega\ell}^{(e/o)}$, that is, from the function $\phi_{s\omega,\lambda-1/2}^{(e/o)}$. It is worth mentioning that the Chandrasekhar-Detweiler relation~\eqref{Relation_Chandra-Detweiler} leads, in the CAM plane, to
\begin{subequations}\label{CAM_Rel}
\begin{equation}\label{CAM_Rel_1}
A^{(-,\text{e})}_{s\omega,\lambda-1/2} = A^{(-,\text{o})}_{s\omega,\lambda-1/2} \equiv A^{(-)}_{s\omega,\lambda-1/2},
\end{equation}
and
\begin{eqnarray}\label{CAM_Rel_2}
\left[\right.& &\left.(\lambda^2-1/4)(\lambda^2-9/4)-2i\omega q_{s}\right] A^{(+,\text{e})}_{s\omega,\lambda-1/2} = \nonumber\\
              & & \left[(\lambda^2-1/4)(\lambda^2-9/4)+2i\omega\right. \left.q_{s}\right] A^{(+,\text{o})}_{s\omega,\lambda-1/2}.
\end{eqnarray}
\end{subequations}

Because the residues of the matrices $S_{\lambda-1/2}^{(e,11)}$ and $S_{\lambda-1/2}^{(o,11)}(\omega)$ at the RPs play a crucial role in the CAM construction, it is important to define them correctly. To achieve this, we first express the matrices in a way that separates the contributions of RPs $\lambda_{n}^{(1)}$ associated with electromagnetic-type perturbations ($s=1$) from those associated with gravitational-type perturbations ($s=2$) $\lambda_{n}^{(2)}$. Thus, we obtain
\begin{equation}\label{Matrice_S_EW_CAM_bis}
 S_{\lambda-1/2}^{(e/o,11)} =  S_{1,\lambda-1/2}^{(e/o,11)} + S_{2,\lambda-1/2}^{(e/o,11)},
\end{equation}
where
\begin{equation}
\label{Matrice_S_EW_CAM_bis_s1}
S_{1,\lambda-1/2}^{(e/o,11)} =\cos^2\psi \,  e^{i(\lambda+1/2)\pi}   \left[\frac{A_{1\omega,\lambda-1/2}^{(+,\text{e/o})}}{A_{1\omega,\lambda-1/2}^{(-)}}\right],
\end{equation}
and
\begin{equation}
\label{Matrice_S_EW_CAM_bis_s2}
S_{2,\lambda-1/2}^{(e/o,11)} =  \sin^2\psi \, e^{i(\lambda+1/2)\pi}   \left[\frac{A_{2\omega,\lambda-1/2}^{(+,\text{e/o})}}{A_{2\omega,\lambda-1/2}^{(-)}}\right].
\end{equation}
We then define the associated residues for each contribution by
\begin{equation}\label{residues_EW_rn1}
     r^{(e/o,11)}_{1 n} = \cos^2\psi \,  r^{(e/o)}_{1 n},
\end{equation}
and
\begin{equation}\label{residues_EW_rn2}
     r^{(e/o,11)}_{2 n} =  \sin^2\psi\, r^{(e/o)}_{2 n},
\end{equation}
where
\begin{equation}\label{residues_rn_s}
     r^{(e/o)}_{s n} = e^{i \pi(\lambda_n^{(s)}(\omega)+1/2)} \left[\frac{A_{s\omega,\lambda-1/2}^{(+,\text{e/o})}}{\frac{d}{d\lambda} A_{s\omega,\lambda-1/2}^{(-)}} \right]_{\lambda = \lambda_{n}^{(s)}(\omega)}.
\end{equation}
\vspace{2pt}

Now, in order to collect, by using Cauchy's theorem, the contributions of all RPs, we must first shift the contour $\mathcal{C}'$ to the left so that it coincides with $\mathcal{C}=]+\infty +i\epsilon, +i\epsilon] \cup
[+i\epsilon,-i\epsilon] \cup [-i\epsilon, +\infty -i\epsilon[$ (see Fig~\ref{Fig:Contour_Shift_s1_s2}). However, we then introduce a spurious double pole at $\lambda =1/2$ (i.e., at $\ell=0$) corresponding to the term $1/[(\lambda-1/2)\cos(\pi\lambda)]$ which we should remove by subtracting the contribution of the associated residue. Thus, we have
\begin{widetext}
\begin{eqnarray}\label{SW_EW_Scattering_amp_bis_1}
& &\tilde{\mathfrak{f}}^{(\pm,11)}(\theta) =  \frac{1}{2 \omega} \left\{ \int_{\cal C} d\lambda \, \frac{\lambda}{(\lambda^2-1/4)\cos (\pi \lambda)}
   \left[ \frac{1}{2} \left(S^{(e,11)}_{\lambda -1/2} \pm S^{(o,11)}_{\lambda -1/2} \right) - \left( \frac{1 \pm 1}{2} \right) \right] \, P_{\lambda -1/2} (-\cos\theta)\right. \nonumber\\
    & & \left.-2i\pi \underset{\lambda \to 1/2 }{\lim} \frac{d}{d\lambda}\left[\left(\lambda-1/2\right)^2 \frac{\lambda}{(\lambda^2-1/4)\cos (\pi \lambda)}
   \left[ \frac{1}{2} \left(S^{(e,11)}_{\lambda -1/2} \pm S^{(o,11)}_{\lambda -1/2}\right) - \left( \frac{1 \pm 1}{2} \right) \right] \, P_{\lambda -1/2} (-\cos\theta)\right] \right\}.
\end{eqnarray}
By using
\begin{equation}\label{SP_Residue_1}
  \left[\frac{d}{d\lambda}P_{\lambda-1/2}(-x)\right]_{\lambda=1/2}= \ln\left(\frac{1-x}{2}\right),
\end{equation}
we can show that
\begin{eqnarray}\label{SW_EW_Scattering_amp_bis_2}
\tilde{\mathfrak{f}}^{(\pm,11)}(\theta) & = & \frac{1}{2 \omega}  \int_{\cal C} d\lambda \, \frac{\lambda}{(\lambda^2-1/4)\cos (\pi \lambda)}
   \left[ \frac{1}{2} \left(S^{(e,11)}_{\lambda -1/2} \pm S^{(o,11)}_{\lambda -1/2} \right) - \left( \frac{1 \pm 1}{2} \right) \right] \, P_{\lambda -1/2} (-\cos\theta)\nonumber \\
   & &+\frac{i}{2\omega}  \left[ \frac{1}{2} \left(S^{(e,11)}_0 \pm S^{(o,11)}_0 \right) - \left( \frac{1 \pm 1}{2} \right) \right] \ln\left[\frac{1}{2}\left(1-\cos\theta\right)\right] \nonumber\\
   & &+ \text{terms independent of}\,\,\theta.
\end{eqnarray}
\end{widetext}

The terms introduced in Eq~\eqref{SW_EW_Scattering_amp_bis_2} to neutralize the spurious modes can be evaluated numerically. However, they do not necessarily contribute to scattering amplitudes~\eqref{fpm_1_1}. Indeed, by applying the differential operator given by~\eqref{Opera_Diff_EW}, we obtain the following expression
\begin{eqnarray}\label{SW_EW_Scattering_amp_bis_3}
\mathfrak{f}^{(\pm,11)}& &(\theta) =\mathfrak{D}_\theta^{(\pm,11)}\left\{ \frac{1}{2\omega} \int_{\cal{C}} d \lambda\, \frac{\lambda}{(\lambda^2-1/4)\cos(\pi\lambda)}\right. \nonumber\\
& &\times \left[ \frac{1}{2} \left(S^{(e,11)}_{\lambda -1/2} \pm S^{(o,11)}_{\lambda -1/2} \right) - \left( \frac{1 \pm 1}{2} \right) \right] \nonumber\\
& &\left. \times P_{\lambda -1/2} (-\cos\theta)\vphantom{\left[ \frac{1}{2} \left(S^{(e,11)}_{\lambda -1/2}(\omega) \pm S^{(o,11)}_{\lambda -1/2}(\omega) \right) - \left( \frac{1 \pm 1}{2} \right) \right]} \right\} + \mathfrak{f}_{\text{SP}}^{(\pm,11)}(\theta),
\end{eqnarray}
where the amplitudes $\mathfrak{f}_{\text{SP}}^{(\pm,11)}(\theta)$ denote the ``shift corrections'' introduced to neutralize the spurious double pole at $\lambda =1/2$, and are given by
\begin{subequations}\label{Shift_Corrections_EW}
\begin{equation}\label{fsp_+}
\mathfrak{f}_{\text{SP}}^{(+,11)}(\theta)=0,
\end{equation}
and
\begin{equation}\label{fsp_-}
\mathfrak{f}_{\text{SP}}^{(-,11)}(\theta)=\frac{i}{\omega}\,\frac{S_0^{(o,11)}}{\left(1-\cos\theta\right)}.
\end{equation}
\end{subequations}
Here, it should be noted that we have simplified the last expression by using the fact that
\begin{equation}\label{Simplification_EW}
  S_0^{(e,11)}+ S_0^{(o,11)} = 0.
\end{equation}

We then deform the integration contour $\mathcal{C}$ in Eq.~\eqref{SW_EW_Scattering_amp_bis_3} with the aim of collecting the RPs in mind (see Fig~\ref{Fig:Contour_Shift_s1_s2}), and we obtain
\begin{equation}\label{CAM_EW_Scattering_amp_tot}
\mathfrak{f}^{(\pm,11)}(\theta) = \mathfrak{f}^{(\pm,11)}_\text{\tiny{B}} (\theta) +  \mathfrak{f}^{(\pm,11)}_\text{\tiny{RP}} (\theta) + \mathfrak{f}_\text{\tiny{SP}}^{(\pm,11)}(\theta).
\end{equation}
Here, the background integral contribution is
\begin{subequations}\label{CAM_EW_Scattering_amp_decomp}
\begin{equation}\label{CAM_EW_Scattering_amp_decomp_Background}
\mathfrak{f}^{(\pm,11)}_\text{\tiny{B}} (\theta) = \mathfrak{f}^{(\pm,11)}_\text{\tiny{B},\tiny{Re}} (\theta) + \mathfrak{f}^{(\pm,11)}_\text{\tiny{B},\tiny{Im}} (\theta),
\end{equation}
with
\begin{eqnarray}\label{CAM_EW_Scattering_amp_decomp_Background_a}
& & \mathfrak{f}^{(\pm,11)}_\text{\tiny{B},\tiny{Re}} (\theta) = \mathfrak{D}^{(\pm,11)}_{\theta} \left\lbrace \frac{1}{2\pi \omega} \int_{{\cal C}_{-}} d\lambda \, \frac{\lambda}{(\lambda^2-1/4)}  \right. \nonumber \\
& &   \left. \phantom{\int_{{\cal C}_{-}}}  \times \left[S^{(e,11)}_{\lambda -1/2}\pm S^{(o,11)}_{\lambda -1/2} \right] \,Q_{\lambda -1/2}(\theta +i0)  \right\rbrace, \nonumber \\
\end{eqnarray}
and
\begin{eqnarray}\label{CAM_EW_Scattering_amp_decomp_Background_b}
& & \mathfrak{f}^{(\pm,11)}_\text{\tiny{B},\tiny{Im}} (\theta) = \mathfrak{D}^{(\pm,11)}_{\theta} \left\lbrace \frac{1}{2\pi \omega} \int_{+i\infty}^0 d\lambda \, \frac{\lambda}{(\lambda^2-1/4)}  \right. \nonumber \\
& &   \left. \phantom{\int_{\infty}^0}  \times \left[S^{(e,11)}_{\lambda -1/2} \pm S^{(o,11)}_{\lambda -1/2} \right] \,Q_{\lambda -1/2}(\theta +i0)  \right\rbrace, \nonumber \\
\end{eqnarray}
\end{subequations}
where ${\mathcal C}_{-}= [0,-i\epsilon] \cup [-i\epsilon, +\infty -i\epsilon[$ with $\epsilon \to 0_+$). The Regge pole contribution is
\begin{eqnarray}
\label{CAM_EW_Scattering_amp_decomp_RP}
&&\mathfrak{f}^{(\pm,1 1)}_\text{\tiny{RP}} (\theta)= \mathfrak{D}^{(\pm, 1 1)}_{\theta}  \nonumber\\
&&\left\lbrace
-\frac{i \pi}{2\omega}    \sum_{n=1}^{+\infty}  \sum_{s=1}^{2} \frac{ \lambda_n^{(s)}(\omega) \left[ r^{(e,11)}_{s n} \pm r^{(o,1 1)}_{s n} \right]}{[\lambda_n^{(s)}(\omega)^2-1/4] \cos[\pi \lambda_n^{(s)}(\omega)]} \right. \nonumber\\
&&\times \left. P_{\lambda_n^{(s)}(\omega) -1/2} (-\cos\theta) \vphantom{\sum_{s=1}^{2} \frac{ \lambda_n^{(s)}(\omega) \left[ r^{(e,11)}_{s n}(\omega) \pm r^{(o,1 1)}_{s n}(\omega) \right]}{[\lambda_n^{(s)}(\omega)^2-1/4] \cos[\pi \lambda_n^{(s)}(\omega)]}}  \right\rbrace,
\end{eqnarray}
which involves a sum over the RPs $\lambda_n^{(1)}(\omega)$ and $\lambda_n^{(2)}(\omega)$ lying in the first quadrant of the CAM plane and their associated residues (see Eq.~\eqref{residues_EW_rn1} and \eqref{residues_EW_rn2}). 

In Eqs.~\eqref{CAM_EW_Scattering_amp_decomp_Background_a} and \eqref{CAM_EW_Scattering_amp_decomp_Background_b}, $Q_{\lambda-1/2}(z)$ is the Legendre function of second kind and is related to $P_{\lambda-1/2}(z)$ by \cite{AS65}
\begin{eqnarray}\label{Legendre_function_second_kind}
& & Q_{\lambda -1/2}(x +i0) = \frac{\pi}{2 \cos(\pi \lambda)} \left[ P_{\lambda -1/2}(-x) \right. \nonumber \\
& & \qquad \qquad \left. - e^{-i\pi (\lambda-1/2)} P_{\lambda -1/2}(+x )\right].
\end{eqnarray}

It is important to note that the scattering amplitude $\mathfrak{f}^{(\pm)}(\theta)$, given by Eqs.~\eqref{CAM_EW_Scattering_amp_tot}--\eqref{CAM_EW_Scattering_amp_decomp_RP} and \eqref{Shift_Corrections_EW}, is equivalent to the initial partial wave expansions defined by~\eqref{fpm_1_1}--\eqref{Opera_Diff_EW}.


\subsubsection{Scattering amplitude of $GW \to GW$}
\label{sec_4_2_2}

In order to obtain the CAM representation of the scattering amplitude for GWs~\eqref{fpm_2_2}~and~\eqref{fpm_tild_GW_to_GW}, we will extend our treatment  of scattering of EWs (cf. Sec.~\ref{sec_4_2_1}) and provide the necessary key steps (see also Sec.II of Ref.~\cite{Folacci:2019vtt}).

By using the Sommerfeld-Watson transformation
\begin{equation}\label{Sommerfeld_Watson_2}
\sum_{\ell=1}^{+\infty}(-1)^{\ell}F(\ell) = \frac{i}{2}\int_{C''}d\lambda\frac{F(\lambda-1/2)}{\cos(\pi\lambda)},
\end{equation}
we replace the discrete sum over the ordinary angular momentum $\ell$ in Eq.~\eqref{fpm_tild_GW_to_GW} with a contour integration in the complex $\lambda$ plane. We then have
\begin{eqnarray}\label{SW_Grav_Scattering_amp}
\tilde{\mathfrak{f}}^{(\pm,22)}(\theta) &=&  \frac{1}{2 \omega}  \int_{\cal C''} d\lambda \, \frac{\lambda}{(\lambda^2-1/4) (\lambda^2-9/4)\cos (\pi \lambda)} \nonumber \\
&&     \times \left[ \frac{1}{2} \left(S^{(e,22)}_{\lambda -1/2} \pm S^{(o,22)}_{\lambda -1/2} \right) - \left( \frac{1 \pm 1}{2} \right) \right]\nonumber \\ && \times \, P_{\lambda -1/2} (-\cos\theta),
\end{eqnarray}
with  ${\mathcal C''}=]+\infty +i\epsilon,2+i\epsilon] \cup [2+i\epsilon,2-i\epsilon] \cup [2-i\epsilon, +\infty -i\epsilon[$ where $\epsilon\to 0_+$ (see Fig.~\ref{Fig:Contour_Shift_s1_s2}).

Here the matrices $S^{(e,22)}_{\lambda -1/2}$ and $S^{(o,22)}_{\lambda -1/2}$ are the analytical extensions of $S^{(e,22)}_{\ell}$ and $S^{(o,22)}_{\ell}$ which can also be defined from the complex amplitudes $A_{s\omega,\lambda-1/2}^{(\pm,e/o)}$ [cf.~\eqref{Matrice_S_22} and \eqref{reflec_coeffs}]
\begin{eqnarray}\label{Matrice_S_Grav_CAM}
& & S_{\lambda-1/2}^{(e/o,22)} =  e^{i(\lambda+1/2)\pi} \nonumber\\
& &\quad\quad \times\left[\frac{A_{2\omega,\lambda-1/2}^{(+,\text{e/o})}}{A_{2\omega,\lambda-1/2}^{(-)}} \cos^2\psi + \frac{A_{1\omega,\lambda-1/2}^{(+,\text{e/o})}}{A_{1\omega,\lambda-1/2}^{(-)}}\sin^2\psi\right].
\end{eqnarray}

It should be noted that the $S_{\lambda-1/2}^{(e,22)}$ and $S_{\lambda-1/2}^{(o,22)}$ matrices have the same RPs [cf.~Eq.\eqref{CAM_Rel_1}] and are identical to those previously defined for the electromagnetic field, specifically those associated with matrices~\eqref{Matrice_S_EW_CAM}.

Similarly, we separate the contributions associated with $s=1$ (i.e., $\lambda_n^{(1)}(\omega)$) from those associated with $s=2$ (i.e., $\lambda_n^{(2)}(\omega)$) in order to determine the associated residues of the $S_{\lambda-1/2}^{(e,22)}$ and $S_{\lambda-1/2}^{(o,22)}$ matrices
\begin{equation}\label{Matrice_S_Grav_CAM_bis}
 S_{\lambda-1/2}^{(e/o,22)}=  S_{1,\lambda-1/2}^{(e/o,22)} +  S_{2,\lambda-1/2}^{(e/o,22)},
\end{equation}
where
\begin{equation}
\label{Matrice_S_Grav_CAM_bis_s1}
S_{1,\lambda-1/2}^{(e/o,11)} =    \sin^2\psi \, e^{i(\lambda+1/2)\pi} \left[\frac{A_{1\omega,\lambda-1/2}^{(+,\text{e/o})}}{A_{1\omega,\lambda-1/2}^{(-)}}\right],
\end{equation}
and
\begin{equation}
\label{Matrice_S_Grav_CAM_bis_s2}
S_{2,\lambda-1/2}^{(e/o,11)} =     \cos^2\psi \, e^{i(\lambda+1/2)\pi} \left[\frac{A_{2\omega,\lambda-1/2}^{(+,\text{e/o})}}{A_{2\omega,\lambda-1/2}^{(-)}}\right],
\end{equation}
and then we define the associated residues by
\begin{equation}\label{residues_GW_rn1}
r^{(e/o,22)}_{1 n} = \sin^2\psi \, r^{(e/o)}_{1 n},
\end{equation}
and
\begin{equation}\label{residues_GW_rn2}
r^{(e/o,22)}_{2 n} = \cos^2\psi \, r^{(e/o)}_{2 n}.
\end{equation}

Taking into account the collection of RPs $\lambda_n^{(1)}(\omega)$ and $\lambda_n^{(2)}(\omega)$, it is necessary to shift the contour $\mathcal{C''}$ to the left until it coincides with the contour $\mathcal{C}$, as shown in Fig.~\ref{Fig:Contour_Shift_s1_s2}. However, this introduces two spurious double poles. The first double pole is located at $\lambda=1/2$ (that is, at $\ell=0$), arising from the term $1/\left[(\lambda-1/2)\cos(\pi\lambda)\right]$. The second double pole is located at $\lambda=3/2$ (that is, at $\ell=1$), which comes from the term $1/\left[(\lambda-3/2)\cos(\pi\lambda)\right]$. The contribution of the associated residues must therefore be removed, which gives the following expression

\begin{widetext}
\begin{eqnarray}\label{SW_Grav_Scattering_amp_bis_1}
& &  \tilde {\mathfrak{ f}}^{(\pm,22)}(x) = \frac{1}{2\omega} \left(\int_{\cal C} d\lambda \, \frac{\lambda}{(\lambda^2-1/4)(\lambda^2-9/4)\cos (\pi \lambda)} \left[ \frac{1}{2} \left(S^{(e,22)}_{\lambda -1/2} \pm S^{(o,22)}_{\lambda -1/2} \right) - \left( \frac{1 \pm 1}{2} \right) \right] P_{\lambda -1/2} (-x) \right.\nonumber \\
&&    - 2i\pi \lim_{\lambda \to 1/2} \frac{d}{d\lambda} \left\lbrace (\lambda-1/2)^2 \times \frac{\lambda}{(\lambda^2-1/4)(\lambda^2-9/4)\cos (\pi \lambda)} \left[ \frac{1}{2} \left(S^{(e,22)}_{\lambda -1/2} \pm S^{(o,22)}_{\lambda -1/2} \right) - \left( \frac{1 \pm 1}{2} \right) \right] P_{\lambda -1/2} (-x) \right\rbrace  \nonumber \\
&&    \left.- 2i\pi \lim_{\lambda \to 3/2} \frac{d}{d\lambda} \left\lbrace (\lambda-3/2)^2 \times \frac{\lambda}{(\lambda^2-1/4)(\lambda^2-9/4)\cos (\pi \lambda)} \left[ \frac{1}{2} \left(S^{(e,22)}_{\lambda -1/2} \pm S^{(o,22)}_{\lambda -1/2} \right) - \left( \frac{1 \pm 1}{2} \right) \right] P_{\lambda -1/2} (-x)\right\rbrace \right). \nonumber \\
& &
\end{eqnarray}

Using, in particular~\eqref{SP_Residue_1} and 
\begin{equation}\label{SP_Residue_2}
\left[\frac{d}{d\lambda} P_{\lambda-1/2}(-x)\right]_{\lambda=3/2} = -(1+x) -x \ln \left( \frac{1-x}{2} \right),
\end{equation}
we can explicitly evaluate the terms that neutralize the contributions of the spurious poles, giving us
\begin{eqnarray}\label{SW_Grav_Scattering_amp_bis_2}
& & \tilde{\mathfrak{f}}^{(\pm,22)}(x) = \frac{1}{2\omega} \int_{\cal C} d\lambda \, \frac{\lambda}{(\lambda^2-1/4)(\lambda^2-9/4)\cos (\pi \lambda)} \left[ \frac{1}{2} \left(S^{(e,22)}_{\lambda -1/2} \pm S^{(o,22)}_{\lambda -1/2} \right) - \left( \frac{1 \pm 1}{2} \right) \right] P_{\lambda -1/2} (-x) \nonumber \\
&&  \qquad\qquad - \frac{i}{4\omega} \left[ \frac{1}{2} \left(S^{(e,22)}_{0} \pm S^{(o,22)}_{0} \right) - \left( \frac{1 \pm 1}{2} \right) \right] \ln \left( \frac{1-x}{2} \right)   \nonumber \\
&&  \qquad\qquad + \frac{i}{4\omega} \left[ \frac{1}{2} \left(S^{(e,22)}_{1} \pm S^{(o,22)}_{1} \right) - \left( \frac{1 \pm 1}{2} \right) \right] \left[(1+x) + x \ln \left( \frac{1-x}{2} \right)\right]   \nonumber \\
&&  \qquad\qquad - \frac{i}{4\omega} \left\lbrace \frac{7}{6}\left[ \frac{1}{2} \left(S^{(e,22)}_{1} \pm S^{(o,22)}_{1}\right) - \left( \frac{1 \pm 1}{2} \right) \right] -  \frac{d}{d\lambda}\left[\frac{1}{2}\left(S^{(e,22)}_{\lambda -1/2} \pm S^{(o,22)}_{\lambda -1/2} \right)  \right]_{\lambda=3/2}  \right\rbrace x  \nonumber \\
&&    \qquad\qquad
+ \mathrm{terms \,\, independent \,\, of \,\,} x.
\end{eqnarray}
\end{widetext}

Applying the differential operator $\mathfrak{D}^{(\pm,22)}_x$ to~\eqref{SW_Grav_Scattering_amp_bis_2}, we can show that the terms introduced to neutralize the spurious modes do not necessarily contribute to the scattering amplitudes~\eqref{fpm_2_2}. Thus, we obtain
\begin{eqnarray}\label{SW_Grav_Scattering_amp_fin}
& & \mathfrak{f}^{(\pm,22)}(x) = \mathfrak{D}^{(\pm,22)}_{x} \left\lbrace \frac{1}{2 \omega} \right. \nonumber\\
& & \qquad\qquad\times\left. \int_{\cal C} d\lambda \, \frac{\lambda}{(\lambda^2-1/4)(\lambda^2-9/4)\cos (\pi \lambda)} \right. \nonumber \\
& &  \qquad\qquad\times \left[ \frac{1}{2} \left(S^{(e,22)}_{\lambda -1/2} \pm S^{(o,22)}_{\lambda -1/2} \right) - \left( \frac{1 \pm 1}{2} \right) \right] \nonumber \\
& & \qquad\qquad \left. P_{\lambda -1/2} (-x) \vphantom{ \left[ \frac{1}{2} \left(S^{(e,gg)}_{\lambda -1/2}(\omega) \pm S^{(o,gg)}_{\lambda -1/2}(\omega) \right) - \left( \frac{1 \pm 1}{2} \right) \right]} \right\rbrace  + \mathfrak{f}_\text{\tiny{SP}}^{(\pm,22)}(x),
\end{eqnarray}
where the amplitudes $ \mathfrak{f}_\text{\tiny{SP}}^{(\pm,22)}(x)$ denote the shift corrections and are given by
\begin{subequations}\label{Shift_Corrections_Grav}
\begin{equation}\label{fsp_+_Grav}
   \mathfrak{f}_{\text{SP}}^{(+,22)}(x)=0,
\end{equation}
and
\begin{equation}\label{fsp_-_Grav}
   \mathfrak{f}_{\text{SP}}^{(-,22)}(x)=\frac{2 i}{\omega} \left[3 S_1^{(o,22)}-S_0^{(o,22)}(2+x)\right] \frac{1}{(1-x)^2}.
\end{equation}
\end{subequations}
Here, it should be noted that we have simplified the last expression by using the fact that
\begin{equation}\label{Simplification_Grav}
  S_\ell^{(e,22)}+ S_\ell^{(o,22)} = 0 \quad\text{for}\quad \ell =0 \quad \text{and} \quad \ell =1.
\end{equation}

We now deform the contour $\mathcal{C}$ in Eq.~\eqref{SW_Grav_Scattering_amp_fin} in order to collect the Regge pole contributions, using Cauchy's residue theorem (see Fig.~\ref{Fig:Contour_Shift_s1_s2}), and we obtain
\begin{equation}\label{CAM_Grav_Scattering_amp_tot}
\mathfrak{f}^{(\pm,22)}(x) = \mathfrak{f}^{(\pm,22)}_\text{\tiny{B}} (x) +  \mathfrak{f}^{(\pm,22)}_\text{\tiny{RP}} (x) +\mathfrak{f}_\text{\tiny{SP}}^{(\pm,22)}(x),
\end{equation}
where
\begin{subequations}\label{CAM_Grav_Scattering_amp_decomp}
\begin{equation}\label{CAM_Grav_Scattering_amp_decomp_Background}
\mathfrak{f}^{(\pm,22)}_\text{\tiny{B}} (x) = \mathfrak{f}^{(\pm,22)}_\text{\tiny{B},\tiny{Re}} (x)+ \mathfrak{f}^{(\pm,22)}_\text{\tiny{B},\tiny{Im}} (x),
\end{equation}
with
\begin{eqnarray}\label{CAM_Grav_Scattering_amp_decomp_Background_a}
& & \mathfrak{f}^{(\pm,22)}_\text{\tiny{B},\tiny{Re}} (x) = \mathfrak{D}^{(\pm,22)}_{x} \left\lbrace \frac{1}{2\pi \omega} \int_{{\cal C}_{-}} d\lambda \, \frac{\lambda}{(\lambda^2-1/4)(\lambda^2-9/4)}  \right. \nonumber \\
& &   \left. \phantom{\int_{{\cal C}_{-}}}  \times \left[S^{(e,22)}_{\lambda -1/2} \pm S^{(o,22)}_{\lambda -1/2} \right] \,Q_{\lambda -1/2}(x +i0)  \right\rbrace,
\end{eqnarray}
and
\begin{eqnarray}\label{CAM_Grav_Scattering_amp_decomp_Background_b}
& & \mathfrak{f}^{(\pm,22)}_\text{\tiny{B},\tiny{Im}} (x) = \mathfrak{D}^{(\pm,22)}_{x} \left\lbrace \frac{1}{2\pi \omega} \int_{+i\infty}^0 d\lambda \, \frac{\lambda}{(\lambda^2-1/4)(\lambda^2-9/4)}  \right. \nonumber \\
& &   \left. \phantom{\int_{\infty}^0}  \times \left[S^{(e,22)}_{\lambda -1/2}\pm S^{(o,22)}_{\lambda -1/2} \right] \,Q_{\lambda -1/2}(x +i0)  \right\rbrace
\end{eqnarray}
\end{subequations}
is a background integral contribution, and where
\begin{eqnarray}\label{CAM_Grav_Scattering_amp_decomp_RP}
& &\mathfrak{f}^{(\pm,2 2)}_\text{\tiny{RP}} (x) = \mathfrak{D}^{(\pm,22)}_{x} \left\lbrace
-\frac{i \pi}{2\omega}    \sum_{n=1}^{+\infty} \sum_{s=1}^{2} \right. \nonumber\\
& & \left.\frac{ \lambda_n^{(s)}(\omega) \left[ r^{(e,22)}_{sn} \pm r^{(o,22)}_{sn} \right]}{[\lambda_n^{(s)}(\omega)^2-1/4][\lambda_n^{(s)}(\omega)^2-9/4] \cos[\pi \lambda_n^{(s)}(\omega)]}\right. \nonumber \\
& &\left. \phantom{\sum_{n=1}^{+\infty}} \times P_{\lambda_n^{(s)}(\omega) -1/2} (-x)  \right\rbrace
\end{eqnarray}
is a sum over the RPs $\lambda_n^{(1)}(\omega)$ and $\lambda_n^{(2)}(\omega)$ lying in the first quadrant of the CAM plane involving the associated residues~\eqref{residues_GW_rn1} and \eqref{residues_GW_rn2}. It is also important to note that the scattering amplitude  $\mathfrak{f}^{(\pm,22)}(x)$ given by \eqref{CAM_Grav_Scattering_amp_tot}--\eqref{CAM_Grav_Scattering_amp_decomp_RP} and \eqref{Shift_Corrections_Grav} is equivalent to the initial partial wave expansions defined by~\eqref{fpm_2_2}--\eqref{Opera_Diff_Grav}.


\subsubsection{Conversion amplitude of  $EW \to GW$}
\label{sec_4_2_3}

By following the steps in Sec.~\ref{sec_4_2_2} and using the Sommerfeld-Watson transformation~\eqref{Sommerfeld_Watson_2}, we can write the sum over the ordinary angular momentum $\ell$ in Eq.~\eqref{fpm_tild_EW_to_GW} as a contour integral in the complex $\lambda$ plane
\begin{eqnarray}\label{SW_EW_to_GW_Scattering_amp}
\tilde{\mathfrak{f}}^{(\pm,12)}(\theta) &=&  -\frac{1}{2 \omega}  \int_{\cal C''} d\lambda \, \frac{\lambda}{(\lambda^2-1/4)\sqrt{(\lambda^2-9/4)}\cos (\pi \lambda)} \nonumber \\
&&    \times \left[ \frac{1}{2} \left(S^{(e,12)}_{\lambda -1/2} \pm S^{(o,12)}_{\lambda -1/2}\right)  \right] \nonumber \\
&&     \times \, P_{\lambda -1/2} (-\cos\theta),
\end{eqnarray}
where
\begin{eqnarray}\label{Matrice_S_EW_to_GW_analy}
S_{\lambda-1/2}^{(e/o,12)} &=& e^{i(\lambda+1/2)\pi} \frac{\sin(2\psi)}{2}  \nonumber\\
& &\times \left[\frac{A_{1\omega,\lambda-1/2}^{(+,\text{e/o})}}{A_{1\omega,\lambda-1/2}^{(-)}}- \frac{A_{2\omega,\lambda-1/2}^{(+,\text{e/o})}}{A_{2\omega,\lambda-1/2}^{(-)}}\right]
\end{eqnarray}
are the analytical extensions of $S^{(e,12)}_{\ell}$ and $S^{(o,12)}_{\ell}$ and they have the same RPs [cf. Eq.~\eqref{CAM_Rel_1}]. These are identical to the ones previously defined.

In order to evaluate the integral in \eqref{SW_EW_to_GW_Scattering_amp} by using Cauchy's theorem to collect the RP contributions, we will use the procedure employed in Sec.~\ref{sec_4_2_2}. First, we need to define the associated residues of $S^{(e,12)}_{\lambda-1/2}$ and $S^{(o,12)}_{\lambda-1/2}$. To do this, we will rewrite \eqref{Matrice_S_EW_to_GW_analy} in order to separate the different contributions associated with $s=1$ and $s=2$. We have
\begin{equation}\label{Matrice_S_EW_to_GW_CAM_bis}
 S_{\lambda-1/2}^{(e/o,12)} =  S_{1,\lambda-1/2}^{(e/o,12)} +  S_{2,\lambda-1/2}^{(e/o,12)},
\end{equation}
where
\begin{equation}
\label{Matrice_S_EW_to_GW_CAM_bis_s1}
S_{1,\lambda-1/2}^{(e/o,12)} =\frac{\sin(2\psi)}{2} \, e^{i\pi(\lambda+1/2)}  \left[\frac{A_{1\omega,\lambda-1/2}^{(+,e/o)}}{A_{1,\omega,\lambda-1/2}^{(-)}(\omega)}\right],
\end{equation}
and
\begin{equation}
\label{Matrice_S_EW_to_GW_CAM_bis_s2}
S_{2,\lambda-1/2}^{(e/o,12)} =  - \frac{\sin(2\psi)}{2} \, e^{i\pi(\lambda+1/2)} \left[\frac{A_{2\omega,\lambda-1/2}^{(+,e/o)}}{A_{2\omega,\lambda-1/2}^{(-)}}\right],
\end{equation}
and then the associated residues are given by
\begin{equation}\label{rn_EW_to_GW_s1}
 r_{1 n}^{(e/o,1 2)}=  \frac{\sin(2\psi)}{2} \, r^{(e/o)}_{1 n},
\end{equation}
and
\begin{equation}\label{rn_EW_to_GW_s2}
 r_{2 n}^{(e/o,1 2)} = -\frac{\sin(2\psi)}{2} \, r^{(e/o)}_{2 n}.
\end{equation}

Now, before applying Cauchy's theorem, we must shift the contour integration ${\mathcal C''}$ to the left so that it coincides with ${\mathcal C}$ (see Fig.~\ref{Fig:Contour_Shift_s1_s2}). However, this introduces two spurious double poles at $\lambda = 1/2$ (i.e., at $\ell = 0$) corresponding to the term $1/[(\lambda-1/2)\cos(\pi\lambda)]$ and at $\lambda = 3/2$ (i.e., at $\ell=1$) which comes from the term $1/\left[(\lambda-3/2)^{3/4}\cos(\pi\lambda)\right]$ and must be removed. We then have
\begin{widetext}
\begin{eqnarray}\label{SW_EW_to_GW_Scattering_amp_bis_2}
& &  \tilde {\mathfrak{f}}^{(\pm,12)}(\theta) = -\frac{1}{2 \omega} \left(\int_{\cal C} d\lambda \, \frac{\lambda}{(\lambda^2-1/4)\sqrt{(\lambda^2-9/4)}\cos (\pi \lambda)} \left[ \frac{1}{2} \left(S^{(e,12)}_{\lambda -1/2} \pm S^{(o,12)}_{\lambda -1/2} \right)\right] P_{\lambda -1/2} (-\cos\theta) \right.\nonumber \\
&&    - 2i\pi \lim_{\lambda \to 1/2} \frac{d}{d\lambda} \left\lbrace (\lambda-1/2)^2 \times \frac{\lambda}{(\lambda^2-1/4)\sqrt{(\lambda^2-9/4)}\cos (\pi \lambda)} \left[ \frac{1}{2} \left(S^{(e,12)}_{\lambda -1/2} \pm S^{(o,12)}_{\lambda -1/2} \right) \right] P_{\lambda -1/2} (-\cos\theta) \right\rbrace  \nonumber \\
&&    \left.- 2i\pi \lim_{\lambda \to 3/2} \frac{d}{d\lambda} \left\lbrace (\lambda-3/2)^{3/2} \times \frac{\lambda}{(\lambda^2-1/4)\sqrt{(\lambda^2-9/4)}\cos (\pi \lambda)} \left[ \frac{1}{2} \left(S^{(e,12)}_{\lambda -1/2} \pm S^{(o,12)}_{\lambda -1/2} \right)  \right] P_{\lambda -1/2} (-\cos\theta)\right\rbrace \right). \nonumber \\
& &
\end{eqnarray}
By using in particular~\eqref{SP_Residue_1} and \eqref{SP_Residue_2}, we can write
\begin{eqnarray}\label{SW_EW_to_GW_Scattering_amp_bis_2}
& & \tilde{\mathfrak{f}}^{(\pm,12)}(\theta) = -\frac{1}{2 \omega}  \int_{\cal C} d\lambda \, \frac{\lambda}{(\lambda^2-1/4)\sqrt{(\lambda^2-9/4)}\cos (\pi \lambda)} \left[ \frac{1}{2} \left(S^{(e,12)}_{\lambda -1/2} \pm S^{(o,12)}_{\lambda -1/2} \right)  \right] P_{\lambda -1/2} (-\cos\theta) \nonumber \\
&&   \qquad\qquad - \frac{1}{2\sqrt{2}\omega} \left[ \frac{1}{2} \left(S^{(e,12)}_{0} \pm S^{(o,12)}_{0}\right) \right] \ln \left( \frac{1-\cos\theta}{2} \right)   \nonumber \\
&&    \qquad\qquad - \frac{i\sqrt{3}}{4\omega} \left[ \frac{1}{2} \left(S^{(e,12)}_{1} \pm S^{(o,12)}_{1}\right)  \right] \left[(1+\cos\theta) + \cos\theta \ln \left( \frac{1-\cos\theta}{2} \right)\right]   \nonumber \\
&&    \qquad\qquad + \frac{i\sqrt{3}}{4\omega} \left\lbrace \left[\frac{1}{2} \left(S^{(e,12)}_{1} \pm S^{(o,12)}_{1} \right) \right] -  \frac{d}{d\lambda}\left[\frac{1}{2}\left(S^{(e,12)}_{\lambda -1/2} \pm S^{(o,12)}_{\lambda -1/2} \right)  \right]_{\lambda=3/2}  \right\rbrace \cos\theta  \nonumber \\
&&    \qquad\qquad
+ \mathrm{terms \,\, independent \,\, of \,\,} \theta.
\end{eqnarray}
\end{widetext}

By applying the differential operator $\mathfrak{D}^{(\pm,12)}_\theta$ to \eqref{SW_EW_to_GW_Scattering_amp_bis_2} and taking into account that $S_0^{(e,12)} - S_0^{(o,12)} = 0$ and $S_1^{(e/o,12)} = 0$, we can show that the terms  introduced to neutralize the spurious modes do not necessarily contribute to the conversion scattering amplitude \eqref{fpm_1_2} and we obtain
\begin{eqnarray}\label{SW_EW_to_GW_Scattering_amp_bis_3}
\mathfrak{f}^{(\pm,12)}(\theta)&=& \mathfrak{D}^{(\pm,12)}_\theta \left\{ -\frac{1}{2\omega} \right. \nonumber \\
& & \left.\times \int_{\cal{C}} d \lambda\, \frac{\lambda}{(\lambda^2-1/4)\sqrt{(\lambda^2-9/4)}\cos(\pi\lambda)}\right. \nonumber\\
& & \times \left[ \frac{1}{2} \left(S^{(e,12)}_{\lambda -1/2} \pm S^{(o,12)}_{\lambda -1/2} \right) \right] \nonumber \\
& & \left. \times P_{\lambda -1/2} (-\cos\theta)\vphantom{\left[ \frac{1}{2} \left(S^{(e,\gamma g)}_{\lambda -1/2}(\omega) \pm S^{(o,\gamma g)}_{\lambda -1/2}(\omega) \right) - \left( \frac{1 \pm 1}{2} \right) \right]} \right\} + \mathfrak{f}_{\text{SP}}^{(\pm,12)}(\theta),
\end{eqnarray}
where the ``shift corrections'' disappear
\begin{equation}\label{Shift_Corrections_EW_to_GW}
   \mathfrak{f}_{\text{SP}}^{(\pm,12)}(\theta)=0.
\end{equation}

We then deform the integration contour $\mathcal{C}$ in Eq.~\eqref{SW_EW_to_GW_Scattering_amp_bis_3} to collect the contributions of the RPs, and thus, we obtain
\begin{equation}\label{CAM_EW_to_GW_Scattering_amp_tot}
\mathfrak{f}^{(\pm,12)}(\theta) = \mathfrak{f}^{(\pm,12)}_\text{\tiny{B}} (\theta) +  \mathfrak{f}^{(\pm,12)}_\text{\tiny{RP}} (\theta),
\end{equation}
where
\begin{subequations}\label{CAM_EW_to_GW_Scattering_amp_decomp}
\begin{equation}\label{CAM_EW_to_GW_Scattering_amp_decomp_Background}
\mathfrak{f}^{(\pm,12)}_\text{\tiny{B}} (\theta) = \mathfrak{f}^{(\pm,12)}_\text{\tiny{B},\tiny{Re}} (\theta)+ \mathfrak{f}^{(\pm,12)}_\text{\tiny{B},\tiny{Im}} (\theta),
\end{equation}
with
\begin{eqnarray}\label{CAM_EW_to_GW_Scattering_amp_decomp_Background_a}
 \mathfrak{f}^{(\pm,12)}_\text{\tiny{B},\tiny{Re}} (\theta) &=& \mathfrak{D}^{(\pm,12)}_{\theta} \left\lbrace -\frac{1}{2\pi \omega} \right.\nonumber \\
&\times&\left.\int_{{\cal C}_{-}} d\lambda \, \frac{\lambda}{(\lambda^2-1/4)\sqrt{(\lambda^2-9/4)}}  \right. \nonumber \\
&\times& \left. \vphantom{\int_{{\cal C}_{-}}}  \left[S^{(e,12)}_{\lambda -1/2} \pm S^{(o,12)}_{\lambda -1/2} \right] \,Q_{\lambda -1/2}(\theta +i0)  \right\rbrace, \nonumber \\
\end{eqnarray}
and
\begin{eqnarray}\label{CAM_EW_to_GW_Scattering_amp_decomp_Background_b}
 \mathfrak{f}^{(\pm,12)}_\text{\tiny{B},\tiny{Im}} (\theta) &=& \mathfrak{D}^{(\pm,12)}_{\theta} \left\lbrace -\frac{1}{2\pi \omega} \right.\nonumber \\
&\times& \left.\int_{+i\infty}^0 d\lambda \, \frac{\lambda}{(\lambda^2-1/4)\sqrt{(\lambda^2-9/4)}}  \right. \nonumber \\
& \times& \left.  \left[S^{(e,12)}_{\lambda -1/2} \pm S^{(o,12)}_{\lambda -1/2} \right] \,Q_{\lambda -1/2}(\theta +i0)  \vphantom{\int_{\infty}^0}\right\rbrace, \nonumber \\
\end{eqnarray}
\end{subequations}
is a background integral contribution, and where
\begin{eqnarray}\label{CAM_EW_to_GW_Scattering_amp_decomp_RP}
& & \mathfrak{f}^{(\pm,12)}_\text{\tiny{RP}} (\theta) = \mathfrak{D}^{(\pm,12)}_{\theta} \left\lbrace
\frac{i \pi}{2\omega}    \sum_{n=1}^{+\infty}  \sum_{s=1}^{2} \right. \nonumber \\
& &  \frac{ \lambda_n^{(s)}(\omega) \left[ r^{(e,12)}_{s n} \pm r^{(o,12)}_{s n} \right]}{\left[\lambda_n^{(s)}(\omega)^2-1/4\right]\left[\sqrt{\lambda_n^{(s)}(\omega)^2-9/4}\right] \cos[\pi \lambda_n^{(s)}(\omega)]}  \nonumber\\
& &  \left. \phantom{\sum_{n=1}^{+\infty}} \times    P_{\lambda_n^{(s)}(\omega) -1/2} (-\cos\theta)  \right\rbrace
\end{eqnarray}
is a sum over the RPs $\lambda_n^{(1)}(\omega)$ and $\lambda_n^{(2)}(\omega)$ lying in the first quadrant of the CAM plane involving their associated residues. It is also important to note that the scattering amplitude $\mathfrak{f}^{(\pm,12)}(\theta)$ given by \eqref{CAM_EW_to_GW_Scattering_amp_tot}--\eqref{CAM_EW_to_GW_Scattering_amp_decomp_RP} and \eqref{Shift_Corrections_EW_to_GW} is equivalent to the partial wave expansions defined by~\eqref{fpm_1_2}--\eqref{Opera_Diff_EW_to_GW}.

\subsection{Analytic Regge pole approximation of scattering and conversion amplitudes}
\label{sec_4_3}

To construct an analytical approximation for the RP contribution, i.e., for the helicity-preserving and helicity-reversing of scattering and conversion amplitudes~\eqref{CAM_EW_Scattering_amp_decomp_RP}, \eqref{CAM_Grav_Scattering_amp_decomp_RP}~and~\eqref{CAM_EW_to_GW_Scattering_amp_decomp_RP}, we need to determine the asymptotic expressions of the RPs $\lambda_n^{(s)}(\omega)$ and their associated residues $r_{sn}^{(e/o)}$.

An analytical expression for the residues $r_{sn}^{(o)}$ corresponding to the lowest RPs of the RN BH can be derived by extending the calculations used by Dolan and Ottewill in their work on obtaining an analytical expression for the excitation factors of the QNMs of the Schwarzschild BH~\cite{Dolan:2011fh}. This has allowed us, after tedious calculations, to obtain 
\begin{widetext}
\begin{equation}\label{rn_odd_s}
  r^{(o)}_{s n} = \frac{\sqrt{\alpha}(\alpha+1)}{2\sqrt{\pi}\sqrt{\alpha+3}\, (n-1)!} \frac{(\alpha-1)\mp
   \sqrt{(\alpha-3)(\alpha+1)}}{\left(\sqrt{\alpha(\alpha-3)} \mp (\alpha-1)\right)^2}
   \left[-i \frac{64\sqrt{2\alpha^5(\alpha+3)}}{(\alpha+1)^2(\sqrt{\alpha}+\sqrt{\alpha+1})^2} \, b_c \omega\right]^{n-\frac{1}{2}}
   e^{2 i M\omega \zeta(\alpha)} e^{i\pi\lambda_n^{(s)}},
\end{equation}
with
\begin{eqnarray}\label{zeta}
&&  \zeta(\alpha) = \frac{(\alpha+3)(\sqrt{\alpha+1}-2\sqrt{\alpha})}{2\sqrt{\alpha+1}}
  + \ln\left[\frac{8(\alpha+1)}{(\alpha+3)(2\alpha+1+2\sqrt{\alpha(\alpha+1)})^2}\right] \nonumber\\
  &+& \frac{\alpha^2+7}{4\sqrt{2}\sqrt{\alpha^2-1}} \left(\ln\left[\frac{2 - \sqrt{2 -\frac{2}{\alpha}}}{2 + \sqrt{2 -\frac{2}{\alpha}}}\right]\right.+ \left.\ln\left[\frac{1+3\sqrt{\alpha}\sqrt{\alpha+1}+\alpha^{3/2}\sqrt{\alpha+1}+\alpha(\alpha+2)-\sqrt{2}\sqrt{\alpha^2-1}}
  {1+3\sqrt{\alpha}\sqrt{\alpha+1}+\alpha^{3/2}\sqrt{\alpha+1}+\alpha(\alpha+2)+\sqrt{2}\sqrt{\alpha^2-1}}\right] \vphantom{\ln\left[\frac{2 - \sqrt{2 -\frac{2}{\alpha}}}{2 + \sqrt{2 -\frac{2}{\alpha}}}\right]} \right).
\end{eqnarray}

As for the even residues, they can be obtained by [cf. Eq.~\eqref{CAM_Rel_2}]
\begin{eqnarray}\label{Rel_2}
 r_{sn}^{(e)}(\omega) =   \frac{\left[({\lambda_n^{(s)}}^2-1/4)({\lambda_n^{(s)}}^2-9/4)+2i\omega q_{s}\right]}{ \left[({\lambda_n^{(s)}}^2-1/4)({\lambda_n^{(s)}}^2-9/4)-2i\omega q_{s}\right]}  r_{sn}^{(o)}(\omega).
\end{eqnarray}
\end{widetext}

It should be noted that setting $Q=0$ (i.e., $\alpha=3$) in the expressions \eqref{rn_odd_s} and \eqref{zeta} yields the same expressions as those derived in Refs.~\cite{Folacci:2019cmc,Folacci:2019vtt} for the Schwarzschild BH case.

Because the concentric event horizons become degenerate in the extreme case ($Q=M$), it is not possible to obtain the tortoise coordinate necessary for the residue calculations by taking the limit $r_+ \to M$ in~\eqref{Coord_RW_RN_bis}. To address this issue, we chose to handle the situation separately by using the tortoise coordinate~\eqref{Coord_RW_RN_Extrem}. Similarly, for the extremal charged RN BH, we also used the Dolan and Ottewill method~\cite{Dolan:2011fh} and obtained
\begin{equation}\label{rn_odd_s_extrem}
  r_{sn}^{(o)} = \pm i \frac{\left[i 32 (4-3\sqrt{2}) b_c\omega\right]^{n-\frac{1}{2}}}{2\sqrt{\pi} (n-1)!} e^{i2M\omega y} e^{i \pi \lambda_n^{(s)}},
\end{equation}
with
\begin{equation}\label{y}
  y = 3-4\sqrt{2}-4\ln[1+\sqrt{2}]-2\ln[M].
\end{equation}
Here, the upper symbol of $\mp$ ($\pm$) in expressions~\eqref{rn_odd_s}~and~\eqref{rn_odd_s_extrem} is associated with electromagnetic-type perturbations ($s=1$), while the lower one with gravitational-type perturbations ($s=2$).

It is worth noting that setting $\lambda_n^{(2)} = \lambda_n^{(1)}+1$ in~\eqref{rn_odd_s_extrem} yields
\begin{equation}\label{rn_rel}
   r_{2n}^{(o)}  =  r_{1n}^{(o)}.
\end{equation}

Using~\eqref{Relation_PRs}, \eqref{rn_rel}, and, for $Q=M$,
\begin{subequations}\label{rel_angle_psi}
\begin{equation}\label{rel_angle_psi_1}
\cos^2\psi =\frac{1}{2}+\frac{3}{4\lambda_n^{(s)}}
\end{equation}
\begin{equation}\label{rel_angle_psi_2}
\sin^2\psi =\frac{1}{2}-\frac{3}{4\lambda_n^{(s)}}
\end{equation}
\end{subequations}
we have shown, after some algebraic calculations, that
\begin{equation}\label{equality_EW_GW}
\mathfrak{f}^{(\pm,22)}(\theta) = \mathfrak{f}^{(\pm,11)}(\theta)
\end{equation}
for the extremal charged RN BH, which means that the gravitational and electromagnetic scattering cross sections are identical.

\section{Results}
\label{sec_5}

\subsection{Numerical computational method}
\label{sec_5_1}

To construct the differential cross sections~\eqref{Diff_Cross_Sec_si_sf}--\eqref{fpm_si_sf} and their RP contributions~\eqref{CAM_EW_Scattering_amp_decomp_RP}, \eqref{CAM_Grav_Scattering_amp_decomp_RP} and \eqref{CAM_EW_to_GW_Scattering_amp_decomp_RP} involving scattering and conversion amplitudes of various processes ($\text{EW} \to \text{EW}$, $\text{GW} \to \text{GW}$ and $\text{EW} \to \text{GW}$), we need (i) the coefficients $A_{s\omega\ell}^{(\pm,e/o)}$, reflection coefficients $\mathcal{R}_s^{(e/o)}$, and the $S$-matrix elements $S_{\ell}^{(e/o, s_is_f)}$ for the partial wave expansion representation, and (ii) the RP spectrum, as well as their associated residues for the RP contribution. To do this:

\begin{enumerate}[label=(\roman*)]

\item \textit{Partial wave expansion case}: we numerically solved the  problem~\eqref{Z-M_R-W_homogene_RN}--\eqref{Potentiel_ZM_RN} using the Runge-Kutta method. This enabled us to compute the functions $\phi_{s \omega\ell}^{(e/o)}(r)$, the coefficients  $A_{s\omega\ell}^{(\pm,e/o)}$, the reflection coefficients $\mathcal{R}_s^{(e/o)}$, and the matrix elements $S_{\ell}^{(e/o, s_is_f)}$. We then summed over $120$ modes to obtain the scattering and conversion amplitudes for the various processes (see Ref.~\cite{OuldElHadj:2021fqi} for more details). It should be noted that the series representation of the scattering and conversion amplitudes suffers from a lack of convergence due to the long-range nature of the electromagnetic and gravitational fields propagating on the RN BH. To handle this, we used the algorithm presented in the appendix of Ref.~\cite{Folacci:2019cmc} to accelerate the convergence of the mode sum.

\item \textit{The RP contribution case}: We first obtained the RP spectrum $\lambda_n^{(s)}$ using the algorithm presented in Sec.~\ref{sec_3}. Then, we constructed the associated residues $r_{sn}^{(e/o, s_is_f)}$ using the same procedure as for the \emph{partial wave expansion case} to determine numerically the coefficient $A_{s\omega,\lambda-1/2}^{(\pm,e/o)}$. All numerical calculations were performed using Mathematica software.

\end{enumerate}

\subsection{Numerical results and comments}
\label{sec_5_2}

\begin{figure*}[htbp]
 \includegraphics[scale=0.60]{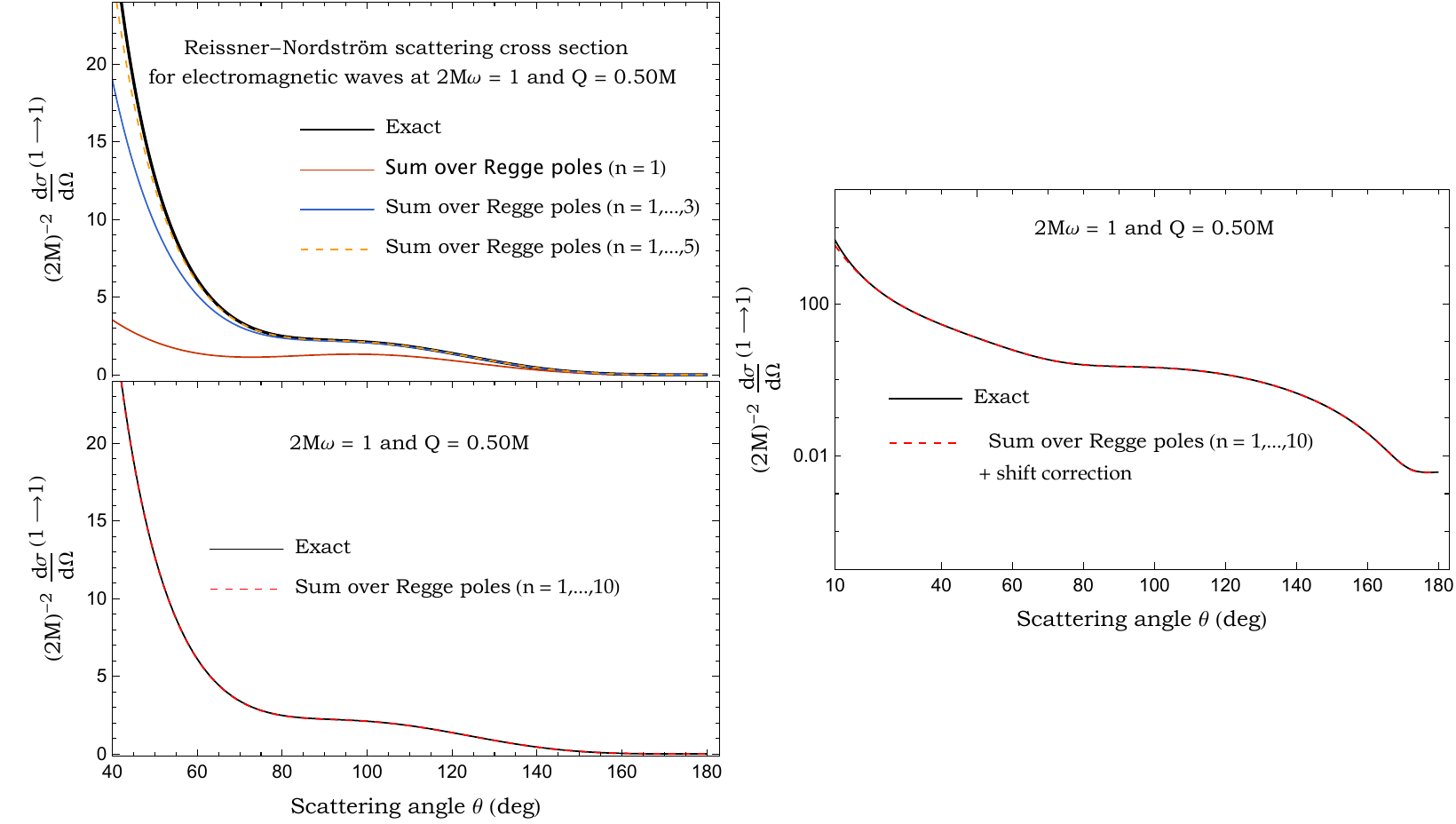}
\caption{\label{Fig:EW_2Mw_1_Q_050M_CAM_vs_Exact} Scattering cross section of a RN BH for EWs ($2M\omega=1$ and $Q =0.50M$). We compare the exact cross section defined by~\eqref{fpm_1_1}--\eqref{Opera_Diff_EW} with its Regge pole contribution constructed from \eqref{CAM_EW_Scattering_amp_decomp_RP}.}
\end{figure*}
\begin{figure*}[htbp]
 \includegraphics[scale=0.60]{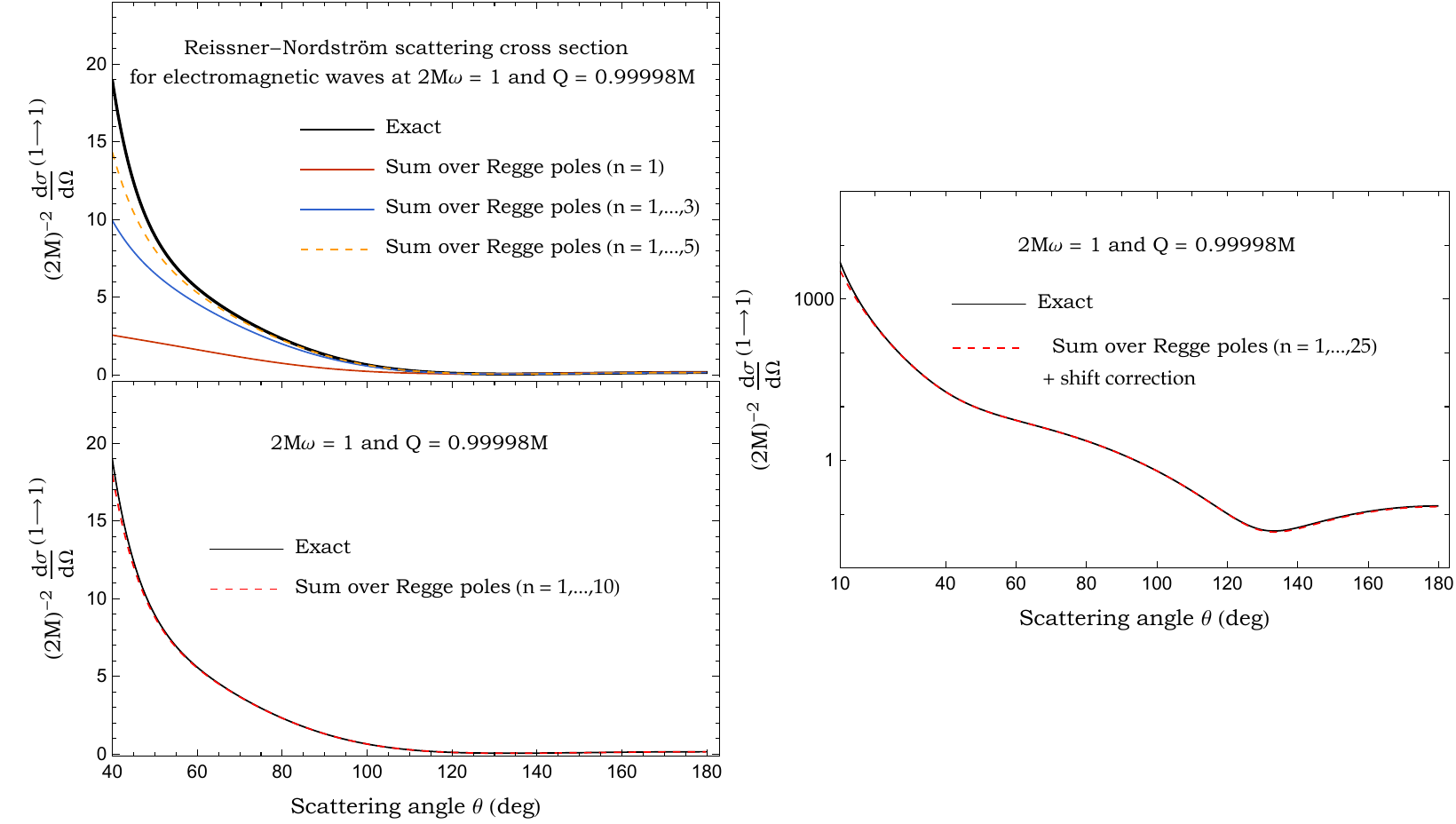}
\caption{\label{Fig:EW_2Mw_1_Q_099998M_CAM_vs_Exact} Scattering cross section of a RN BH for EWs ($2M\omega=1$ and $Q =0.99998M$). We compare the exact cross section defined by~\eqref{fpm_1_1}--\eqref{Opera_Diff_EW} with its Regge pole contribution constructed from \eqref{CAM_EW_Scattering_amp_decomp_RP}.}
\end{figure*}
%

\begin{figure*}[htbp]
 \includegraphics[scale=0.60]{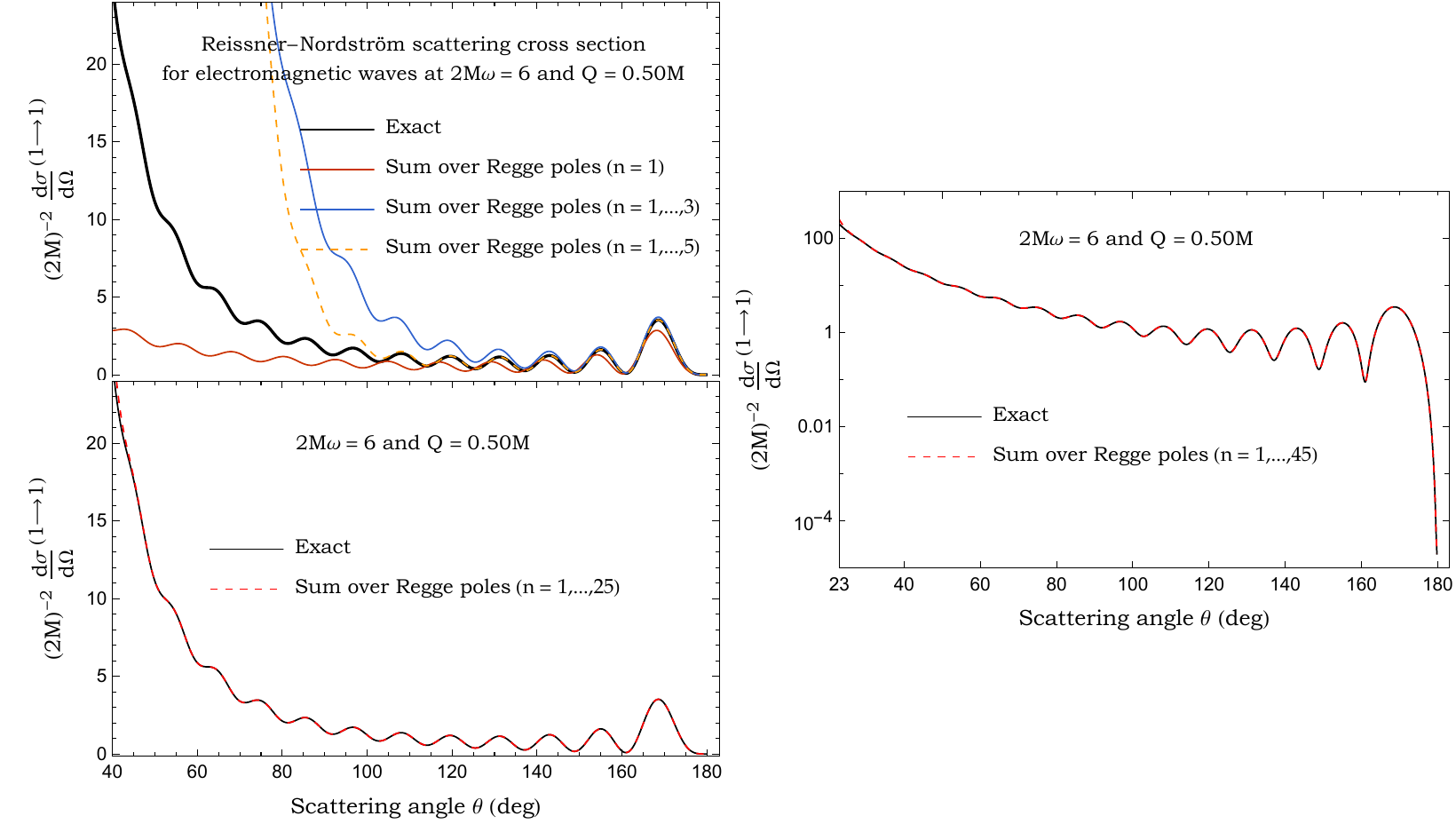}
\caption{\label{Fig:EW_2Mw_6_Q_050M_CAM_vs_Exact} Scattering cross section of a RN BH for EWs ($2M\omega=6$ and $Q =0.50M$). We compare the exact cross section defined by~\eqref{fpm_1_1}--\eqref{Opera_Diff_EW} with its Regge pole approximation constructed from \eqref{CAM_EW_Scattering_amp_decomp_RP}.}
\end{figure*}
\begin{figure*}[htbp]
 \includegraphics[scale=0.60]{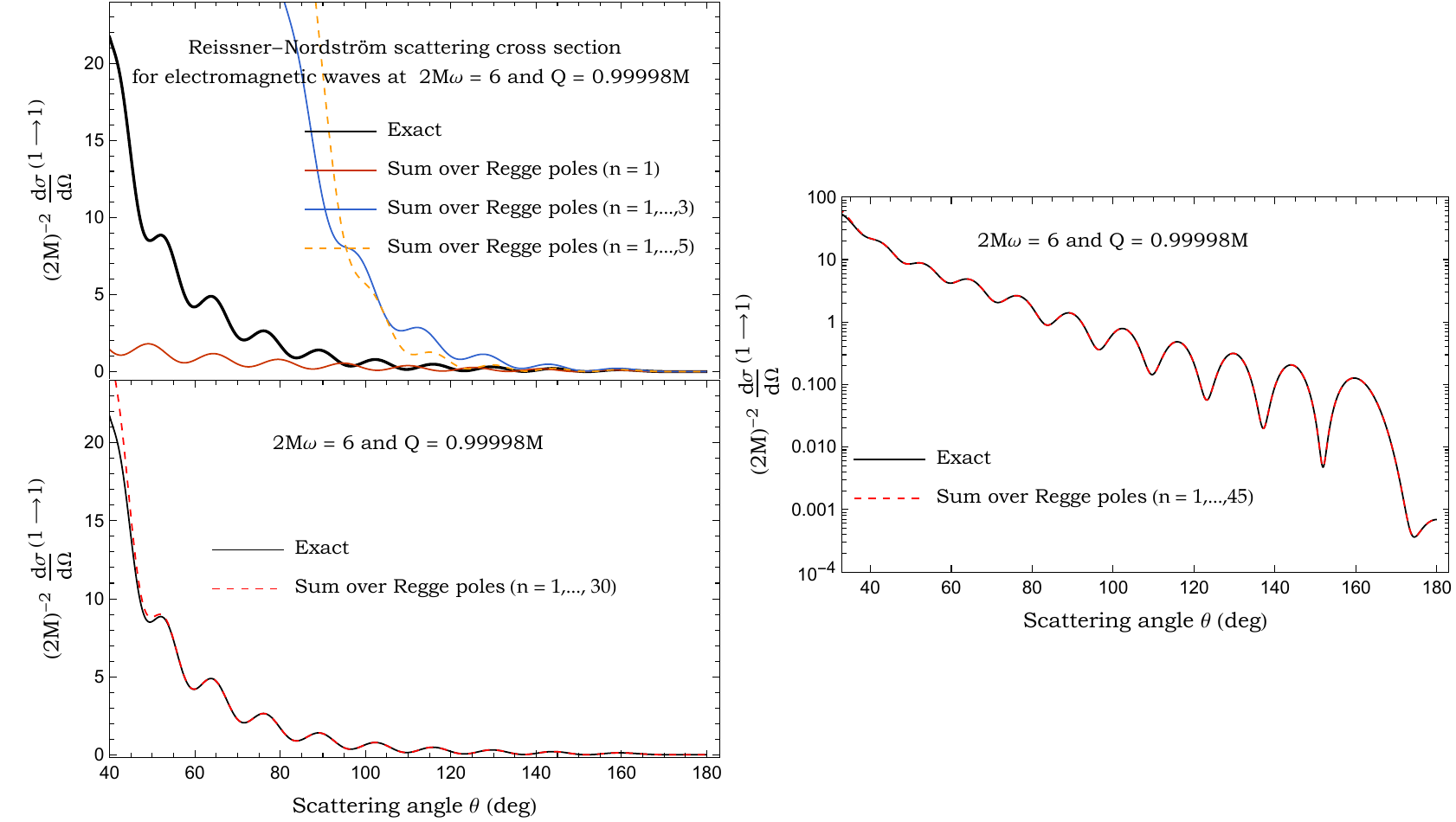}
\caption{\label{Fig:EW_2Mw_6_Q_099998M_CAM_vs_Exact} Scattering cross section of a RN BH for EWs ($2M\omega=6$ and $Q =0.99998M$). We compare the exact cross section defined by~\eqref{fpm_1_1}--\eqref{Opera_Diff_EW} with its Regge pole contribution constructed from \eqref{CAM_EW_Scattering_amp_decomp_RP}.}
\end{figure*}
%
\begin{figure*}[htbp]
 \includegraphics[scale=0.60]{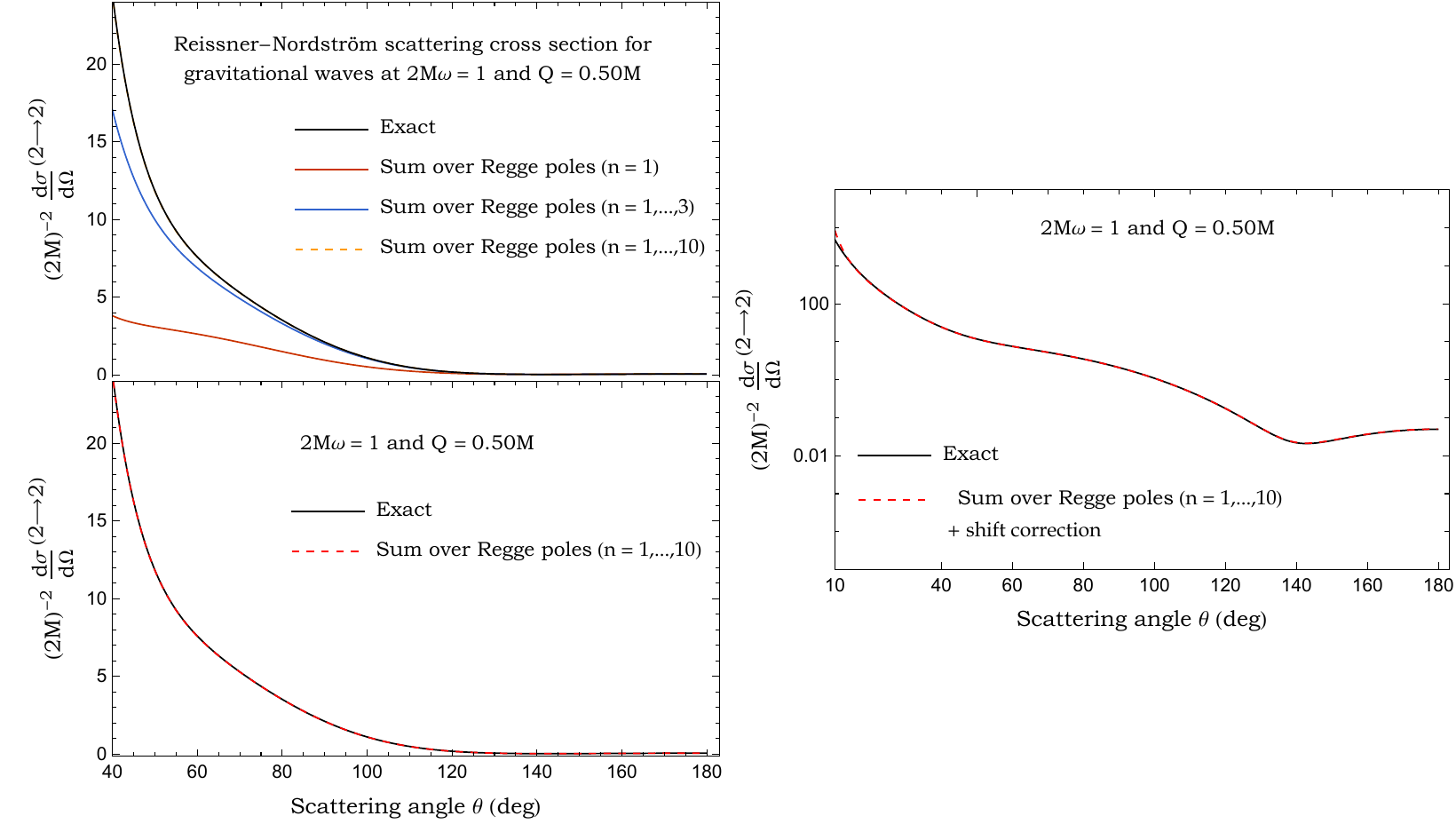}
\caption{\label{Fig:Grav_2Mw_1_Q_050M_CAM_vs_Exact} Scattering cross section of a RN BH for GWs ($2M\omega=1$ and $Q =0.50M$). We compare the exact cross section defined by~\eqref{fpm_2_2}--\eqref{Opera_Diff_Grav} with its Regge pole contribution constructed from \eqref{CAM_Grav_Scattering_amp_decomp_RP}.}
\end{figure*}
\begin{figure*}[htbp]
 \includegraphics[scale=0.60]{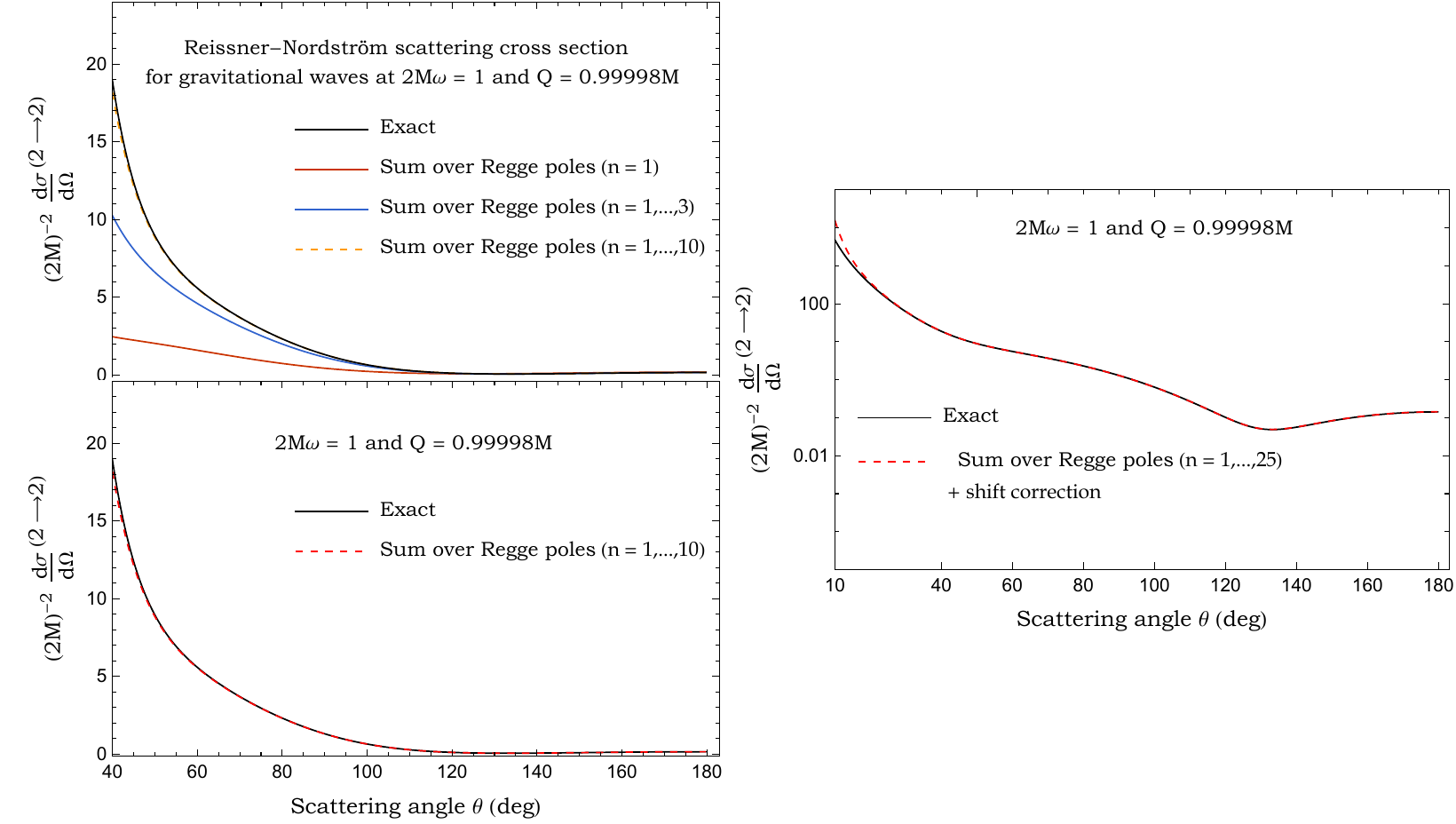}
\caption{\label{Fig:Grav_2Mw_1_Q_099998M_CAM_vs_Exact} Scattering cross section of a RN BH for GWs ($2M\omega=1$ and $Q =0.99998M$). We compare the exact cross section defined by~\eqref{fpm_2_2}--\eqref{Opera_Diff_Grav} with its Regge pole contribution constructed from \eqref{CAM_Grav_Scattering_amp_decomp_RP}.}
\end{figure*}
%

\begin{figure*}[htbp]
 \includegraphics[scale=0.60]{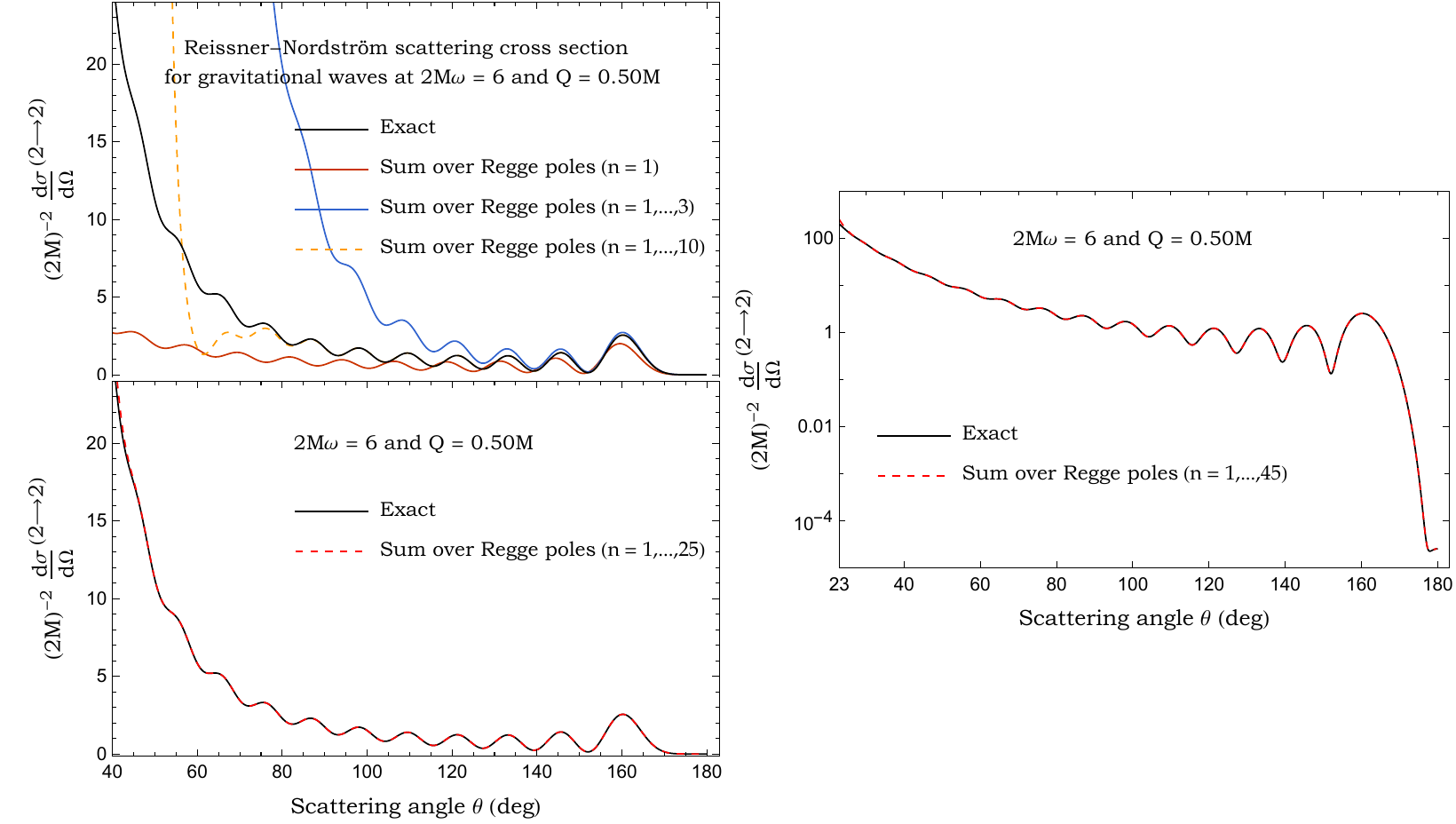}
\caption{\label{Fig:Grav_2Mw_6_Q_050M_CAM_vs_Exact} Scattering cross section of a RN BH for GWs ($2M\omega=6$ and $Q =0.50M$). We compare the exact cross section defined by~\eqref{fpm_2_2}--\eqref{Opera_Diff_Grav} with its Regge pole contribution constructed from \eqref{CAM_Grav_Scattering_amp_decomp_RP}.}
\end{figure*}
\begin{figure*}[htbp]
 \includegraphics[scale=0.60]{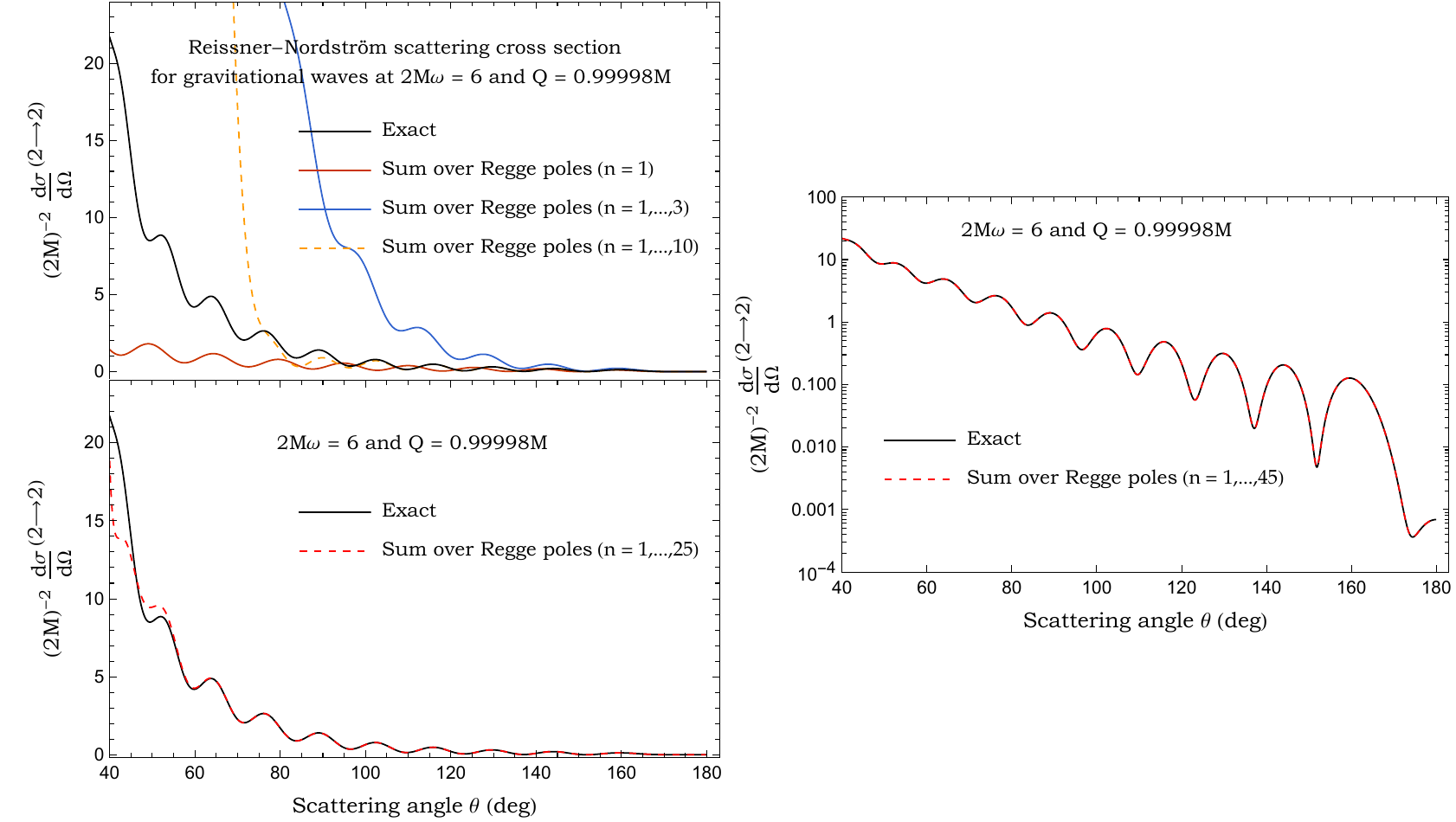}
\caption{\label{Fig:Grav_2Mw_6_Q_099998M_CAM_vs_Exact} Scattering cross section of a RN BH for GWs ($2M\omega=6$ and $Q =0.99998M$). We compare the exact cross section defined by~\eqref{fpm_2_2}--\eqref{Opera_Diff_Grav} with its Regge pole contribution constructed from \eqref{CAM_Grav_Scattering_amp_decomp_RP}.}
\end{figure*}
%
\begin{figure*}[htbp]
 \includegraphics[scale=0.60]{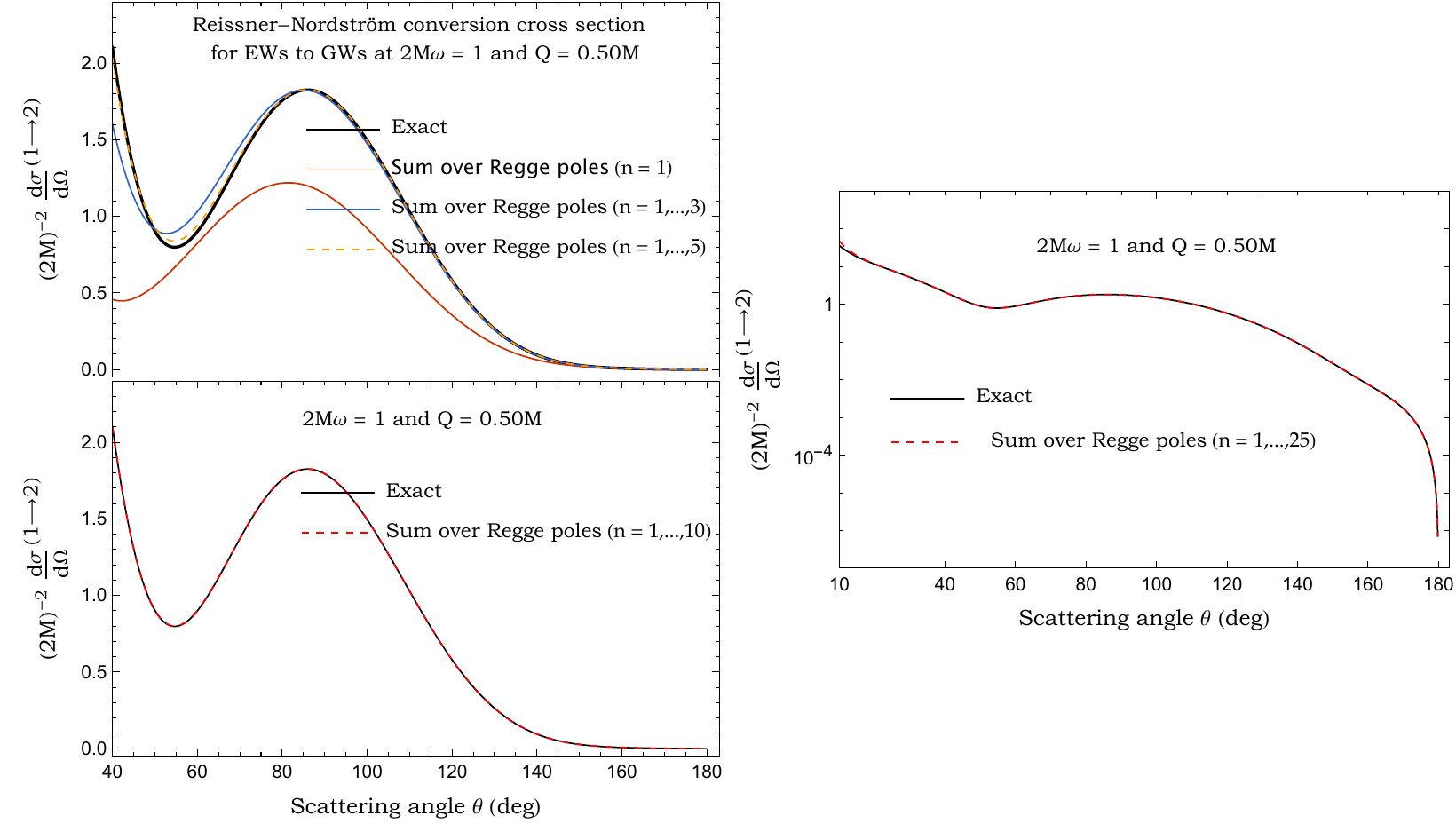}
\caption{\label{Fig:Conv_EW_to_GW_2Mw_1_Q_050M_CAM_vs_Exact} Conversion cross section of a RN BH for EWs to GWs ($2M\omega=1$ and $Q =0.50M$). We compare the exact cross section defined by~\eqref{fpm_1_2}--\eqref{Opera_Diff_EW_to_GW} with its Regge pole contribution constructed from \eqref{CAM_EW_to_GW_Scattering_amp_decomp_RP}.}
\end{figure*}

\begin{figure*}[htbp]
 \includegraphics[scale=0.60]{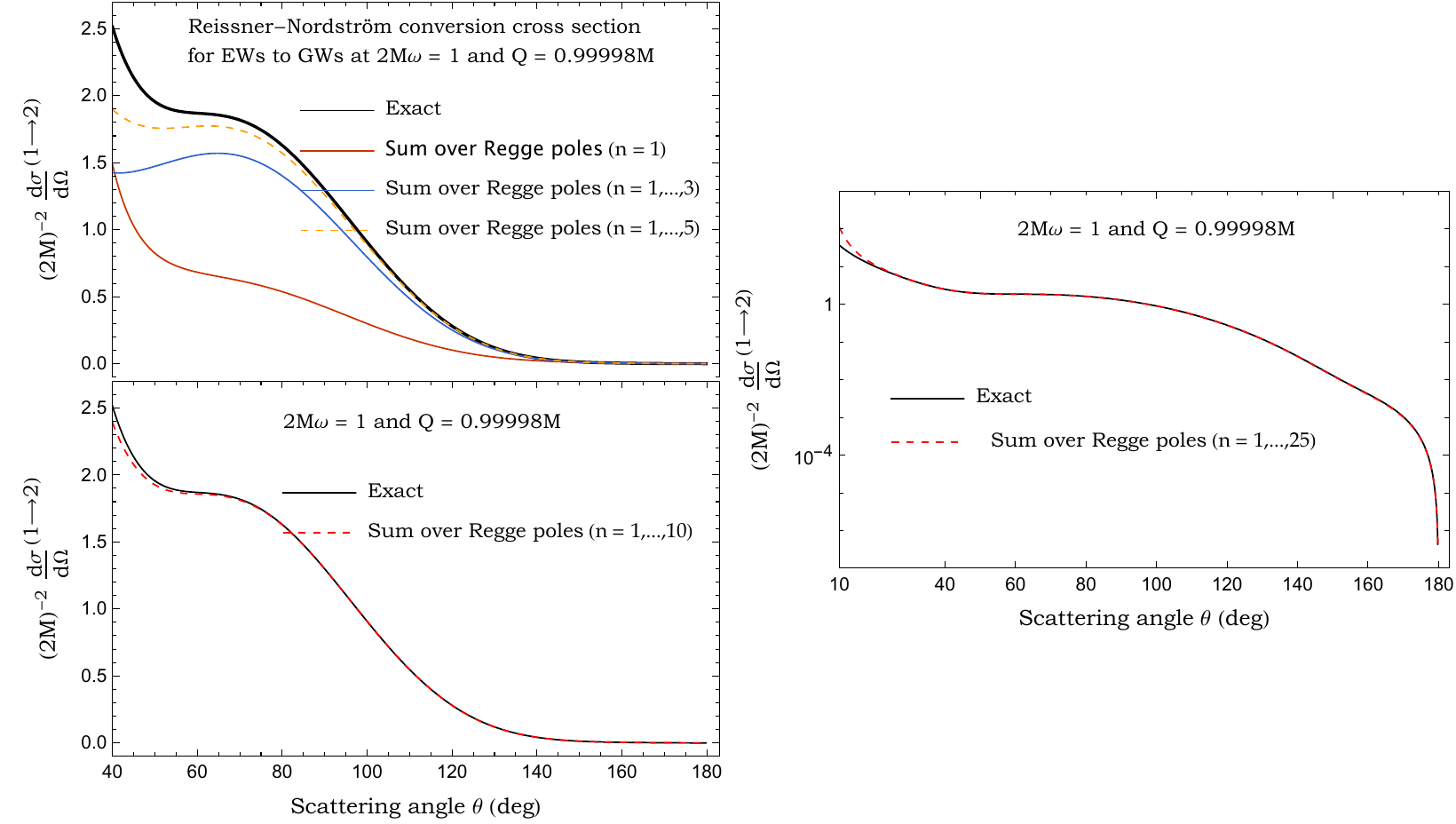}
\caption{\label{Fig:Conv_EW_to_GW_2Mw_1_Q_099998M_CAM_vs_Exact} Conversion cross section of a RN BH for EWs to GWs ($2M\omega=1$ and $Q =0.99998M$). We compare the exact cross section defined by~\eqref{fpm_1_2}--\eqref{Opera_Diff_EW_to_GW} with its Regge pole contribution constructed from \eqref{CAM_EW_to_GW_Scattering_amp_decomp_RP}.}
\end{figure*}

\begin{figure*}[htbp]
 \includegraphics[scale=0.60]{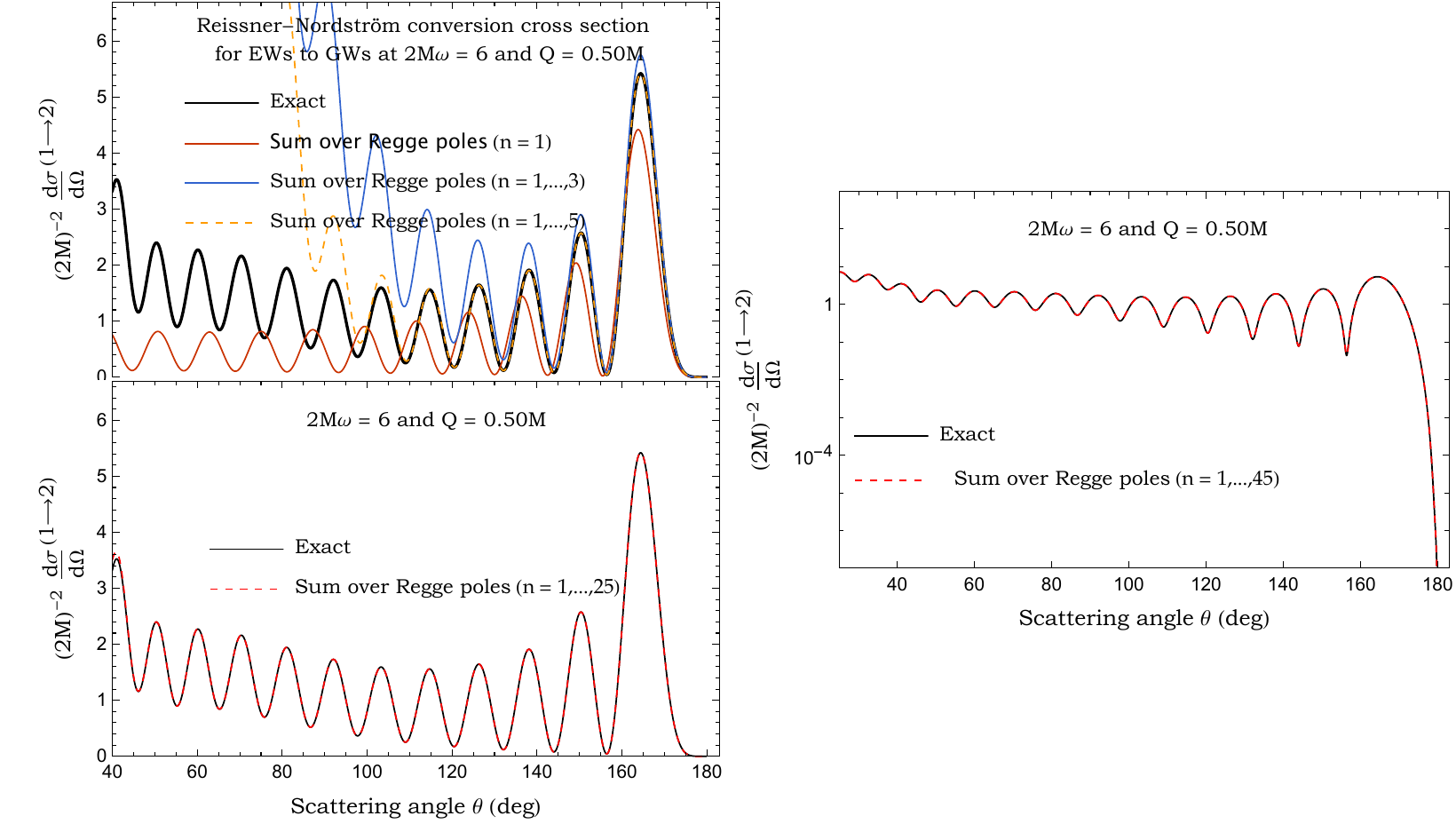}
\caption{\label{Fig:Conv_EW_to_GW_2Mw_6_Q_050M_CAM_vs_Exact} Conversion cross section of a RN BH for EWs to GWs ($2M\omega=6$ and $Q =0.50M$). We compare the exact cross section defined by~\eqref{fpm_1_2}--\eqref{Opera_Diff_EW_to_GW} with its Regge pole contribution constructed from \eqref{CAM_EW_to_GW_Scattering_amp_decomp_RP}.}
\end{figure*}

\begin{figure*}[htbp]
 \includegraphics[scale=0.60]{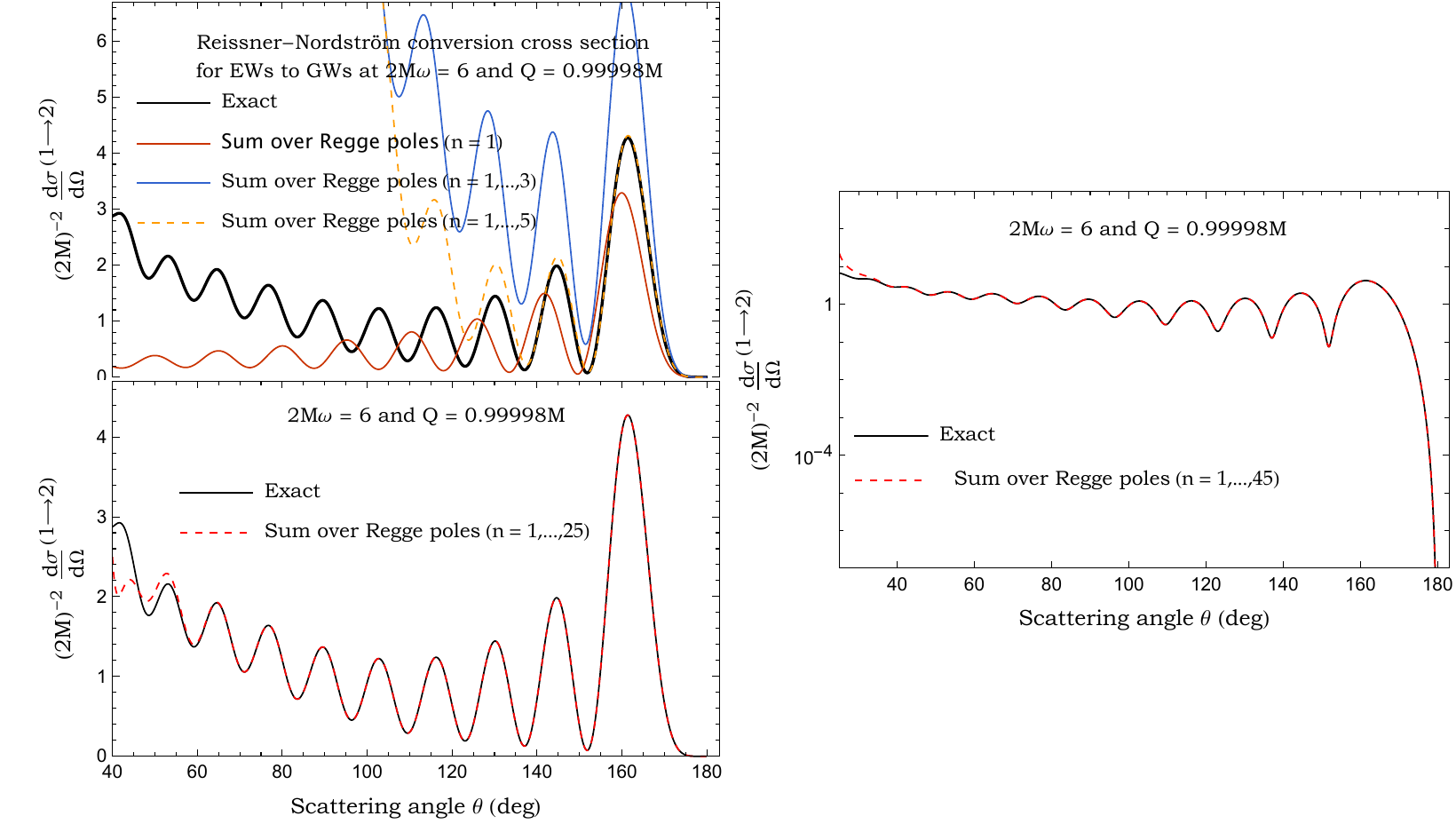}
\caption{\label{Fig:Conv_EW_to_GW_2Mw_6_Q_099998M_CAM_vs_Exact} Conversion cross section of a RN BH for EWs to GWs ($2M\omega=6$ and $Q =0.99998M$). We compare the exact cross section defined by~\eqref{fpm_1_2}--\eqref{Opera_Diff_EW_to_GW} with its Regge pole contribution constructed from \eqref{CAM_EW_to_GW_Scattering_amp_decomp_RP}.}
\end{figure*}

\begin{figure*}[htbp]
 \includegraphics[scale=0.60]{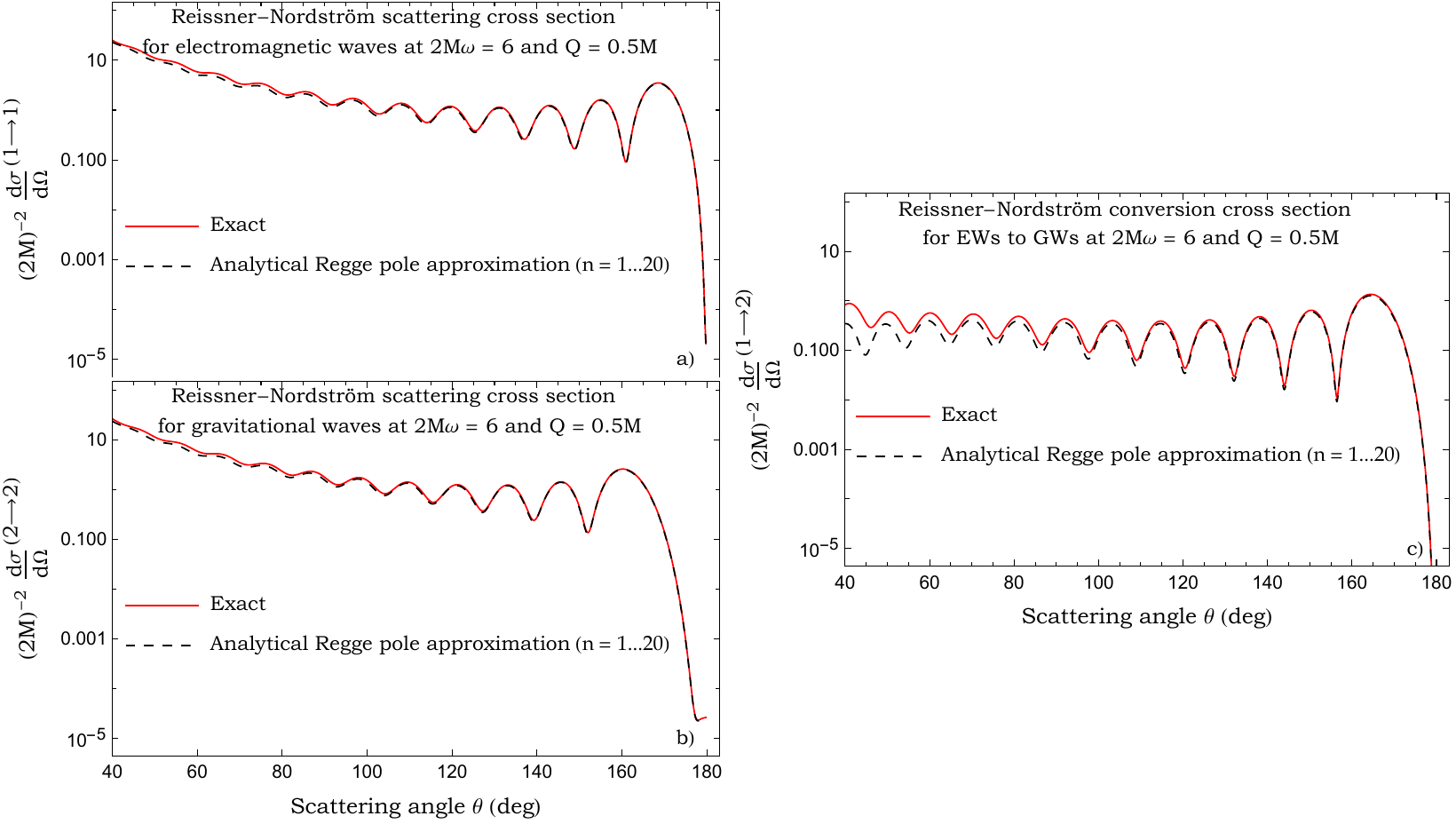}
\caption{\label{Fig:Exact_vs_Asympt_2Mw_6_Q_050M_CAM_vs_Exact}  The differential cross section of a RN BH at $2M\omega=6$ and $Q =0.50M$ a) for electromagnetic waves b) gravitational waves and c) for a conversion process from electromagnetic to gravitational waves . We compare the exact results given in Sec.~\ref{sec_4} with those obtained from the analytical Regge pole approximations constructed in Sec.~\ref{sec_5}.}
 \end{figure*}
\begin{figure*}[htbp]
 \includegraphics[scale=0.60]{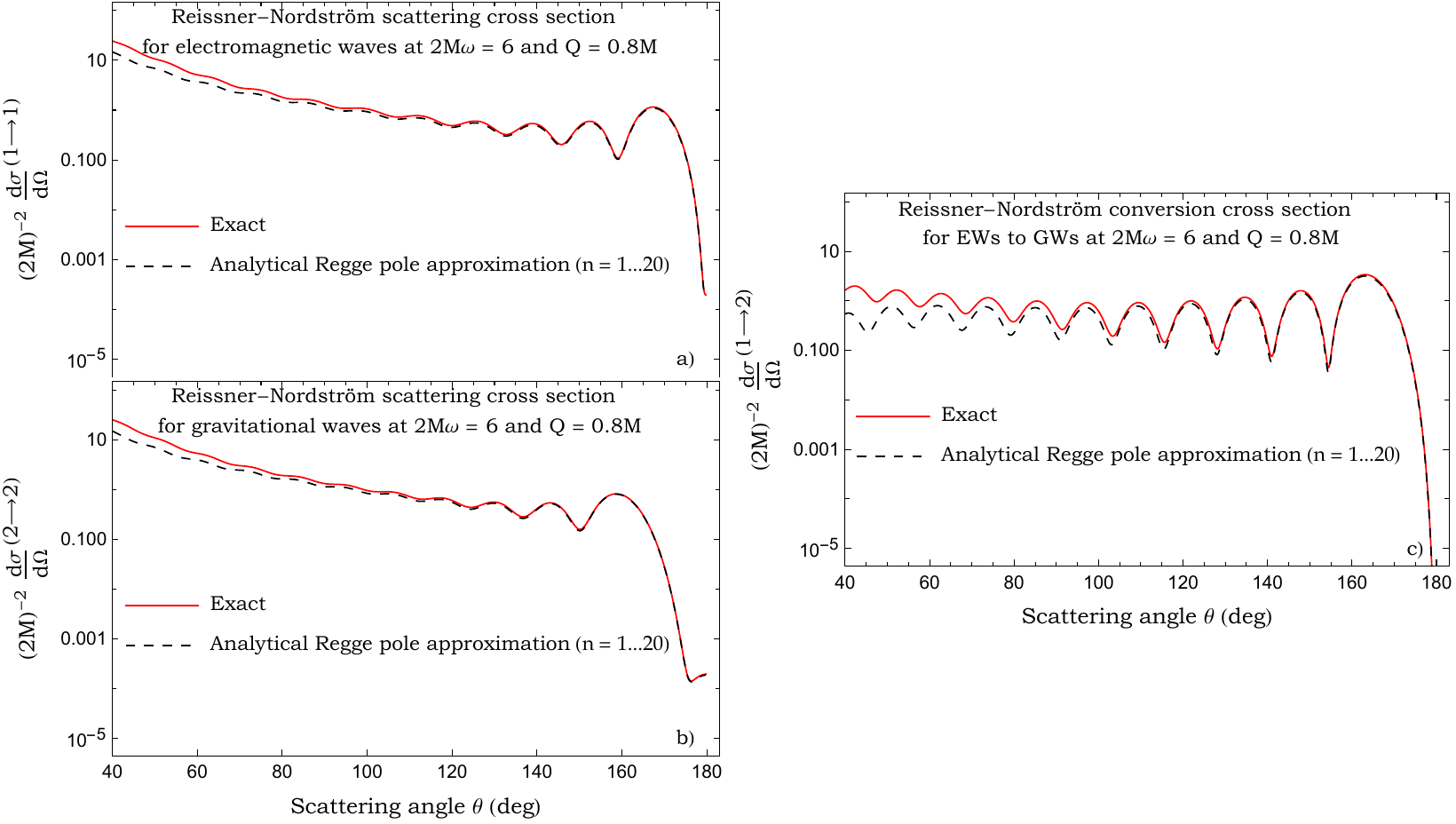}
\caption{\label{Fig:Exact_vs_Asympt_2Mw_6_Q_080M_CAM_vs_Exact}  The differential cross section of a RN BH at $2M\omega=6$ and $Q =0.80M$ a) for electromagnetic waves b) for gravitational waves and c) for a conversion process from electromagnetic to gravitational waves . We compare the exact results given in Sec.~\ref{sec_4} with those obtained from the analytical Regge pole approximations constructed in Sec.~\ref{sec_5}.}
 \end{figure*}
\begin{figure}[htbp]
 \includegraphics[scale=0.60]{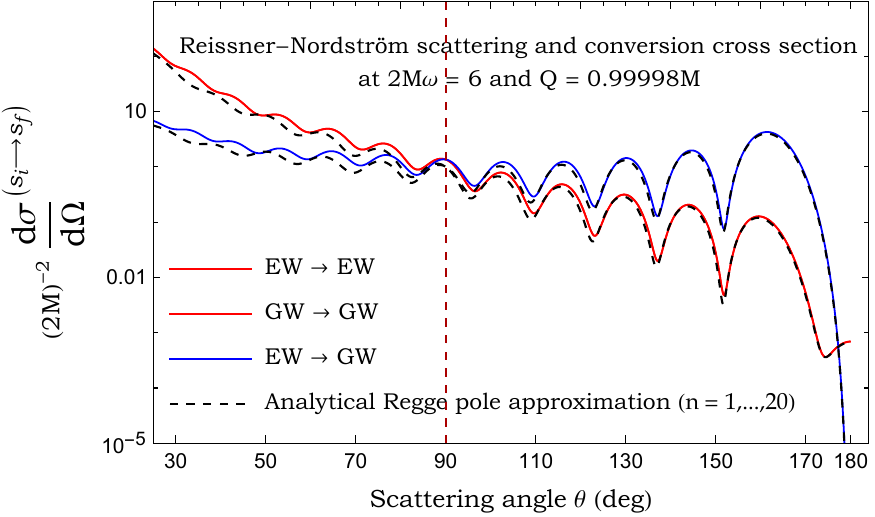}
\caption{\label{Fig:Exact_vs_Asympt_2Mw_6_Q_099998M_CAM_vs_Exact}  The differential cross section of a RN BH at $2M\omega=6$ and $Q =0.99998M$  for electromagnetic and gravitational waves (red line) and for a conversion process from electromagnetic to gravitational waves (blue line) . We compare the exact results given in Sec.~\ref{sec_4} with those obtained from the analytical Regge pole approximation for $Q=M$ constructed in Sec.~\ref{sec_5} (black dashed line).}
 \end{figure}

In Figs.~\ref{Fig:EW_2Mw_1_Q_050M_CAM_vs_Exact}--\ref{Fig:Conv_EW_to_GW_2Mw_6_Q_099998M_CAM_vs_Exact}, we display the scattering and conversion cross sections for different processes. The results are obtained using the RP contribution and compared with those obtained from the partial wave expansion method. The comparisons are performed for the reduced frequencies $2M\omega=1$ and $6$, as well as for the charge-to-mass ratios $Q/M=0.5$ and $0.99998$.

In Figs.~\ref{Fig:EW_2Mw_1_Q_050M_CAM_vs_Exact}--\ref{Fig:EW_2Mw_6_Q_099998M_CAM_vs_Exact}, we focus on the
differential scattering cross section of the EWs and we compare the results
obtained from the partial wave expansion~\eqref{fpm_1_1}--\eqref{Opera_Diff_EW} with its RP contribution constructed from~\eqref{CAM_EW_Scattering_amp_decomp_RP}. In Figs.~\ref{Fig:Grav_2Mw_1_Q_050M_CAM_vs_Exact}--\ref{Fig:Grav_2Mw_6_Q_099998M_CAM_vs_Exact}, we illustrate the differential scattering cross section of the GWs and we compare the results obtained from the partial wave expansion~\eqref{fpm_2_2}--\eqref{Opera_Diff_Grav} with its RP
contribution constructed from~\eqref{CAM_Grav_Scattering_amp_decomp_RP}. In Figs.~\ref{Fig:Conv_EW_to_GW_2Mw_1_Q_050M_CAM_vs_Exact}--\ref{Fig:Conv_EW_to_GW_2Mw_6_Q_099998M_CAM_vs_Exact}, we show the conversion of the EWs to GWs and we compare the differential conversion cross section constructed from the partial wave expansion~\eqref{fpm_1_2}--\eqref{Opera_Diff_EW_to_GW} with its RP contribution obtained from~\eqref{CAM_EW_to_GW_Scattering_amp_decomp_RP}.

In the intermediate reduced frequency regime ($2M\omega=1$) and for different charge-to-mass ratios $Q/M$, we can accurately describe the differential and conversion cross sections for scattering angle $\theta > 10^{\circ}$ for various processes by summing over a small number of RPs in the Regge pole contribution. Figs.\ref{Fig:EW_2Mw_1_Q_050M_CAM_vs_Exact}--\ref{Fig:EW_2Mw_1_Q_099998M_CAM_vs_Exact} show the results for $\text{EW} \to \text{EW}$, Figs.\ref{Fig:Grav_2Mw_1_Q_050M_CAM_vs_Exact}--\ref{Fig:Grav_2Mw_1_Q_099998M_CAM_vs_Exact} illustrate the results for $\text{GW}\to \text{GW}$, and Figs.~\ref{Fig:Conv_EW_to_GW_2Mw_1_Q_050M_CAM_vs_Exact}--\ref{Fig:Conv_EW_to_GW_2Mw_1_Q_099998M_CAM_vs_Exact} display the results for the conversion process of $\text{EW} \to \text{GW}$. However, it should be noted that in the case of nearly extremal RN BH ($Q/M=0.99998$), additional RPs are required to accurately describe the scattering and conversion cross sections at small scattering angles. It should also be noted that we did not consider background integrals in the construction of the RP contribution.

In the ``high''-frequency regime ($2M\omega = 6$) and for different charge-to-mass ratios $Q/M$, we can observe that the RP contribution involving a small number of RPs allows us to construct the differential scattering and conversion cross sections very well for intermediate and large scattering angle $\theta$ of different processes (see
Figs.~\ref{Fig:EW_2Mw_6_Q_050M_CAM_vs_Exact}--\ref{Fig:EW_2Mw_6_Q_099998M_CAM_vs_Exact} for EW $\to$ EW,
Figs.~\ref{Fig:Grav_2Mw_6_Q_050M_CAM_vs_Exact}--\ref{Fig:Grav_2Mw_6_Q_099998M_CAM_vs_Exact} for GW $\to$ GW, and
Figs.~\ref{Fig:Conv_EW_to_GW_2Mw_6_Q_050M_CAM_vs_Exact}--\ref{Fig:Conv_EW_to_GW_2Mw_6_Q_099998M_CAM_vs_Exact}
for EW $\to$ GW ). Summing over a larger number of RPs improves the Regge pole contribution, allowing us to describe the whole scattering and conversion differential cross sections, including small scattering angles, without considering the background integral. Furthermore, using the RP sum, we were able to overcome the challenges associated with the lack of convergence that characterizes the partial wave expansion defining the cross sections, which arises from the long range nature of the fields propagating on a RN background. It also enabled us to derive analytical approximations, in the short-wavelength regime, that describe both the BH glory and a large part of the orbiting oscillations, thus decoding the physical information hidden into the partial wave expansions.

\subsection{Asymptotic results and comments}
\label{sec_5_3}

We now have analytical approximations available to calculate the scattering and conversion cross sections for different processes, which are formally valid for $2M\omega \to \infty$. Specifically, we used the analytical expressions for PRs~\eqref{RP_s} and \eqref{rn_odd_s}--\eqref{zeta} for the associated $r_{sn}^{(o)}$, as well as the relation~\eqref{Rel_2} for $r_{sn}^{(e)}$, to obtain results for the case $Q<M$ (see Figs.~\ref{Fig:Exact_vs_Asympt_2Mw_6_Q_050M_CAM_vs_Exact}~and~\ref{Fig:Exact_vs_Asympt_2Mw_6_Q_080M_CAM_vs_Exact}). In the case of an extremal charged RN BH ($Q=M$), we achieved this by using asymptotic approximations for the lowest Regge poles [cf.~\eqref{PR_s_extrem}] 
\begin{align}\label{PR_s_extrem_bis}
  \lambda_n^{(s)}(\omega) & = b_c \,\omega +\left(\frac{i(2n-1)}{2\sqrt{2}}\mp \frac{1}{2}\right) \nonumber \\
   & \frac{6n(n-1)+35}{256}\frac{1}{(M\omega)}
   + \underset{\omega \to \infty}{\mathcal{O}}\left(\frac{1}{(2M\omega)^2}\right)
\end{align}
and the associated residues $r_{sn}^{(o)}$~\eqref{rn_odd_s_extrem}--\eqref{y}, as well as $r_{sn}^{(e)}$ obtained from  the relation~\eqref{Rel_2} (see Fig.~\ref{Fig:Exact_vs_Asympt_2Mw_6_Q_099998M_CAM_vs_Exact}).

In Figs.~\ref{Fig:Exact_vs_Asympt_2Mw_6_Q_050M_CAM_vs_Exact}--\ref{Fig:Exact_vs_Asympt_2Mw_6_Q_099998M_CAM_vs_Exact}, we present results comparing exact scattering and conversion cross sections with their analytical RP approximations. The comparisons are made for reduced frequency $2M\omega = 6$ and different charge-to-mass ratios $Q/M = 0.50$, $0.80$, and $0.99998$, with the summations performed over $20$ RPs. The analytical RP approximations are found to reproduce the glory cross section and a significant part of the orbiting cross section with high accuracy. In fact, from a purely wave point of view, which characterizes the RP approach, the glory and orbiting oscillation effects are not fundamentally different. Indeed, we consider that both phenomena are generated by the excitation of surface waves propagating close to the RN BH photon (graviton) sphere and are a consequence of diffraction effects due to this hypersurface~\cite{Andersson1994bis, Decanini2003, Dolan:2009, Decanini2010, Decanini2010bis, Folacci:2019cmc, Folacci:2019vtt}. It should be noted that despite the inaccuracy of the higher RP approximations, using analytical RP sums involving $20$ terms enables us to describe the cross sections over a wide range of scattering angles.

In Fig.~\ref{Fig:Exact_vs_Asympt_2Mw_6_Q_099998M_CAM_vs_Exact}, we first  observe that the scattering cross sections for EWs and GWs coincide. This is because the behavior of the RPs and their associated residues converge towards those for $Q=M$, in which case the scattering cross sections for EWs and GWs are equal. In fact, this also confirms the fact that there are no other branches of RPs when $Q=M$, as discussed in Sec.~\ref{sec_3_2_2}. Moreover, we find that for nearly extremal charged RN BH with a charge-to-mass ratio of $Q/M=0.99998$ and beyond $\theta=90^\circ$, the conversion process of EWs into GWs (or vice versa) becomes larger than the scattering process. In other words, the converted flux exceeds the scattered flux. This effect is confirmed by the analytic RP approximation in the ``high''-frequency regime (short-wavelength). It is important to note that this effect has already been shown using the geometric-optics approximation~\cite{OuldElHadj:2021fqi}.

\section{Conclusion and perspectives}
\label{sec_6}

In this article, we used an alternative approach based on the analytic extension of the $S$-matrix in the complex angular momentum plane and its associated Regge poles to compute the scattering and conversion cross sections for planar electromagnetic and gravitational waves interacting with a Reissner-Nordström BH. To achieve this, we focused on the resonance spectra of the $S$-matrix in the CAM plane and calculated its Regge poles for various configurations of the charge-to-mass ratio $Q/M$ in both intermediate- and high-reduced frequency regimes. We employed Leaver's method~\cite{Leaver:1985ax,leaver1986solutions,Leaver:1990zz}, specifically, its modified Hill determinant approach~\cite{mp} for $Q<M$, and the Onozawa \emph{et al.} algorithm~\cite{Onozawa:1995vu} for the extremal charged RN case $Q=M$.

In the intermediate and short-wavelength regimes, we numerically reconstructed the scattering and conversion cross sections for the different processes (EW $\to$ EW, GW $\to$ GW and EW $\to$ GW) from the Regge pole sums and their associated residues, and we shown a perfect agreement with those obtained using the partial wave expansion method. This was achieved for different charge-to-mass ratios $Q/M$.

Finally, in the short-wavelength regime, we derived an analytical Regge pole approximation from the asymptotic expressions of Regge poles and their associated residues, which allowed us to accurately describe both the glory and the orbiting oscillations of the RN BH. Furthermore, we illustrated that for sufficiently charged BH, the converted flux exceeds the scattered flux at large angles. This effect is particularly pronounced for extremal charged RN BH ($Q=M$), where the angle at which the converted flux exceeds the scattered flux reaches a minimum of $90^{\circ}$. This has already been demonstrated in our previous article using the geometric-optics approximation \cite{OuldElHadj:2021fqi}.

We hope to extend our study in our next works to the scattering of a scalar field by a rotating charged black hole to explore various implications \cite{Leite:2019eis}. Extending our study to the scattering and conversion of electromagnetic and gravitational waves by a Kerr-Newman black hole is a more challenging task \cite{Pani:2013ija}. We also hope to make progress in this direction.

\begin{acknowledgments}
We are grateful to Sam R. Dolan for the valuable conversations and comments regarding this work, and we would like to thank him.
\end{acknowledgments}

\bibliography{RN_CAM}

\begin{thebibliography}{71}%
\makeatletter
\providecommand \@ifxundefined [1]{%
 \@ifx{#1\undefined}
}%
\providecommand \@ifnum [1]{%
 \ifnum #1\expandafter \@firstoftwo
 \else \expandafter \@secondoftwo
 \fi
}%
\providecommand \@ifx [1]{%
 \ifx #1\expandafter \@firstoftwo
 \else \expandafter \@secondoftwo
 \fi
}%
\providecommand \natexlab [1]{#1}%
\providecommand \enquote  [1]{``#1''}%
\providecommand \bibnamefont  [1]{#1}%
\providecommand \bibfnamefont [1]{#1}%
\providecommand \citenamefont [1]{#1}%
\providecommand \href@noop [0]{\@secondoftwo}%
\providecommand \href [0]{\begingroup \@sanitize@url \@href}%
\providecommand \@href[1]{\@@startlink{#1}\@@href}%
\providecommand \@@href[1]{\endgroup#1\@@endlink}%
\providecommand \@sanitize@url [0]{\catcode `\\12\catcode `\$12\catcode
  `\&12\catcode `\#12\catcode `\^12\catcode `\_12\catcode `\%12\relax}%
\providecommand \@@startlink[1]{}%
\providecommand \@@endlink[0]{}%
\providecommand \url  [0]{\begingroup\@sanitize@url \@url }%
\providecommand \@url [1]{\endgroup\@href {#1}{\urlprefix }}%
\providecommand \urlprefix  [0]{URL }%
\providecommand \Eprint [0]{\href }%
\providecommand \doibase [0]{https://doi.org/}%
\providecommand \selectlanguage [0]{\@gobble}%
\providecommand \bibinfo  [0]{\@secondoftwo}%
\providecommand \bibfield  [0]{\@secondoftwo}%
\providecommand \translation [1]{[#1]}%
\providecommand \BibitemOpen [0]{}%
\providecommand \bibitemStop [0]{}%
\providecommand \bibitemNoStop [0]{.\EOS\space}%
\providecommand \EOS [0]{\spacefactor3000\relax}%
\providecommand \BibitemShut  [1]{\csname bibitem#1\endcsname}%
\let\auto@bib@innerbib\@empty
\bibitem [{\citenamefont {Zerilli}(1974)}]{Zerilli:1974ai}%
  \BibitemOpen
  \bibfield  {author} {\bibinfo {author} {\bibfnamefont {F.~J.}\ \bibnamefont
  {Zerilli}},\ }\bibfield  {title} {\bibinfo {title} {{Perturbation analysis
  for gravitational and electromagnetic radiation in a Reissner-Nordstr\"om
  geometry}},\ }\href {https://doi.org/10.1103/PhysRevD.9.860} {\bibfield
  {journal} {\bibinfo  {journal} {Phys. Rev. D}\ }\textbf {\bibinfo {volume}
  {9}},\ \bibinfo {pages} {860} (\bibinfo {year} {1974})}\BibitemShut {NoStop}%
\bibitem [{\citenamefont {Moncrief}(1974{\natexlab{a}})}]{Moncrief-1974a}%
  \BibitemOpen
  \bibfield  {author} {\bibinfo {author} {\bibfnamefont {V.}~\bibnamefont
  {Moncrief}},\ }\bibfield  {title} {\bibinfo {title} {Odd-parity stability of
  a {Reissner-Nordstr\"om} black hole},\ }\href
  {https://doi.org/10.1103/PhysRevD.9.2707} {\bibfield  {journal} {\bibinfo
  {journal} {Phys. Rev. D}\ }\textbf {\bibinfo {volume} {9}},\ \bibinfo {pages}
  {2707} (\bibinfo {year} {1974}{\natexlab{a}})}\BibitemShut {NoStop}%
\bibitem [{\citenamefont {Moncrief}(1974{\natexlab{b}})}]{Moncrief-1974b}%
  \BibitemOpen
  \bibfield  {author} {\bibinfo {author} {\bibfnamefont {V.}~\bibnamefont
  {Moncrief}},\ }\bibfield  {title} {\bibinfo {title} {Stability of
  {Reissner-Nordstr\"om} black holes},\ }\href
  {https://doi.org/10.1103/PhysRevD.10.1057} {\bibfield  {journal} {\bibinfo
  {journal} {Phys. Rev. D}\ }\textbf {\bibinfo {volume} {10}},\ \bibinfo
  {pages} {1057} (\bibinfo {year} {1974}{\natexlab{b}})}\BibitemShut {NoStop}%
\bibitem [{\citenamefont {Olson}\ and\ \citenamefont
  {Unruh}(1974)}]{Olson:1974nk}%
  \BibitemOpen
  \bibfield  {author} {\bibinfo {author} {\bibfnamefont {D.~W.}\ \bibnamefont
  {Olson}}\ and\ \bibinfo {author} {\bibfnamefont {W.~G.}\ \bibnamefont
  {Unruh}},\ }\bibfield  {title} {\bibinfo {title} {{Conversion of
  electromagnetic to gravitational radiation by scattering from a charged black
  hole}},\ }\href {https://doi.org/10.1103/PhysRevLett.33.1116} {\bibfield
  {journal} {\bibinfo  {journal} {Phys. Rev. Lett.}\ }\textbf {\bibinfo
  {volume} {33}},\ \bibinfo {pages} {1116} (\bibinfo {year}
  {1974})}\BibitemShut {NoStop}%
\bibitem [{\citenamefont {Moncrief}(1975)}]{Moncrief-1975}%
  \BibitemOpen
  \bibfield  {author} {\bibinfo {author} {\bibfnamefont {V.}~\bibnamefont
  {Moncrief}},\ }\bibfield  {title} {\bibinfo {title} {Gauge-invariant
  perturbations of {Reissner-Nordstr\"om} black holes},\ }\href
  {https://doi.org/10.1103/PhysRevD.12.1526} {\bibfield  {journal} {\bibinfo
  {journal} {Phys. Rev. D}\ }\textbf {\bibinfo {volume} {12}},\ \bibinfo
  {pages} {1526} (\bibinfo {year} {1975})}\BibitemShut {NoStop}%
\bibitem [{\citenamefont {Chitre}\ \emph {et~al.}(1975)\citenamefont {Chitre},
  \citenamefont {Price},\ and\ \citenamefont {Sandberg}}]{Chitre:1975ew}%
  \BibitemOpen
  \bibfield  {author} {\bibinfo {author} {\bibfnamefont {D.~M.}\ \bibnamefont
  {Chitre}}, \bibinfo {author} {\bibfnamefont {R.~H.}\ \bibnamefont {Price}},\
  and\ \bibinfo {author} {\bibfnamefont {V.~D.}\ \bibnamefont {Sandberg}},\
  }\bibfield  {title} {\bibinfo {title} {{Electromagnetic Radiation Due to
  Space-Time Oscillations}},\ }\href {https://doi.org/10.1103/PhysRevD.11.747}
  {\bibfield  {journal} {\bibinfo  {journal} {Phys. Rev. D}\ }\textbf {\bibinfo
  {volume} {11}},\ \bibinfo {pages} {747} (\bibinfo {year} {1975})}\BibitemShut
  {NoStop}%
\bibitem [{\citenamefont {Chandrasekhar}(1979)}]{Chandrasekhar:1979iz}%
  \BibitemOpen
  \bibfield  {author} {\bibinfo {author} {\bibfnamefont {S.}~\bibnamefont
  {Chandrasekhar}},\ }\bibfield  {title} {\bibinfo {title} {On the equations
  governing the perturbations of the {Reissner-Nordstr\"om} black hole},\
  }\href {https://doi.org/10.1098/rspa.1979.0028} {\bibfield  {journal}
  {\bibinfo  {journal} {Proc. Roy. Soc. Lond. A}\ }\textbf {\bibinfo {volume}
  {365}},\ \bibinfo {pages} {453} (\bibinfo {year} {1979})}\BibitemShut
  {NoStop}%
\bibitem [{\citenamefont {Johnston}\ \emph {et~al.}(1974)\citenamefont
  {Johnston}, \citenamefont {Ruffini},\ and\ \citenamefont
  {Zerilli}}]{Johnston:1974vf}%
  \BibitemOpen
  \bibfield  {author} {\bibinfo {author} {\bibfnamefont {M.}~\bibnamefont
  {Johnston}}, \bibinfo {author} {\bibfnamefont {R.}~\bibnamefont {Ruffini}},\
  and\ \bibinfo {author} {\bibfnamefont {F.}~\bibnamefont {Zerilli}},\
  }\bibfield  {title} {\bibinfo {title} {{Electromagnetically induced
  gravitational radiation}},\ }\href
  {https://doi.org/10.1016/0370-2693(74)90505-X} {\bibfield  {journal}
  {\bibinfo  {journal} {Phys. Lett. B}\ }\textbf {\bibinfo {volume} {49}},\
  \bibinfo {pages} {185} (\bibinfo {year} {1974})}\BibitemShut {NoStop}%
\bibitem [{\citenamefont {Gerlach}(1974)}]{Gerlach:1974zz}%
  \BibitemOpen
  \bibfield  {author} {\bibinfo {author} {\bibfnamefont {U.~H.}\ \bibnamefont
  {Gerlach}},\ }\bibfield  {title} {\bibinfo {title} {Beat frequency
  oscillations near charged black holes and other electrovacuum geometries},\
  }\href {https://doi.org/10.1103/PhysRevLett.32.1023} {\bibfield  {journal}
  {\bibinfo  {journal} {Phys. Rev. Lett.}\ }\textbf {\bibinfo {volume} {32}},\
  \bibinfo {pages} {1023} (\bibinfo {year} {1974})}\BibitemShut {NoStop}%
\bibitem [{\citenamefont {Sibgatullin}(1974)}]{Sibgatullin:1974jq}%
  \BibitemOpen
  \bibfield  {author} {\bibinfo {author} {\bibfnamefont {N.~R.}\ \bibnamefont
  {Sibgatullin}},\ }\bibfield  {title} {\bibinfo {title} {{Interaction between
  short gravitational and electromagnetic waves in arbitrary external
  electromagnetic fields}},\ }\href@noop {} {\bibfield  {journal} {\bibinfo
  {journal} {Zh. Eksp. Teor. Fiz.}\ }\textbf {\bibinfo {volume} {66}},\
  \bibinfo {pages} {1187} (\bibinfo {year} {1974})}\BibitemShut {NoStop}%
\bibitem [{\citenamefont {Matzner}(1976)}]{Matzner:1976kj}%
  \BibitemOpen
  \bibfield  {author} {\bibinfo {author} {\bibfnamefont {R.~A.}\ \bibnamefont
  {Matzner}},\ }\bibfield  {title} {\bibinfo {title} {Low frequency limit
  conversion cross-sections for charged black holes},\ }\href
  {https://doi.org/10.1103/PhysRevD.14.3274} {\bibfield  {journal} {\bibinfo
  {journal} {Phys. Rev. D}\ }\textbf {\bibinfo {volume} {14}},\ \bibinfo
  {pages} {3274} (\bibinfo {year} {1976})}\BibitemShut {NoStop}%
\bibitem [{\citenamefont {Fabbri}(1977)}]{Fabbri:1977}%
  \BibitemOpen
  \bibfield  {author} {\bibinfo {author} {\bibfnamefont {R.}~\bibnamefont
  {Fabbri}},\ }\bibfield  {title} {\bibinfo {title} {Electromagnetic and
  gravitational waves in the background of a {Reissner-Nordstr\"om} black
  hole},\ }\href@noop {} {\bibfield  {journal} {\bibinfo  {journal} {Il Nuovo
  Cimento B (1971-1996)}\ }\textbf {\bibinfo {volume} {40}},\ \bibinfo {pages}
  {311} (\bibinfo {year} {1977})}\BibitemShut {NoStop}%
\bibitem [{\citenamefont {De~Logi}\ and\ \citenamefont
  {Mickelson}(1977)}]{DeLogi:1977qe}%
  \BibitemOpen
  \bibfield  {author} {\bibinfo {author} {\bibfnamefont {W.~K.}\ \bibnamefont
  {De~Logi}}\ and\ \bibinfo {author} {\bibfnamefont {A.~R.}\ \bibnamefont
  {Mickelson}},\ }\bibfield  {title} {\bibinfo {title} {Electrogravitational
  conversion cross-sections in static electromagnetic fields},\ }\href
  {https://doi.org/10.1103/PhysRevD.16.2915} {\bibfield  {journal} {\bibinfo
  {journal} {Phys. Rev. D}\ }\textbf {\bibinfo {volume} {16}},\ \bibinfo
  {pages} {2915} (\bibinfo {year} {1977})}\BibitemShut {NoStop}%
\bibitem [{\citenamefont {Breuer}\ \emph {et~al.}(1981)\citenamefont {Breuer},
  \citenamefont {Rosenbaum}, \citenamefont {Ryan},\ and\ \citenamefont
  {Matzner}}]{Breuer:1981kd}%
  \BibitemOpen
  \bibfield  {author} {\bibinfo {author} {\bibfnamefont {R.~A.}\ \bibnamefont
  {Breuer}}, \bibinfo {author} {\bibfnamefont {M.}~\bibnamefont {Rosenbaum}},
  \bibinfo {author} {\bibfnamefont {M.~P.}\ \bibnamefont {Ryan}},\ and\
  \bibinfo {author} {\bibfnamefont {R.~A.}\ \bibnamefont {Matzner}},\
  }\bibfield  {title} {\bibinfo {title} {Gravitational electromagnetic
  conversion scattering on fixed charges in the {B}orn approximation},\ }\href
  {https://doi.org/10.1103/PhysRevD.23.305} {\bibfield  {journal} {\bibinfo
  {journal} {Phys. Rev. D}\ }\textbf {\bibinfo {volume} {23}},\ \bibinfo
  {pages} {305} (\bibinfo {year} {1981})}\BibitemShut {NoStop}%
\bibitem [{\citenamefont {Gunter}(1980)}]{Gunter:1980}%
  \BibitemOpen
  \bibfield  {author} {\bibinfo {author} {\bibfnamefont {D.~L.}\ \bibnamefont
  {Gunter}},\ }\bibfield  {title} {\bibinfo {title} {A study of the coupled
  gravitational and electromagnetic perturbations to the reissner-nordstr{\"o}m
  black hole: The scattering matrix, energy conversion, and quasi-normal
  modes},\ }\href@noop {} {\bibfield  {journal} {\bibinfo  {journal}
  {Philosophical Transactions of the Royal Society of London. Series A,
  Mathematical and Physical Sciences}\ }\textbf {\bibinfo {volume} {296}},\
  \bibinfo {pages} {497} (\bibinfo {year} {1980})}\BibitemShut {NoStop}%
\bibitem [{\citenamefont {de~Castillo}(1987)}]{Castillo:1987}%
  \BibitemOpen
  \bibfield  {author} {\bibinfo {author} {\bibfnamefont {G.~F.~T.}\
  \bibnamefont {de~Castillo}},\ }\bibfield  {title} {\bibinfo {title}
  {Gravitational and electromagnetic perturbations of the
  {Reissner-Nordstr\"om} solution},\ }\href@noop {} {\bibfield  {journal}
  {\bibinfo  {journal} {Classical and Quantum Gravity}\ }\textbf {\bibinfo
  {volume} {4}},\ \bibinfo {pages} {1133} (\bibinfo {year} {1987})}\BibitemShut
  {NoStop}%
\bibitem [{\citenamefont {Castillo}\ and\ \citenamefont
  {Cartas-Fuentevilla}(1996)}]{Castillo:1996jm}%
  \BibitemOpen
  \bibfield  {author} {\bibinfo {author} {\bibfnamefont {G.~F. T.~d.}\
  \bibnamefont {Castillo}}\ and\ \bibinfo {author} {\bibfnamefont
  {R.}~\bibnamefont {Cartas-Fuentevilla}},\ }\bibfield  {title} {\bibinfo
  {title} {Scattering by a {Reissner-Nordstr\"om} black hole},\ }\href
  {https://doi.org/10.1103/PhysRevD.54.4886} {\bibfield  {journal} {\bibinfo
  {journal} {Phys. Rev. D}\ }\textbf {\bibinfo {volume} {54}},\ \bibinfo
  {pages} {4886} (\bibinfo {year} {1996})}\BibitemShut {NoStop}%
\bibitem [{\citenamefont {Gertsenshtein}(1962)}]{gertsenshtein1962wave}%
  \BibitemOpen
  \bibfield  {author} {\bibinfo {author} {\bibfnamefont {M.~E.}\ \bibnamefont
  {Gertsenshtein}},\ }\bibfield  {title} {\bibinfo {title} {Wave resonance of
  light and gravitational waves},\ }\href@noop {} {\bibfield  {journal}
  {\bibinfo  {journal} {Sov Phys JETP}\ }\textbf {\bibinfo {volume} {14}},\
  \bibinfo {pages} {84} (\bibinfo {year} {1962})}\BibitemShut {NoStop}%
\bibitem [{\citenamefont {Zel'dovich}(1973)}]{zel1973electromagnetic}%
  \BibitemOpen
  \bibfield  {author} {\bibinfo {author} {\bibfnamefont {Y.~B.}\ \bibnamefont
  {Zel'dovich}},\ }\bibfield  {title} {\bibinfo {title} {Electromagnetic and
  gravitational waves in a stationary magnetic field},\ }\href@noop {}
  {\bibfield  {journal} {\bibinfo  {journal} {Zh. Eksp. Teor. Fiz}\ }\textbf
  {\bibinfo {volume} {65}},\ \bibinfo {pages} {1311} (\bibinfo {year}
  {1973})}\BibitemShut {NoStop}%
\bibitem [{\citenamefont {Domcke}\ and\ \citenamefont
  {Garcia-Cely}(2021)}]{Domcke:2020yzq}%
  \BibitemOpen
  \bibfield  {author} {\bibinfo {author} {\bibfnamefont {V.}~\bibnamefont
  {Domcke}}\ and\ \bibinfo {author} {\bibfnamefont {C.}~\bibnamefont
  {Garcia-Cely}},\ }\bibfield  {title} {\bibinfo {title} {{Potential of radio
  telescopes as high-frequency gravitational wave detectors}},\ }\href
  {https://doi.org/10.1103/PhysRevLett.126.021104} {\bibfield  {journal}
  {\bibinfo  {journal} {Phys. Rev. Lett.}\ }\textbf {\bibinfo {volume} {126}},\
  \bibinfo {pages} {021104} (\bibinfo {year} {2021})},\ \Eprint
  {https://arxiv.org/abs/2006.01161} {arXiv:2006.01161 [astro-ph.CO]}
  \BibitemShut {NoStop}%
\bibitem [{\citenamefont {Marklund}\ \emph {et~al.}(2000)\citenamefont
  {Marklund}, \citenamefont {Brodin},\ and\ \citenamefont
  {Dunsby}}]{Marklund:1999sp}%
  \BibitemOpen
  \bibfield  {author} {\bibinfo {author} {\bibfnamefont {M.}~\bibnamefont
  {Marklund}}, \bibinfo {author} {\bibfnamefont {G.}~\bibnamefont {Brodin}},\
  and\ \bibinfo {author} {\bibfnamefont {P.~K.~S.}\ \bibnamefont {Dunsby}},\
  }\bibfield  {title} {\bibinfo {title} {{Radio wave emissions due to
  gravitational radiation}},\ }\href {https://doi.org/10.1086/308957}
  {\bibfield  {journal} {\bibinfo  {journal} {Astrophys. J.}\ }\textbf
  {\bibinfo {volume} {536}},\ \bibinfo {pages} {875} (\bibinfo {year}
  {2000})},\ \Eprint {https://arxiv.org/abs/astro-ph/9907350}
  {arXiv:astro-ph/9907350} \BibitemShut {NoStop}%
\bibitem [{\citenamefont {Dolgov}\ and\ \citenamefont
  {Ejlli}(2012)}]{Dolgov:2012be}%
  \BibitemOpen
  \bibfield  {author} {\bibinfo {author} {\bibfnamefont {A.~D.}\ \bibnamefont
  {Dolgov}}\ and\ \bibinfo {author} {\bibfnamefont {D.}~\bibnamefont {Ejlli}},\
  }\bibfield  {title} {\bibinfo {title} {{Conversion of relic gravitational
  waves into photons in cosmological magnetic fields}},\ }\href
  {https://doi.org/10.1088/1475-7516/2012/12/003} {\bibfield  {journal}
  {\bibinfo  {journal} {JCAP}\ }\textbf {\bibinfo {volume} {12}},\ \bibinfo
  {pages} {003}},\ \Eprint {https://arxiv.org/abs/1211.0500} {arXiv:1211.0500
  [gr-qc]} \BibitemShut {NoStop}%
\bibitem [{\citenamefont {Fujita}\ \emph {et~al.}(2020)\citenamefont {Fujita},
  \citenamefont {Kamada},\ and\ \citenamefont {Nakai}}]{Fujita:2020rdx}%
  \BibitemOpen
  \bibfield  {author} {\bibinfo {author} {\bibfnamefont {T.}~\bibnamefont
  {Fujita}}, \bibinfo {author} {\bibfnamefont {K.}~\bibnamefont {Kamada}},\
  and\ \bibinfo {author} {\bibfnamefont {Y.}~\bibnamefont {Nakai}},\ }\bibfield
   {title} {\bibinfo {title} {Gravitational waves from primordial magnetic
  fields via photon-graviton conversion},\ }\href
  {https://doi.org/10.1103/PhysRevD.102.103501} {\bibfield  {journal} {\bibinfo
   {journal} {Phys. Rev. D}\ }\textbf {\bibinfo {volume} {102}},\ \bibinfo
  {pages} {103501} (\bibinfo {year} {2020})},\ \Eprint
  {https://arxiv.org/abs/2002.07548} {arXiv:2002.07548 [astro-ph.CO]}
  \BibitemShut {NoStop}%
\bibitem [{\citenamefont {Kushwaha}\ \emph {et~al.}(2022)\citenamefont
  {Kushwaha}, \citenamefont {Malik},\ and\ \citenamefont
  {Shankaranarayanan}}]{Kushwaha:2022twx}%
  \BibitemOpen
  \bibfield  {author} {\bibinfo {author} {\bibfnamefont {A.}~\bibnamefont
  {Kushwaha}}, \bibinfo {author} {\bibfnamefont {S.}~\bibnamefont {Malik}},\
  and\ \bibinfo {author} {\bibfnamefont {S.}~\bibnamefont
  {Shankaranarayanan}},\ }\bibfield  {title} {\bibinfo {title}
  {{Gertsenshtein-Zel$'$dovich effect explains the origin of Fast Radio
  Bursts}},\ }\href@noop {} {\  (\bibinfo {year} {2022})},\ \Eprint
  {https://arxiv.org/abs/2202.00032} {arXiv:2202.00032 [astro-ph.HE]}
  \BibitemShut {NoStop}%
\bibitem [{\citenamefont {Crispino}\ \emph
  {et~al.}(2009{\natexlab{a}})\citenamefont {Crispino}, \citenamefont {Dolan},\
  and\ \citenamefont {Oliveira}}]{Crispino:2009ki}%
  \BibitemOpen
  \bibfield  {author} {\bibinfo {author} {\bibfnamefont {L.~C.~B.}\
  \bibnamefont {Crispino}}, \bibinfo {author} {\bibfnamefont {S.~R.}\
  \bibnamefont {Dolan}},\ and\ \bibinfo {author} {\bibfnamefont {E.~S.}\
  \bibnamefont {Oliveira}},\ }\bibfield  {title} {\bibinfo {title} {{Scattering
  of massless scalar waves by Reissner-Nordstr\"om black holes}},\ }\href
  {https://doi.org/10.1103/PhysRevD.79.064022} {\bibfield  {journal} {\bibinfo
  {journal} {Phys. Rev. D}\ }\textbf {\bibinfo {volume} {79}},\ \bibinfo
  {pages} {064022} (\bibinfo {year} {2009}{\natexlab{a}})},\ \Eprint
  {https://arxiv.org/abs/0904.0999} {arXiv:0904.0999 [gr-qc]} \BibitemShut
  {NoStop}%
\bibitem [{\citenamefont {Crispino}\ \emph {et~al.}(2014)\citenamefont
  {Crispino}, \citenamefont {Dolan}, \citenamefont {Higuchi},\ and\
  \citenamefont {de~Oliveira}}]{Crispino:2014eea}%
  \BibitemOpen
  \bibfield  {author} {\bibinfo {author} {\bibfnamefont {L.~C.~B.}\
  \bibnamefont {Crispino}}, \bibinfo {author} {\bibfnamefont {S.~R.}\
  \bibnamefont {Dolan}}, \bibinfo {author} {\bibfnamefont {A.}~\bibnamefont
  {Higuchi}},\ and\ \bibinfo {author} {\bibfnamefont {E.~S.}\ \bibnamefont
  {de~Oliveira}},\ }\bibfield  {title} {\bibinfo {title} {{Inferring black hole
  charge from backscattered electromagnetic radiation}},\ }\href
  {https://doi.org/10.1103/PhysRevD.90.064027} {\bibfield  {journal} {\bibinfo
  {journal} {Phys. Rev. D}\ }\textbf {\bibinfo {volume} {90}},\ \bibinfo
  {pages} {064027} (\bibinfo {year} {2014})},\ \Eprint
  {https://arxiv.org/abs/1409.4803} {arXiv:1409.4803 [gr-qc]} \BibitemShut
  {NoStop}%
\bibitem [{\citenamefont {Crispino}\ \emph {et~al.}(2015)\citenamefont
  {Crispino}, \citenamefont {Dolan}, \citenamefont {Higuchi},\ and\
  \citenamefont {de~Oliveira}}]{Crispino:2015gua}%
  \BibitemOpen
  \bibfield  {author} {\bibinfo {author} {\bibfnamefont {L.~C.~B.}\
  \bibnamefont {Crispino}}, \bibinfo {author} {\bibfnamefont {S.~R.}\
  \bibnamefont {Dolan}}, \bibinfo {author} {\bibfnamefont {A.}~\bibnamefont
  {Higuchi}},\ and\ \bibinfo {author} {\bibfnamefont {E.~S.}\ \bibnamefont
  {de~Oliveira}},\ }\bibfield  {title} {\bibinfo {title} {{Scattering from
  charged black holes and supergravity}},\ }\href
  {https://doi.org/10.1103/PhysRevD.92.084056} {\bibfield  {journal} {\bibinfo
  {journal} {Phys. Rev. D}\ }\textbf {\bibinfo {volume} {92}},\ \bibinfo
  {pages} {084056} (\bibinfo {year} {2015})},\ \Eprint
  {https://arxiv.org/abs/1507.03993} {arXiv:1507.03993 [gr-qc]} \BibitemShut
  {NoStop}%
\bibitem [{\citenamefont {Cotaescu}\ \emph {et~al.}(2016)\citenamefont
  {Cotaescu}, \citenamefont {Crucean},\ and\ \citenamefont
  {Sporea}}]{Cotaescu:2016aty}%
  \BibitemOpen
  \bibfield  {author} {\bibinfo {author} {\bibfnamefont {I.~I.}\ \bibnamefont
  {Cotaescu}}, \bibinfo {author} {\bibfnamefont {C.}~\bibnamefont {Crucean}},\
  and\ \bibinfo {author} {\bibfnamefont {C.}~\bibnamefont {Sporea}},\
  }\bibfield  {title} {\bibinfo {title} {{Partial wave analysis of the Dirac
  fermions scattered from Reissner\textendash{}Nordstr\"om charged black
  holes}},\ }\href {https://doi.org/10.1140/epjc/s10052-016-4260-0} {\bibfield
  {journal} {\bibinfo  {journal} {Eur. Phys. J. C}\ }\textbf {\bibinfo {volume}
  {76}},\ \bibinfo {pages} {413} (\bibinfo {year} {2016})},\ \Eprint
  {https://arxiv.org/abs/1601.03673} {arXiv:1601.03673 [gr-qc]} \BibitemShut
  {NoStop}%
\bibitem [{\citenamefont {Sporea}(2017)}]{Sporea:2017zxe}%
  \BibitemOpen
  \bibfield  {author} {\bibinfo {author} {\bibfnamefont {C.~A.}\ \bibnamefont
  {Sporea}},\ }\bibfield  {title} {\bibinfo {title} {{Scattering of massless
  fermions by Schwarzschild and Reissner-Nordstr\"om black holes}},\ }\href
  {https://doi.org/10.1088/1674-1137/41/12/123101} {\bibfield  {journal}
  {\bibinfo  {journal} {Chin. Phys. C}\ }\textbf {\bibinfo {volume} {41}},\
  \bibinfo {pages} {123101} (\bibinfo {year} {2017})},\ \Eprint
  {https://arxiv.org/abs/1707.08374} {arXiv:1707.08374 [gr-qc]} \BibitemShut
  {NoStop}%
\bibitem [{\citenamefont {Ould El~Hadj}\ and\ \citenamefont
  {Dolan}(2022)}]{OuldElHadj:2021fqi}%
  \BibitemOpen
  \bibfield  {author} {\bibinfo {author} {\bibfnamefont {M.}~\bibnamefont {Ould
  El~Hadj}}\ and\ \bibinfo {author} {\bibfnamefont {S.~R.}\ \bibnamefont
  {Dolan}},\ }\bibfield  {title} {\bibinfo {title} {{Conversion of
  electromagnetic and gravitational waves by a charged black hole}},\ }\href
  {https://doi.org/10.1103/PhysRevD.106.044002} {\bibfield  {journal} {\bibinfo
   {journal} {Phys. Rev. D}\ }\textbf {\bibinfo {volume} {106}},\ \bibinfo
  {pages} {044002} (\bibinfo {year} {2022})},\ \Eprint
  {https://arxiv.org/abs/2106.09731} {arXiv:2106.09731 [gr-qc]} \BibitemShut
  {NoStop}%
\bibitem [{\citenamefont {Futterman}\ \emph {et~al.}(2012)\citenamefont
  {Futterman}, \citenamefont {Handler},\ and\ \citenamefont
  {Matzner}}]{Futterman:1988ni}%
  \BibitemOpen
  \bibfield  {author} {\bibinfo {author} {\bibfnamefont {J.~A.~H.}\
  \bibnamefont {Futterman}}, \bibinfo {author} {\bibfnamefont {F.~A.}\
  \bibnamefont {Handler}},\ and\ \bibinfo {author} {\bibfnamefont {R.~A.}\
  \bibnamefont {Matzner}},\ }\href {https://doi.org/10.1017/CBO9780511735615}
  {\emph {\bibinfo {title} {Scattering from black holes}}},\ Cambridge
  Monographs on Mathematical Physics\ (\bibinfo  {publisher} {Cambridge
  University Press},\ \bibinfo {year} {2012})\BibitemShut {NoStop}%
\bibitem [{\citenamefont {Folacci}\ and\ \citenamefont {Ould
  El~Hadj}(2019{\natexlab{a}})}]{Folacci:2019cmc}%
  \BibitemOpen
  \bibfield  {author} {\bibinfo {author} {\bibfnamefont {A.}~\bibnamefont
  {Folacci}}\ and\ \bibinfo {author} {\bibfnamefont {M.}~\bibnamefont {Ould
  El~Hadj}},\ }\bibfield  {title} {\bibinfo {title} {{Regge pole description of
  scattering of scalar and electromagnetic waves by a Schwarzschild black
  hole}},\ }\href {https://doi.org/10.1103/PhysRevD.99.104079} {\bibfield
  {journal} {\bibinfo  {journal} {Phys. Rev.}\ }\textbf {\bibinfo {volume}
  {D99}},\ \bibinfo {pages} {104079} (\bibinfo {year} {2019}{\natexlab{a}})},\
  \Eprint {https://arxiv.org/abs/1901.03965} {arXiv:1901.03965 [gr-qc]}
  \BibitemShut {NoStop}%
\bibitem [{\citenamefont {Folacci}\ and\ \citenamefont {Ould
  El~Hadj}(2019{\natexlab{b}})}]{Folacci:2019vtt}%
  \BibitemOpen
  \bibfield  {author} {\bibinfo {author} {\bibfnamefont {A.}~\bibnamefont
  {Folacci}}\ and\ \bibinfo {author} {\bibfnamefont {M.}~\bibnamefont {Ould
  El~Hadj}},\ }\bibfield  {title} {\bibinfo {title} {{Regge pole description of
  scattering of gravitational waves by a Schwarzschild black hole}},\ }\href
  {https://doi.org/10.1103/PhysRevD.100.064009} {\bibfield  {journal} {\bibinfo
   {journal} {Phys. Rev.}\ }\textbf {\bibinfo {volume} {D100}},\ \bibinfo
  {pages} {064009} (\bibinfo {year} {2019}{\natexlab{b}})},\ \Eprint
  {https://arxiv.org/abs/1906.01441} {arXiv:1906.01441 [gr-qc]} \BibitemShut
  {NoStop}%
\bibitem [{\citenamefont {Folacci}\ and\ \citenamefont {Ould
  El~Hadj}(2018)}]{Folacci:2018sef}%
  \BibitemOpen
  \bibfield  {author} {\bibinfo {author} {\bibfnamefont {A.}~\bibnamefont
  {Folacci}}\ and\ \bibinfo {author} {\bibfnamefont {M.}~\bibnamefont {Ould
  El~Hadj}},\ }\bibfield  {title} {\bibinfo {title} {{Alternative description
  of gravitational radiation from black holes based on the Regge poles of the
  ${\cal S}$-matrix and the associated residues}},\ }\href
  {https://doi.org/10.1103/PhysRevD.98.064052} {\bibfield  {journal} {\bibinfo
  {journal} {Phys.\ Rev.\ D}\ }\textbf {\bibinfo {volume} {98}},\ \bibinfo
  {pages} {064052} (\bibinfo {year} {2018})},\ \Eprint
  {https://arxiv.org/abs/1807.09056} {arXiv:1807.09056 [gr-qc]} \BibitemShut
  {NoStop}%
\bibitem [{\citenamefont {Ould El~Hadj}\ \emph {et~al.}(2020)\citenamefont
  {Ould El~Hadj}, \citenamefont {Stratton},\ and\ \citenamefont
  {Dolan}}]{OuldElHadj:2019kji}%
  \BibitemOpen
  \bibfield  {author} {\bibinfo {author} {\bibfnamefont {M.}~\bibnamefont {Ould
  El~Hadj}}, \bibinfo {author} {\bibfnamefont {T.}~\bibnamefont {Stratton}},\
  and\ \bibinfo {author} {\bibfnamefont {S.~R.}\ \bibnamefont {Dolan}},\
  }\bibfield  {title} {\bibinfo {title} {{Scattering from compact objects:
  Regge poles and the complex angular momentum method}},\ }\href
  {https://doi.org/10.1103/PhysRevD.101.104035} {\bibfield  {journal} {\bibinfo
   {journal} {Phys. Rev. D}\ }\textbf {\bibinfo {volume} {101}},\ \bibinfo
  {pages} {104035} (\bibinfo {year} {2020})},\ \Eprint
  {https://arxiv.org/abs/1912.11348} {arXiv:1912.11348 [gr-qc]} \BibitemShut
  {NoStop}%
\bibitem [{\citenamefont {Torres}\ \emph {et~al.}(2023)\citenamefont {Torres},
  \citenamefont {Ould El~Hadj}, \citenamefont {Hu},\ and\ \citenamefont
  {Gregory}}]{Torres:2022fyf}%
  \BibitemOpen
  \bibfield  {author} {\bibinfo {author} {\bibfnamefont {T.}~\bibnamefont
  {Torres}}, \bibinfo {author} {\bibfnamefont {M.}~\bibnamefont {Ould
  El~Hadj}}, \bibinfo {author} {\bibfnamefont {S.-Q.}\ \bibnamefont {Hu}},\
  and\ \bibinfo {author} {\bibfnamefont {R.}~\bibnamefont {Gregory}},\
  }\bibfield  {title} {\bibinfo {title} {{Regge pole description of scattering
  by dirty black holes}},\ }\href {https://doi.org/10.1103/PhysRevD.107.064028}
  {\bibfield  {journal} {\bibinfo  {journal} {Phys. Rev. D}\ }\textbf {\bibinfo
  {volume} {107}},\ \bibinfo {pages} {064028} (\bibinfo {year} {2023})},\
  \Eprint {https://arxiv.org/abs/2211.17147} {arXiv:2211.17147 [gr-qc]}
  \BibitemShut {NoStop}%
\bibitem [{\citenamefont {Andersson}(1994)}]{Andersson1994bis}%
  \BibitemOpen
  \bibfield  {author} {\bibinfo {author} {\bibfnamefont {N.}~\bibnamefont
  {Andersson}},\ }\bibfield  {title} {\bibinfo {title} {Complex angular momenta
  and the black-hole glory},\ }\href
  {https://doi.org/10.1088/0264-9381/11/12/014} {\bibfield  {journal} {\bibinfo
   {journal} {Classical and Quantum Gravity}\ }\textbf {\bibinfo {volume}
  {11}},\ \bibinfo {pages} {3003} (\bibinfo {year} {1994})}\BibitemShut
  {NoStop}%
\bibitem [{\citenamefont {D\'ecanini}\ \emph {et~al.}(2003)\citenamefont
  {D\'ecanini}, \citenamefont {Folacci},\ and\ \citenamefont
  {Jensen}}]{Decanini2003}%
  \BibitemOpen
  \bibfield  {author} {\bibinfo {author} {\bibfnamefont {Y.}~\bibnamefont
  {D\'ecanini}}, \bibinfo {author} {\bibfnamefont {A.}~\bibnamefont
  {Folacci}},\ and\ \bibinfo {author} {\bibfnamefont {B.}~\bibnamefont
  {Jensen}},\ }\bibfield  {title} {\bibinfo {title} {Complex angular momentum
  in black hole physics and quasinormal modes},\ }\href
  {https://doi.org/10.1103/PhysRevD.67.124017} {\bibfield  {journal} {\bibinfo
  {journal} {Phys. Rev. D}\ }\textbf {\bibinfo {volume} {67}},\ \bibinfo
  {pages} {124017} (\bibinfo {year} {2003})}\BibitemShut {NoStop}%
\bibitem [{\citenamefont {D\'ecanini}\ and\ \citenamefont
  {Folacci}(2010)}]{Decanini2010}%
  \BibitemOpen
  \bibfield  {author} {\bibinfo {author} {\bibfnamefont {Y.}~\bibnamefont
  {D\'ecanini}}\ and\ \bibinfo {author} {\bibfnamefont {A.}~\bibnamefont
  {Folacci}},\ }\bibfield  {title} {\bibinfo {title} {Regge poles of the
  schwarzschild black hole: A wkb approach},\ }\href
  {https://doi.org/10.1103/PhysRevD.81.024031} {\bibfield  {journal} {\bibinfo
  {journal} {Phys. Rev. D}\ }\textbf {\bibinfo {volume} {81}},\ \bibinfo
  {pages} {024031} (\bibinfo {year} {2010})}\BibitemShut {NoStop}%
\bibitem [{\citenamefont {Dolan}\ and\ \citenamefont
  {Ottewill}(2009)}]{Dolan:2009}%
  \BibitemOpen
  \bibfield  {author} {\bibinfo {author} {\bibfnamefont {S.~R.}\ \bibnamefont
  {Dolan}}\ and\ \bibinfo {author} {\bibfnamefont {A.~C.}\ \bibnamefont
  {Ottewill}},\ }\bibfield  {title} {\bibinfo {title} {On an expansion method
  for black hole quasinormal modes and regge poles},\ }\href
  {https://doi.org/10.1088/0264-9381/26/22/225003} {\bibfield  {journal}
  {\bibinfo  {journal} {Classical and Quantum Gravity}\ }\textbf {\bibinfo
  {volume} {26}},\ \bibinfo {pages} {225003} (\bibinfo {year}
  {2009})}\BibitemShut {NoStop}%
\bibitem [{\citenamefont {D\'ecanini}\ \emph {et~al.}(2010)\citenamefont
  {D\'ecanini}, \citenamefont {Folacci},\ and\ \citenamefont
  {Raffaelli}}]{Decanini2010bis}%
  \BibitemOpen
  \bibfield  {author} {\bibinfo {author} {\bibfnamefont {Y.}~\bibnamefont
  {D\'ecanini}}, \bibinfo {author} {\bibfnamefont {A.}~\bibnamefont
  {Folacci}},\ and\ \bibinfo {author} {\bibfnamefont {B.}~\bibnamefont
  {Raffaelli}},\ }\bibfield  {title} {\bibinfo {title} {Unstable circular null
  geodesics of static spherically symmetric black holes, regge poles, and
  quasinormal frequencies},\ }\href
  {https://doi.org/10.1103/PhysRevD.81.104039} {\bibfield  {journal} {\bibinfo
  {journal} {Phys. Rev. D}\ }\textbf {\bibinfo {volume} {81}},\ \bibinfo
  {pages} {104039} (\bibinfo {year} {2010})}\BibitemShut {NoStop}%
\bibitem [{\citenamefont {D\'ecanini}\ and\ \citenamefont
  {Folacci}(2009)}]{Decanini:2009}%
  \BibitemOpen
  \bibfield  {author} {\bibinfo {author} {\bibfnamefont {Y.}~\bibnamefont
  {D\'ecanini}}\ and\ \bibinfo {author} {\bibfnamefont {A.}~\bibnamefont
  {Folacci}},\ }\bibfield  {title} {\bibinfo {title} {Quasinormal modes of the
  btz black hole are generated by surface waves supported by its boundary at
  infinity},\ }\href {https://doi.org/10.1103/PhysRevD.79.044021} {\bibfield
  {journal} {\bibinfo  {journal} {Phys. Rev. D}\ }\textbf {\bibinfo {volume}
  {79}},\ \bibinfo {pages} {044021} (\bibinfo {year} {2009})}\BibitemShut
  {NoStop}%
\bibitem [{\citenamefont {D\'ecanini}\ \emph
  {et~al.}(2011{\natexlab{a}})\citenamefont {D\'ecanini}, \citenamefont
  {Folacci},\ and\ \citenamefont {Raffaelli}}]{Decanini:2011}%
  \BibitemOpen
  \bibfield  {author} {\bibinfo {author} {\bibfnamefont {Y.}~\bibnamefont
  {D\'ecanini}}, \bibinfo {author} {\bibfnamefont {A.}~\bibnamefont
  {Folacci}},\ and\ \bibinfo {author} {\bibfnamefont {B.}~\bibnamefont
  {Raffaelli}},\ }\bibfield  {title} {\bibinfo {title} {Resonance and
  absorption spectra of the schwarzschild black hole for massive scalar
  perturbations: A complex angular momentum analysis},\ }\href
  {https://doi.org/10.1103/PhysRevD.84.084035} {\bibfield  {journal} {\bibinfo
  {journal} {Phys. Rev. D}\ }\textbf {\bibinfo {volume} {84}},\ \bibinfo
  {pages} {084035} (\bibinfo {year} {2011}{\natexlab{a}})}\BibitemShut
  {NoStop}%
\bibitem [{\citenamefont {Folacci}\ and\ \citenamefont
  {Tamar}(2021)}]{Folacci:2021uld}%
  \BibitemOpen
  \bibfield  {author} {\bibinfo {author} {\bibfnamefont {A.}~\bibnamefont
  {Folacci}}\ and\ \bibinfo {author} {\bibfnamefont {A.}~\bibnamefont
  {Tamar}},\ }\bibfield  {title} {\bibinfo {title} {{Quasinormal mode
  frequencies of Kerr black holes from Regge trajectories}},\ }\href@noop {} {\
   (\bibinfo {year} {2021})},\ \Eprint {https://arxiv.org/abs/2103.01258}
  {arXiv:2103.01258 [gr-qc]} \BibitemShut {NoStop}%
\bibitem [{\citenamefont {D\'ecanini}\ \emph
  {et~al.}(2011{\natexlab{b}})\citenamefont {D\'ecanini}, \citenamefont
  {Esposito-Far\`ese},\ and\ \citenamefont {Folacci}}]{Decanini:2011bis}%
  \BibitemOpen
  \bibfield  {author} {\bibinfo {author} {\bibfnamefont {Y.}~\bibnamefont
  {D\'ecanini}}, \bibinfo {author} {\bibfnamefont {G.}~\bibnamefont
  {Esposito-Far\`ese}},\ and\ \bibinfo {author} {\bibfnamefont
  {A.}~\bibnamefont {Folacci}},\ }\bibfield  {title} {\bibinfo {title}
  {Universality of high-energy absorption cross sections for black holes},\
  }\href {https://doi.org/10.1103/PhysRevD.83.044032} {\bibfield  {journal}
  {\bibinfo  {journal} {Phys. Rev. D}\ }\textbf {\bibinfo {volume} {83}},\
  \bibinfo {pages} {044032} (\bibinfo {year} {2011}{\natexlab{b}})}\BibitemShut
  {NoStop}%
\bibitem [{\citenamefont {Dolan}\ \emph {et~al.}(2012)\citenamefont {Dolan},
  \citenamefont {Oliveira},\ and\ \citenamefont {Crispino}}]{Dolan:2012}%
  \BibitemOpen
  \bibfield  {author} {\bibinfo {author} {\bibfnamefont {S.~R.}\ \bibnamefont
  {Dolan}}, \bibinfo {author} {\bibfnamefont {L.~A.}\ \bibnamefont
  {Oliveira}},\ and\ \bibinfo {author} {\bibfnamefont {L.~C.~B.}\ \bibnamefont
  {Crispino}},\ }\bibfield  {title} {\bibinfo {title} {Resonances of a rotating
  black hole analogue},\ }\href {https://doi.org/10.1103/PhysRevD.85.044031}
  {\bibfield  {journal} {\bibinfo  {journal} {Phys. Rev. D}\ }\textbf {\bibinfo
  {volume} {85}},\ \bibinfo {pages} {044031} (\bibinfo {year}
  {2012})}\BibitemShut {NoStop}%
\bibitem [{\citenamefont {Dolan}\ and\ \citenamefont
  {Ottewill}(2011)}]{Dolan:2011fh}%
  \BibitemOpen
  \bibfield  {author} {\bibinfo {author} {\bibfnamefont {S.~R.}\ \bibnamefont
  {Dolan}}\ and\ \bibinfo {author} {\bibfnamefont {A.~C.}\ \bibnamefont
  {Ottewill}},\ }\bibfield  {title} {\bibinfo {title} {{Wave Propagation and
  Quasinormal Mode Excitation on Schwarzschild Spacetime}},\ }\href
  {https://doi.org/10.1103/PhysRevD.84.104002} {\bibfield  {journal} {\bibinfo
  {journal} {Phys. Rev. D}\ }\textbf {\bibinfo {volume} {84}},\ \bibinfo
  {pages} {104002} (\bibinfo {year} {2011})},\ \Eprint
  {https://arxiv.org/abs/1106.4318} {arXiv:1106.4318 [gr-qc]} \BibitemShut
  {NoStop}%
\bibitem [{\citenamefont {Newton}(1982)}]{Newton:1982qc}%
  \BibitemOpen
  \bibfield  {author} {\bibinfo {author} {\bibfnamefont {R.~G.}\ \bibnamefont
  {Newton}},\ }\href@noop {} {\emph {\bibinfo {title} {{Scattering Theory of
  Waves and Particles}}}},\ \bibinfo {edition} {2nd}\ ed.\ (\bibinfo
  {publisher} {Springer-Verlag, New York},\ \bibinfo {year} {1982})\BibitemShut
  {NoStop}%
\bibitem [{\citenamefont {Watson}(1918)}]{Watson18}%
  \BibitemOpen
  \bibfield  {author} {\bibinfo {author} {\bibfnamefont {G.~N.}\ \bibnamefont
  {Watson}},\ }\bibfield  {title} {\bibinfo {title} {{The diffraction of
  electric waves by the Earth}},\ }\href@noop {} {\bibfield  {journal}
  {\bibinfo  {journal} {Proc.\ R.\ Soc.\ London A}\ }\textbf {\bibinfo {volume}
  {95}},\ \bibinfo {pages} {83} (\bibinfo {year} {1918})}\BibitemShut {NoStop}%
\bibitem [{\citenamefont {Sommerfeld}(1949)}]{Sommerfeld49}%
  \BibitemOpen
  \bibfield  {author} {\bibinfo {author} {\bibfnamefont {A.}~\bibnamefont
  {Sommerfeld}},\ }\href@noop {} {\emph {\bibinfo {title} {Partial Differential
  Equations of Physics}}}\ (\bibinfo  {publisher} {Academic Press, New York},\
  \bibinfo {year} {1949})\BibitemShut {NoStop}%
\bibitem [{\citenamefont {Chandrasekhar}(1983)}]{Chandrasekhar:1985kt}%
  \BibitemOpen
  \bibfield  {author} {\bibinfo {author} {\bibfnamefont {S.}~\bibnamefont
  {Chandrasekhar}},\ }\href@noop {} {\emph {\bibinfo {title} {{The Mathematical
  Theory of Black Holes}}}}\ (\bibinfo  {publisher} {Oxford University Press,
  Oxford},\ \bibinfo {year} {1983})\BibitemShut {NoStop}%
\bibitem [{\citenamefont {Moncrief}(1974{\natexlab{c}})}]{Moncrief:1974gw}%
  \BibitemOpen
  \bibfield  {author} {\bibinfo {author} {\bibfnamefont {V.}~\bibnamefont
  {Moncrief}},\ }\bibfield  {title} {\bibinfo {title} {{Odd-parity stability of
  a Reissner-Nordstrom black hole}},\ }\href
  {https://doi.org/10.1103/PhysRevD.9.2707} {\bibfield  {journal} {\bibinfo
  {journal} {Phys. Rev. D}\ }\textbf {\bibinfo {volume} {9}},\ \bibinfo {pages}
  {2707} (\bibinfo {year} {1974}{\natexlab{c}})}\BibitemShut {NoStop}%
\bibitem [{\citenamefont {Crispino}\ \emph
  {et~al.}(2009{\natexlab{b}})\citenamefont {Crispino}, \citenamefont
  {Higuchi},\ and\ \citenamefont {Oliveira}}]{Crispino:2009zza}%
  \BibitemOpen
  \bibfield  {author} {\bibinfo {author} {\bibfnamefont {L.~C.~B.}\
  \bibnamefont {Crispino}}, \bibinfo {author} {\bibfnamefont {A.}~\bibnamefont
  {Higuchi}},\ and\ \bibinfo {author} {\bibfnamefont {E.~S.}\ \bibnamefont
  {Oliveira}},\ }\bibfield  {title} {\bibinfo {title} {{Electromagnetic
  absorption cross section of Reissner-Nordstrom black holes revisited}},\
  }\href {https://doi.org/10.1103/PhysRevD.80.104026} {\bibfield  {journal}
  {\bibinfo  {journal} {Phys. Rev. D}\ }\textbf {\bibinfo {volume} {80}},\
  \bibinfo {pages} {104026} (\bibinfo {year} {2009}{\natexlab{b}})}\BibitemShut
  {NoStop}%
\bibitem [{\citenamefont {Gunter}\ and\ \citenamefont
  {Chandrasekhar}(1980)}]{Gunter1980}%
  \BibitemOpen
  \bibfield  {author} {\bibinfo {author} {\bibfnamefont {D.~L.}\ \bibnamefont
  {Gunter}}\ and\ \bibinfo {author} {\bibfnamefont {S.}~\bibnamefont
  {Chandrasekhar}},\ }\bibfield  {title} {\bibinfo {title} {A study of the
  coupled gravitational and electromagnetic perturbations to the
  reissner-nordström black hole: the scattering matrix, energy conversion, and
  quasi-normal modes},\ }\href {https://doi.org/10.1098/rsta.1980.0190}
  {\bibfield  {journal} {\bibinfo  {journal} {Philosophical Transactions of the
  Royal Society of London. Series A, Mathematical and Physical Sciences}\
  }\textbf {\bibinfo {volume} {296}},\ \bibinfo {pages} {497} (\bibinfo {year}
  {1980})}\BibitemShut {NoStop}%
\bibitem [{\citenamefont {Kokkotas}\ and\ \citenamefont
  {Schutz}(1988)}]{Kokkotas1988}%
  \BibitemOpen
  \bibfield  {author} {\bibinfo {author} {\bibfnamefont {K.~D.}\ \bibnamefont
  {Kokkotas}}\ and\ \bibinfo {author} {\bibfnamefont {B.~F.}\ \bibnamefont
  {Schutz}},\ }\bibfield  {title} {\bibinfo {title} {Black-hole normal modes: A
  wkb approach. iii. the reissner-nordstr\"om black hole},\ }\href
  {https://doi.org/10.1103/PhysRevD.37.3378} {\bibfield  {journal} {\bibinfo
  {journal} {Phys. Rev. D}\ }\textbf {\bibinfo {volume} {37}},\ \bibinfo
  {pages} {3378} (\bibinfo {year} {1988})}\BibitemShut {NoStop}%
\bibitem [{\citenamefont {Leaver}(1990)}]{Leaver:1990zz}%
  \BibitemOpen
  \bibfield  {author} {\bibinfo {author} {\bibfnamefont {E.~W.}\ \bibnamefont
  {Leaver}},\ }\bibfield  {title} {\bibinfo {title} {{Quasinormal modes of
  Reissner-Nordstrom black holes}},\ }\href
  {https://doi.org/10.1103/PhysRevD.41.2986} {\bibfield  {journal} {\bibinfo
  {journal} {Phys. Rev. D}\ }\textbf {\bibinfo {volume} {41}},\ \bibinfo
  {pages} {2986} (\bibinfo {year} {1990})}\BibitemShut {NoStop}%
\bibitem [{\citenamefont {Andersson}(1993)}]{Andersson1993}%
  \BibitemOpen
  \bibfield  {author} {\bibinfo {author} {\bibfnamefont {N.}~\bibnamefont
  {Andersson}},\ }\bibfield  {title} {\bibinfo {title} {Normal-mode frequencies
  of reissner–nordström black holes},\ }\href
  {https://doi.org/10.1098/rspa.1993.0112} {\bibfield  {journal} {\bibinfo
  {journal} {Proceedings of the Royal Society of London. Series A: Mathematical
  and Physical Sciences}\ }\textbf {\bibinfo {volume} {442}},\ \bibinfo {pages}
  {427} (\bibinfo {year} {1993})}\BibitemShut {NoStop}%
\bibitem [{\citenamefont {Andersson}\ \emph {et~al.}(1994)\citenamefont
  {Andersson}, \citenamefont {Ara\'ujo},\ and\ \citenamefont
  {Schutz}}]{Andersson1994}%
  \BibitemOpen
  \bibfield  {author} {\bibinfo {author} {\bibfnamefont {N.}~\bibnamefont
  {Andersson}}, \bibinfo {author} {\bibfnamefont {M.~E.}\ \bibnamefont
  {Ara\'ujo}},\ and\ \bibinfo {author} {\bibfnamefont {B.~F.}\ \bibnamefont
  {Schutz}},\ }\bibfield  {title} {\bibinfo {title} {Quasinormal modes of
  reissner-nordstr\"om black holes: Phase-integral approach},\ }\href
  {https://doi.org/10.1103/PhysRevD.49.2703} {\bibfield  {journal} {\bibinfo
  {journal} {Phys. Rev. D}\ }\textbf {\bibinfo {volume} {49}},\ \bibinfo
  {pages} {2703} (\bibinfo {year} {1994})}\BibitemShut {NoStop}%
\bibitem [{\citenamefont {Kokkotas}\ and\ \citenamefont
  {Schmidt}(1999)}]{Kokkotas:1999bd}%
  \BibitemOpen
  \bibfield  {author} {\bibinfo {author} {\bibfnamefont {K.~D.}\ \bibnamefont
  {Kokkotas}}\ and\ \bibinfo {author} {\bibfnamefont {B.~G.}\ \bibnamefont
  {Schmidt}},\ }\bibfield  {title} {\bibinfo {title} {{Quasinormal modes of
  stars and black holes}},\ }\href {https://doi.org/10.12942/lrr-1999-2}
  {\bibfield  {journal} {\bibinfo  {journal} {Living Rev. Rel.}\ }\textbf
  {\bibinfo {volume} {2}},\ \bibinfo {pages} {2} (\bibinfo {year} {1999})},\
  \Eprint {https://arxiv.org/abs/gr-qc/9909058} {arXiv:gr-qc/9909058}
  \BibitemShut {NoStop}%
\bibitem [{\citenamefont {Berti}\ and\ \citenamefont
  {Kokkotas}(2003)}]{Berti:2003zu}%
  \BibitemOpen
  \bibfield  {author} {\bibinfo {author} {\bibfnamefont {E.}~\bibnamefont
  {Berti}}\ and\ \bibinfo {author} {\bibfnamefont {K.~D.}\ \bibnamefont
  {Kokkotas}},\ }\bibfield  {title} {\bibinfo {title} {{Asymptotic quasinormal
  modes of Reissner-Nordstrom and Kerr black holes}},\ }\href
  {https://doi.org/10.1103/PhysRevD.68.044027} {\bibfield  {journal} {\bibinfo
  {journal} {Phys. Rev. D}\ }\textbf {\bibinfo {volume} {68}},\ \bibinfo
  {pages} {044027} (\bibinfo {year} {2003})},\ \Eprint
  {https://arxiv.org/abs/hep-th/0303029} {arXiv:hep-th/0303029} \BibitemShut
  {NoStop}%
\bibitem [{\citenamefont {Berti}(2004)}]{Berti:2004md}%
  \BibitemOpen
  \bibfield  {author} {\bibinfo {author} {\bibfnamefont {E.}~\bibnamefont
  {Berti}},\ }\bibfield  {title} {\bibinfo {title} {{Black hole quasinormal
  modes: Hints of quantum gravity?}},\ }\href@noop {} {\bibfield  {journal}
  {\bibinfo  {journal} {Conf. Proc. C}\ }\textbf {\bibinfo {volume}
  {0405132}},\ \bibinfo {pages} {145} (\bibinfo {year} {2004})},\ \Eprint
  {https://arxiv.org/abs/gr-qc/0411025} {arXiv:gr-qc/0411025} \BibitemShut
  {NoStop}%
\bibitem [{\citenamefont {Leaver}(1985)}]{Leaver:1985ax}%
  \BibitemOpen
  \bibfield  {author} {\bibinfo {author} {\bibfnamefont {E.~W.}\ \bibnamefont
  {Leaver}},\ }\bibfield  {title} {\bibinfo {title} {{An Analytic
  representation for the quasi normal modes of Kerr black holes}},\ }\href
  {https://doi.org/10.1098/rspa.1985.0119} {\bibfield  {journal} {\bibinfo
  {journal} {Proc. Roy. Soc. Lond. A}\ }\textbf {\bibinfo {volume} {402}},\
  \bibinfo {pages} {285} (\bibinfo {year} {1985})}\BibitemShut {NoStop}%
\bibitem [{\citenamefont {Leaver}(1986)}]{leaver1986solutions}%
  \BibitemOpen
  \bibfield  {author} {\bibinfo {author} {\bibfnamefont {E.~W.}\ \bibnamefont
  {Leaver}},\ }\bibfield  {title} {\bibinfo {title} {Solutions to a generalized
  spheroidal wave equation: Teukolsky's equations in general relativity, and
  the two-center problem in molecular quantum mechanics},\ }\href@noop {}
  {\bibfield  {journal} {\bibinfo  {journal} {Journal of mathematical physics}\
  }\textbf {\bibinfo {volume} {27}},\ \bibinfo {pages} {1238} (\bibinfo {year}
  {1986})}\BibitemShut {NoStop}%
\bibitem [{\citenamefont {Majumdar}\ and\ \citenamefont
  {Panchapakesan}(1989)}]{mp}%
  \BibitemOpen
  \bibfield  {author} {\bibinfo {author} {\bibfnamefont {B.}~\bibnamefont
  {Majumdar}}\ and\ \bibinfo {author} {\bibfnamefont {N.}~\bibnamefont
  {Panchapakesan}},\ }\bibfield  {title} {\bibinfo {title} {{Schwarzschild
  black-hole normal modes using the Hill determinant}},\ }\href
  {https://doi.org/10.1103/PhysRevD.40.2568} {\bibfield  {journal} {\bibinfo
  {journal} {Phys.\ Rev.\ D}\ }\textbf {\bibinfo {volume} {40}},\ \bibinfo
  {pages} {2568} (\bibinfo {year} {1989})}\BibitemShut {NoStop}%
\bibitem [{\citenamefont {Onozawa}\ \emph {et~al.}(1996)\citenamefont
  {Onozawa}, \citenamefont {Mishima}, \citenamefont {Okamura},\ and\
  \citenamefont {Ishihara}}]{Onozawa:1995vu}%
  \BibitemOpen
  \bibfield  {author} {\bibinfo {author} {\bibfnamefont {H.}~\bibnamefont
  {Onozawa}}, \bibinfo {author} {\bibfnamefont {T.}~\bibnamefont {Mishima}},
  \bibinfo {author} {\bibfnamefont {T.}~\bibnamefont {Okamura}},\ and\ \bibinfo
  {author} {\bibfnamefont {H.}~\bibnamefont {Ishihara}},\ }\bibfield  {title}
  {\bibinfo {title} {{Quasinormal modes of maximally charged black holes}},\
  }\href {https://doi.org/10.1103/PhysRevD.53.7033} {\bibfield  {journal}
  {\bibinfo  {journal} {Phys. Rev. D}\ }\textbf {\bibinfo {volume} {53}},\
  \bibinfo {pages} {7033} (\bibinfo {year} {1996})},\ \Eprint
  {https://arxiv.org/abs/gr-qc/9603021} {arXiv:gr-qc/9603021} \BibitemShut
  {NoStop}%
\bibitem [{\citenamefont {Onozawa}\ \emph {et~al.}(1997)\citenamefont
  {Onozawa}, \citenamefont {Okamura}, \citenamefont {Mishima},\ and\
  \citenamefont {Ishihara}}]{Onozawa:1996ba}%
  \BibitemOpen
  \bibfield  {author} {\bibinfo {author} {\bibfnamefont {H.}~\bibnamefont
  {Onozawa}}, \bibinfo {author} {\bibfnamefont {T.}~\bibnamefont {Okamura}},
  \bibinfo {author} {\bibfnamefont {T.}~\bibnamefont {Mishima}},\ and\ \bibinfo
  {author} {\bibfnamefont {H.}~\bibnamefont {Ishihara}},\ }\bibfield  {title}
  {\bibinfo {title} {{Perturbing supersymmetric black hole}},\ }\href
  {https://doi.org/10.1103/PhysRevD.55.R4529} {\bibfield  {journal} {\bibinfo
  {journal} {Phys. Rev. D}\ }\textbf {\bibinfo {volume} {55}},\ \bibinfo
  {pages} {4529} (\bibinfo {year} {1997})},\ \Eprint
  {https://arxiv.org/abs/gr-qc/9606086} {arXiv:gr-qc/9606086} \BibitemShut
  {NoStop}%
\bibitem [{\citenamefont {Konoplya}(2003)}]{Konoplya:2003dd}%
  \BibitemOpen
  \bibfield  {author} {\bibinfo {author} {\bibfnamefont {R.~A.}\ \bibnamefont
  {Konoplya}},\ }\bibfield  {title} {\bibinfo {title} {{Gravitational
  quasinormal radiation of higher dimensional black holes}},\ }\href
  {https://doi.org/10.1103/PhysRevD.68.124017} {\bibfield  {journal} {\bibinfo
  {journal} {Phys. Rev. D}\ }\textbf {\bibinfo {volume} {68}},\ \bibinfo
  {pages} {124017} (\bibinfo {year} {2003})},\ \Eprint
  {https://arxiv.org/abs/hep-th/0309030} {arXiv:hep-th/0309030} \BibitemShut
  {NoStop}%
\bibitem [{\citenamefont {Decanini}\ \emph {et~al.}(2003)\citenamefont
  {Decanini}, \citenamefont {Folacci},\ and\ \citenamefont
  {Jensen}}]{Decanini:2002ha}%
  \BibitemOpen
  \bibfield  {author} {\bibinfo {author} {\bibfnamefont {Y.}~\bibnamefont
  {Decanini}}, \bibinfo {author} {\bibfnamefont {A.}~\bibnamefont {Folacci}},\
  and\ \bibinfo {author} {\bibfnamefont {B.}~\bibnamefont {Jensen}},\
  }\bibfield  {title} {\bibinfo {title} {{Complex angular momentum in black
  hole physics and the quasinormal modes}},\ }\href
  {https://doi.org/10.1103/PhysRevD.67.124017} {\bibfield  {journal} {\bibinfo
  {journal} {Phys. Rev. D}\ }\textbf {\bibinfo {volume} {67}},\ \bibinfo
  {pages} {124017} (\bibinfo {year} {2003})},\ \Eprint
  {https://arxiv.org/abs/gr-qc/0212093} {arXiv:gr-qc/0212093} \BibitemShut
  {NoStop}%
\bibitem [{\citenamefont {Abramowitz}\ and\ \citenamefont
  {Stegun}(1965)}]{AS65}%
  \BibitemOpen
  \bibfield  {author} {\bibinfo {author} {\bibfnamefont {M.}~\bibnamefont
  {Abramowitz}}\ and\ \bibinfo {author} {\bibfnamefont {I.~A.}\ \bibnamefont
  {Stegun}},\ }\href@noop {} {\emph {\bibinfo {title} {Handbook of Mathematical
  Functions}}}\ (\bibinfo  {publisher} {Dover, New-York},\ \bibinfo {year}
  {1965})\BibitemShut {NoStop}%
\bibitem [{\citenamefont {Leite}\ \emph {et~al.}(2019)\citenamefont {Leite},
  \citenamefont {Benone},\ and\ \citenamefont {Crispino}}]{Leite:2019eis}%
  \BibitemOpen
  \bibfield  {author} {\bibinfo {author} {\bibfnamefont {L.~C.~S.}\
  \bibnamefont {Leite}}, \bibinfo {author} {\bibfnamefont {C.~L.}\ \bibnamefont
  {Benone}},\ and\ \bibinfo {author} {\bibfnamefont {L.~C.~B.}\ \bibnamefont
  {Crispino}},\ }\bibfield  {title} {\bibinfo {title} {{On-axis scattering of
  scalar fields by charged rotating black holes}},\ }\href
  {https://doi.org/10.1016/j.physletb.2019.06.027} {\bibfield  {journal}
  {\bibinfo  {journal} {Phys. Lett. B}\ }\textbf {\bibinfo {volume} {795}},\
  \bibinfo {pages} {496} (\bibinfo {year} {2019})},\ \Eprint
  {https://arxiv.org/abs/1907.04746} {arXiv:1907.04746 [gr-qc]} \BibitemShut
  {NoStop}%
\bibitem [{\citenamefont {Pani}\ \emph {et~al.}(2013)\citenamefont {Pani},
  \citenamefont {Berti},\ and\ \citenamefont {Gualtieri}}]{Pani:2013ija}%
  \BibitemOpen
  \bibfield  {author} {\bibinfo {author} {\bibfnamefont {P.}~\bibnamefont
  {Pani}}, \bibinfo {author} {\bibfnamefont {E.}~\bibnamefont {Berti}},\ and\
  \bibinfo {author} {\bibfnamefont {L.}~\bibnamefont {Gualtieri}},\ }\bibfield
  {title} {\bibinfo {title} {{Gravitoelectromagnetic Perturbations of
  Kerr-Newman Black Holes: Stability and Isospectrality in the Slow-Rotation
  Limit}},\ }\href {https://doi.org/10.1103/PhysRevLett.110.241103} {\bibfield
  {journal} {\bibinfo  {journal} {Phys. Rev. Lett.}\ }\textbf {\bibinfo
  {volume} {110}},\ \bibinfo {pages} {241103} (\bibinfo {year} {2013})},\
  \Eprint {https://arxiv.org/abs/1304.1160} {arXiv:1304.1160 [gr-qc]}
  \BibitemShut {NoStop}%
\end{thebibliography}%

\end{document}